\documentclass[a4paper,11pt]{article}
\usepackage{jheppub} 

\usepackage[utf8]{inputenc}
\usepackage{slashed,cancel}
\usepackage[section]{placeins}
\usepackage{bm}
\usepackage{pifont}
\usepackage{tabularx}    
\usepackage{adjustbox} 
\usepackage[english]{babel}
\usepackage{multirow}
\usepackage[normalem]{ulem} 
\usepackage{booktabs}
\usepackage{braket}
\usepackage{MnSymbol}
\usepackage{graphicx}
\usepackage{float}
\usepackage{subcaption}
\usepackage{multirow}
\usepackage[skins,theorems]{tcolorbox}
\usepackage[capitalise]{cleveref}
\usepackage{orcidlink} 

\tcbset{highlight math style={enhanced,
  colframe=red,colback=white,arc=0pt,boxrule=1pt}}

\definecolor{rossoc}{cmyk}{0,1,1,0.2}
\definecolor{dgreen}{cmyk}{1,0,1,0.3}

\newcommand{\secref}[1]{Section~\ref{#1}}

\newcommand{\ov}[1]{\overline{#1}}
\newcommand{\sign}[1]{\text{sgn}(#1)} 
\newcommand{\Rho}{\mathrm{P}}

\begin{document}
\title{\boldmath Do neutrinos dream in 5D? \\
\Large{Towards a comprehensive extra-dimensional neutrino phenomenology} 
}

\author[a]{Arturo~de Giorgi~\orcidlink{0000-0002-9260-5466}\,,}
\author[a]{Dhruv Pasari~\orcidlink{0009-0007-1283-1492}\,,}
\author[a]{and Jessica~Turner~\orcidlink{0000-0002-9679-5252}}

\affiliation[a]{Institute for Particle Physics Phenomenology, Department of Physics,\\ Durham University, Durham DH1 3LE, U.K.}

\emailAdd{arturo.de-giorgi@durham.ac.uk}
\emailAdd{dhruv.pasari@durham.ac.uk}
\emailAdd{jessica.turner@durham.ac.uk}

\preprint{IPPP/25/84}

\abstract{This paper provides a comprehensive overview of neutrino masses and mixing in Large Extra Dimension scenarios, focusing on the phenomenological impact of a five-dimensional (5D) bulk fermion. In a flat extra dimension compactified on an $S^1/\mathbb{Z}_2$ orbifold, this fermion manifests as a Kaluza-Klein tower of right-handed neutrinos in the 4D effective theory. We systematically investigate four distinct scenarios for mass generation, considering both Dirac and Majorana mass terms originating from either the bulk or the 3-brane. For each case, we analyse the consequences for neutrino oscillations in a vacuum and in matter, deriving the resulting mass spectra and mixing patterns. By comparing these theoretical predictions with experimental data, we explore the constraints on the large extra dimensions' parameters.
}

\maketitle 
\flushbottom 

\section{Introduction}
One of the most compelling pieces of evidence for physics beyond the Standard Model (SM) comes from the neutrino sector. The discovery of neutrino oscillations confirmed that neutrinos have mass and mix \cite{Super-Kamiokande:1998kpq,SNO:2002tuh}, a phenomenon that the minimal SM cannot accommodate. While many theories can generate neutrino mass, the simplest extension of the SM is to introduce right-handed neutrinos. As SM gauge singlets, these particles allow for a Dirac mass term, and light neutrinos can acquire mass via the Higgs mechanism. However, generating the observed sub-eV neutrino masses requires Yukawa couplings as small as \( y_\nu \sim m_\nu/v \sim 10^{-13}\text{--}10^{-12} \), a solution often regarded as contrived.
Because right-handed neutrinos are gauge singlets, no symmetry, beyond lepton number, which is only an accidental symmetry of the SM, protects them against Majorana mass terms. Allowing lepton-number violation leads to the {type-I seesaw} mechanism \cite{Minkowski:1977sc,Gell-Mann:1979vob,Yanagida:1979as,Mohapatra:1980yp}, which naturally yields tiny light-neutrino masses \( m_\nu \simeq -\, m_D^{\mathsf T} M^{-1} m_D \) and active–sterile mixing \( \Theta \simeq m_D M^{-1} \), where \( m_D \) is the Dirac mass matrix and \( M \) the Majorana mass matrix of the right-handed neutrinos. This points to a new mass scale associated with typically heavy right-handed neutrinos. 
The viability of this idea hinges on the right-handed neutrinos being SM gauge singlets, prompting the question: is there a deeper principle that produces such fields?
A compelling geometrical answer is offered by theories with extra spatial dimensions, first proposed in the seminal works of Kaluza and Klein~(KK)~\cite{Kaluza:1921tu,Klein:1926tv} to unify electromagnetism and gravity.  Since then, they have found a revival in the early 2000s due to their ability to elegantly solve several SM puzzles, such as the hierarchy~\cite{Antoniadis:1990ew,Arkani-Hamed:1998jmv,Antoniadis:1998ig,Randall:1999ee,Randall:1999vf} and flavour puzzles \cite{Arkani-Hamed:1999ylh,Kaplan:2001ga}. In many constructions, the SM is localised on a 3-brane in a higher-dimensional spacetime, whereas gravity, and possibly SM-singlet fields, propagate in the bulk. In such setups, any fermion living in the bulk must be an SM singlet, providing a natural origin for the right-handed neutrinos invoked by the type-I seesaw. By contrast, in models like universal extra dimensions, the entire SM propagates in the bulk \cite{Appelquist:2000nn}.
Concrete realisations have been explored in both flat and warped backgrounds. The flat case for Dirac and Majorana neutrino masses has been first studied in \cite{Dienes:1998sb,Arkani-Hamed:1998wuz,Dvali:1999cn,Lukas:2000rg}, while warped geometries, where small neutrino masses arise from wavefunction localisation, were examined in \cite{Grossman:1999ra,Huber:2003sf,Fong:2011xh}.

Although neutrino masses in extra-dimensional models have been explored in specific cases \cite{Appelquist:2000nn,Dienes:1998sb,Arkani-Hamed:1998wuz,Dvali:1999cn,Lukas:2000rg,Grossman:1999ra,Huber:2003sf,Fong:2011xh}, a comprehensive analysis of LEDs with neutrino mass generation is lacking. Here, we fill this gap by examining four distinct scenarios under a unified 5D framework, thereby illustrating how each scenario leads to unique phenomenological signatures. Specifically, we distinguish between Dirac vs. Majorana masses, each of which can reside either in the 5D bulk or on the 4D brane. One could question the possibility of more than one extra dimension, as would arise in a string-inspired framework. It is worth noting that in such scenarios the bounds are generically stronger, since the conversion to KK modes becomes more efficient when additional extra dimensions are present~\cite{Elacmaz:2025ihm}.

The structure of the work is as follows: in Section~\ref{sec:theory}, we begin by introducing the theoretical framework and the notation. We go through the derivation of the wavefunctions and the masses of the KK-modes for the different scenarios, highlighting their differences and possible generalisations to warped geometries. In Section~\ref{sec:phenomenology}, we move to the analysis of the associated phenomenology in neutrino oscillations. We study each case separately and describe the salient features. In each Section, we investigate the impact of extra dimensions, both in vacuum and in matter. In Section~\ref{sec:setup}, we analyse the oscillation data in the context of Daya Bay and MINOS/MINOS+ for each case and derive the qualitative bounds on the relevant parameters in question. Finally, in Section~\ref{sec:conclusions} we draw our conclusions and outline future research directions.
\section{5D Bulk Fermions}
\label{sec:theory}
We begin by reviewing the formalism and notation employed in this work.\footnote{Some useful and complementary details can be found, e.g. in Refs.~\cite{Chang:1999nh,Ponton:2012bi}.}
We work in an extra-dimensional model with $D=5$ whose fifth dimension, $y$, is compactified on $S^1/\mathbb{Z}_2$ with radius $R$. We consider the minimal setup with a fermion $\Psi$ propagating in the bulk. 
We consider the bulk action given by~\cite{Lukas:2000rg}:
\begin{equation}
    \label{eq:bulk-action}S_{\Psi,\text{bulk}}=\int d^4x\int \limits_{-\pi R}^{\pi R}dy\,\sqrt{G}\left[i\ov{\Psi} \Gamma^M\nabla_M\Psi-\sign{y} M_D \ov{\Psi}\Psi-\frac{M_J}{2}\ov{\Psi}\Psi^c\right]\,,
\end{equation}
where $G$ is the determinant of the 5D  metric, $\nabla_M$ is the covariant derivative (needed if spacetime is curved), $M_{J}$ is a Majorana mass, $M_D$ is a Dirac mass, and the $\sign{\cdot}$ function is necessary to make the mass term compatible with the orbifold symmetry. Throughout this work, we will employ capital Latin letters for the $5D$ spacetime indices, e.g. $M=0,1,2,3,5$, and Greek letters for $4D$ spacetime indices, e.g. $\mu=0,1,2,3$.
The $\mathrm{sgn}(y)$ function, which appears in the Dirac term, can be generated by some UV dynamics, for instance, through couplings of the bulk fermion to a pseudoscalar bulk scalar field. 

We adopt the following representation for the gamma matrices:
\begin{align}
    &\Gamma^\mu=\gamma^\mu\,, && \Gamma^5=i\gamma^5\,, &&\{\Gamma^A,\,\Gamma^B\}=2\eta^{AB}\,.
\end{align}
Throughout this work, we will employ the Weyl basis so that $\gamma^\mu$ can be written via $\sigma^\mu\equiv (1,\sigma^i)$ and $\ov{\sigma}^\mu\equiv (1,-\sigma^i)$, where $\sigma^i$ are the Pauli matrices. In particular, one has
\begin{align}
    & \gamma^\mu=\begin{pmatrix}
        0 & \sigma^\mu\\
        \ov{\sigma}^\mu & 0
    \end{pmatrix}\,, &&\gamma^5=\begin{pmatrix}
        -1 & 0\\
        0 & 1
    \end{pmatrix}\,.
\end{align}
In $5$D it is \emph{not} possible to construct an analogue of $\gamma_5$ that anticommutes with all five $\Gamma^A$, and hence no 5D chiral projectors can be defined. 
Nevertheless, chirality can be consistently assigned by making use of the orbifold symmetry, which allows to assign each field an orbifold parity
\begin{equation}
    \label{eq:parity5}\gamma^5 \Psi(x,-y)=\pm\Psi(x,y)\,.
\end{equation}
Without loss of generality, we will assign even orbifold parity ($+$) to $\Psi$.
The bulk fermion $\Psi$ can be decomposed in terms of Weyl spinors
\begin{equation}
    \Psi(x,y)=\begin{pmatrix}
        \psi_L(x,y)\\
        \psi_R(x,y)
    \end{pmatrix}\,.
\end{equation}
As anticipated, the orbifold parity defined in Eq.~\eqref{eq:parity5} allows to distinguish the two components of $\Psi$
\begin{align}
   \label{eq:parity} &\psi_L(x,-y)=-\psi_L(x,y)\,, &\psi_R(x,-y)=+\psi_R(x,y)\,,
\end{align}
thus making the theory chiral.
From the 4D perspective, they can be identified as left- and right-handed fields, respectively.
The orbifold parity conditions defined in Eq.~\eqref{eq:parity} naturally induce vanishing Dirichlet and Neumann boundary conditions on the branes for the odd and even fields, respectively,
\begin{align}
\label{eq:null-boundary}
    &\psi_L(x,0)=\psi_L(x,\pm \pi R)=0\,, &\partial_5\psi_R(x,0)=\partial_5\psi_R(x,\pm \pi R)=0\,.
\end{align}
Consequently, only $\psi_R$ is localised on the branes.

Finally, we consider charge conjugation defined as in four dimensions:
\begin{align}
    &\Psi^c\equiv C(\ov{\Psi})^T=C\gamma^0\Psi^\star\,, &C=i\gamma^2\gamma^0\,.
\end{align}
In the following, we will use the notation $\Psi_{L,R}^c\equiv (\Psi_{L,R})^c$,
where
\begin{align}
    &\Psi_L=\begin{pmatrix}
        \psi_L(x,y)\\
       0
    \end{pmatrix}\,, &&\Psi_R=\begin{pmatrix}
        0\\
       \psi_R(x,y)
    \end{pmatrix}\,,
\end{align}
so that $\Psi_{L,R}^c$ transforms as a 4D right-(left-) handed field, respectively. In the next Section, we will derive the equations of motion and the 4D EFT for the different cases of interest.

\subsection{Kaluza Klein Decomposition in Flat Spacetime}
We first present the general curved spacetime formalism before specialising to flat space. The fermionic Lagrangian in curved spacetime can be written in a torsionless background as
\begin{equation}
    \mathcal{L}\supset i\ov{\Psi}e_A^M \Gamma^A\nabla_M\Psi\,,
\end{equation}
where $e_M^A$ is the vielbein,
\begin{align}
    &\nabla_M\equiv \partial_M +\Omega_M\,, &\Omega_M=\frac{1}{8}\omega^{AB}_M[\Gamma_A,\Gamma_B]\,, &&\left(\omega^a_b\right)_M\equiv e^A_N e^{N}_{B,M}+e^{A}_N e^O_B \Gamma^{N}_{MO}\,,
\end{align}
$\omega^{AB}_M$ is the spin-connection and $\Gamma^{N}_{MO}$ is the Christoffel symbol.
In flat spacetime, the spin connection vanishes, $\Omega_M =0$, and the covariant derivative reduces to the usual four-derivative. The equations of motion (EOM) of the 5D spinor read
\begin{equation}
    i\Gamma^A\partial_A \Psi=\sign{y} M_D\Psi+\frac{M_J}{2}\Psi^c\,.
\end{equation}
Projecting out the 4D chiralities, the following system of EOM is generated
\begin{align}
\label{eq:EOMs-general}
    i\slashed{\partial} \Psi_R - \gamma^5 \partial_5 \Psi_L 
    &= \sign{y} M_D \Psi_L + \frac{M_J}{2} \Psi_R^c \,, \nonumber\\
    i\slashed{\partial} \Psi_L - \gamma^5 \partial_5 \Psi_R 
    &= \sign{y} M_D \Psi_R + \frac{M_J}{2} \Psi_L^c\,.
\end{align}
To recover the 4D effective field theory (EFT), the fifth dimension has to be integrated out. This is achieved by solving the 5D EOMs and substituting the solutions back into the action. Physically, the 5D kinetic term manifests as an effective mass term in the 4D effective theory, such that the EOMs of Eq.~\eqref{eq:EOMs-general} reduce to the standard 4D Dirac equation.
Denoted by $\{\chi_n(y)\}_n$ a set of eigenfunctions (hereafter called ``wavefunctions"~(WFs)) for the fifth dimension, one can directly employ the Kaluza-Klein~(KK) decomposition of the field:
\begin{align}
    \label{eq:KK-decomposition}&\psi_L(x,y)=\frac{1}{\sqrt{V}}\sum\limits_{n=-\infty}^\infty \psi_{L,n}(x) \xi_n(y)\,, &\psi_R(x,y)=\frac{1}{\sqrt{V}}\sum\limits_{n=-\infty}^\infty \psi_{R,n}(x) \chi_n(y)\,,
\end{align}
where \( n\in\mathbb{Z} \) labels the Kaluza–Klein excitation level and $V=2\pi R$ is the volume of the extra dimension whose inclusion is to ensure the correct normalisation of the WFs
\begin{equation}
\label{eq:normalisation}
    \frac{1}{V}\int\limits_{-\pi}^\pi dy\, \xi_{n}(y)\xi_{m}(y)=\frac{1}{V}\int\limits_{-\pi}^\pi dy\, \chi_{n}(y)\chi_{m}(y)=\delta_{nm}\,.
\end{equation}
Technically, consistency of the 4D EFT requires the KK decomposition to be truncated at some $n_\text{max}\equiv N$. KK-modes with larger $n$ are associated with heavier modes, and their masses must not exceed the UV cut-off.
This is expected to be of the order of the extra-dimensional Planck mass, $M_5$,\footnote{An example of explicit computation of the cut-off can be found, e.g. in Ref.~\cite{deGiorgi:2021xvm}.} 
\begin{equation}
    M_5 = \left(\frac{\bar{M}_P^2}{V}\right)^{1/3}\approx 6\times 10^8~\text{GeV}~\left(\frac{\mu\text{m}}{R}\right)^{1/3}\approx  10^9~\text{GeV}~\left(\frac{1/R}{\text{eV}}\right)^{1/3}\,,
\end{equation}
where $\bar{M}_P$ is the reduced 4D Planck mass.
The dominant constraint on the size of the extra dimension comes from studying modifications of Newton's law at different length scales~\cite{Hoskins:1985tn,Bordag:2001qi,Mostepanenko:2001fx,Chiaverini:2002cb,Long:2003dx,Chen:2014oda,Tan:2016vwu,Lee:2020zjt}, leading to $1/R\gtrsim 10^{-2}$~eV.~\footnote{Complementary bounds, yet weaker for this scenario, come from a range of astrophysical probes~\cite{Hannestad:2001jv,Hannestad:2001xi,Hannestad:2003yd,Cembranos:2017vgi,Fiorillo:2025zzx,Hardy:2025ajb,Garcia-Cely:2025ula}.}
In practice, the number of modes that can be included is so large that the limit $N\to\infty$ yields stable results and simplifies the analytical expressions. In the rest of the work, we will derive results keeping $N$ finite and take the limit $N\to\infty$ at the very end. 
\subsubsection{Massless Bulk Fermion}
\label{sec:theory-dirac-brane}
We start by considering a massless bulk fermion in flat spacetime, $M_J=M_D=0$,
\begin{equation}
  S_{\Psi,\text{bulk}}=\int d^4x\int \limits_{-\pi R}^{\pi R}dy\,\sqrt{G}\left[i\ov{\Psi} \Gamma^A\partial_A\Psi\right]\,.
\end{equation}
The wavefunctions can be determined by solving the EOMs of Eq.~\eqref{eq:EOMs-general}
\begin{align}
    &\partial_5 \xi_{n}= -\mu_n\,\chi_{n}\,, &&\partial_5 \chi_{n}= \mu_n\,\xi_{n}\,,
\end{align}
where $\mu_n$ is an effective mass which must be determined.
The equations are equivalent to
\begin{align}
    &\partial_5^2 \xi_{n}= -\mu_n^2 \xi_n\,, &\partial_5^2 \chi_{n}= -\mu_n^2 \chi_n\,,
\end{align}
whose solutions are Fourier modes with frequency $\mu_n$.
Due to the parity of $\psi_{L,R}$, one can unify negative and positive $n,-n$ for $n\neq 0$ modes and define
\begin{align}
    &\xi_{n}=c_{\xi,n}\sin(\mu_n y)\,, &&\chi_{n}=c_{\chi,n}\cos(\mu_n y)\,,
\end{align}
thus restricting the sum to $n>0$. The solutions automatically satisfy the boundary conditions at $y=0$.
The eigenvalues are determined by the BCs at $y=\pm \pi$, yielding
\begin{equation}
    \mu_n=\frac{n}{R}\,.
\end{equation}
Finally, the normalisation of the solutions fixes the remaining free coefficients
\begin{align}
    &\xi_{0}(y)=0\,, &&\xi_{n>0}(y)=\sqrt{2}\,\sin(\mu_n y)\,, &&\chi_{0}(y)=1\,, &&\chi_{n>0}(y)=\sqrt{2}\,\cos(\mu_n y)\,,
\end{align}
where the orbifold parity forces the left-handed zero mode to vanish, 
$\xi_{0}(y)=0$, leaving only the right-handed zero mode, $\chi_{0}(y)=1$. The extra-dimensional kinetic term converts into a 4D tower of Dirac mass terms:
\begin{align}
    \int dy\,\ov{\Psi} i\Gamma^5\partial_5\Psi=-\sum\limits_{n=1}^N\,\mu_n\left[\ov{\psi_{R,n}}\psi_{L,n} +\ov{\psi_{L,n}}\psi_{R,n}\right]\,.
\end{align}
In the following, we examine how the bulk fermion WFs and KK spectrum are modified by a constant Majorana and Dirac mass.
\subsubsection{Majorana Bulk Mass}
\label{sec:theory-majorana-bulk}
We turn now to the case of a Majorana bulk term, $M_J\neq 0$, $M_D=0$,
\begin{equation}
   S_{\Psi,\text{bulk}}=\int d^4x\int \limits_{-\pi R}^{\pi R}dy\,\sqrt{G}\left[i\ov{\Psi} \Gamma^A\partial_A\Psi-\frac{M_J}{2}\ov{\Psi}\Psi^c\right]\,.
\end{equation}
Since the bulk Majorana mass $M_J$ is constant along $y$, the KK WFs are identical to the massless case. 
However, each KK mode now acquires an additional Majorana mass contribution in the
4D effective theory.
The 4D EFT contribution from the Majorana mass term so obtained reads
\begin{align}
    S_{\text{free}}^{\Psi}&\supset-\int d^4x\left[\sum\limits_{n=1}^N\,\mu_n~\ov{\psi_{L,n}}\psi_{R,n}+\frac{M_J}{2}\sum\limits_{n=0}^N\left(\ov{\psi_{L,n}}\psi_{L,n}^c+\ov{\psi_{R,n}}\psi_{R,n}^c\right)+\text{h.c.}\right]\,.
\end{align}
For $n>0$, the KK-mass terms and the Majorana one can be conveniently combined by employing the fields
\begin{align}
    &\psi_{R,n}\equiv \frac{\psi_{1,n}+\psi_{2,n}}{\sqrt{2}}\,, && \psi_{L,n}\equiv \frac{\psi_{1,n}^c-\psi_{2,n}^c}{\sqrt{2}}\,,
\end{align}
where $\psi_{1,2,n}$ are right-handed fields.
In this basis, the KK mass term takes the form:
\begin{align}
    \label{eq:action-smart-n} S\supset-\frac{1}{2}\int d^4x&\left\{M_J\ov{\psi_{R,0}^c}\psi_{R,0}+\sum\limits_{n=1}^N\left[\ov{\psi_{1,n}^c}\psi_{1,n}\left(M_J+\mu_n\right)+\ov{\psi_{2,n}^c}\psi_{2,n}\left(M_J-\mu_n\right)\right]+\text{h.c.}\right\}\,,
\end{align}
where the mass matrix for the KK-modes is diagonal. This will prove very convenient for studying the spectrum of the theory once the active SM neutrino is included. 

\subsubsection{Dirac Bulk Mass}
\label{sec:theory-dirac-bulk}
Next, we consider the case of a Dirac bulk term in a flat spacetime background,\footnote{A detailed analysis for the warped counterpart can be found, e.g. in Ref.~\cite{Grossman:1999ra}.} $M_J= 0$, $M_D\neq0$, whose action reads
\begin{equation}
    S_{\Psi,\text{bulk}}=\int d^4x\int \limits_{-\pi R}^{\pi R}dy\,\left[i\ov{\Psi} \Gamma^A\partial_A\Psi-\sign{y} M_D \ov{\Psi}\Psi\right]\,.
\end{equation}
The EOMs of Eq.~\eqref{eq:EOMs-general} can be reduced to
\begin{align}
    &\partial_5 \chi_n = \mu_{D,n} \xi_n-M_D \sign{y}\chi_n\,,\\
    &\partial_5 \xi_n = -\left(\mu_{D,n}\chi_n-M_D \sign{y}\xi_n\right)\,,
\end{align}
where for later convenience we denoted the KK masses by $\mu_{D,n}$.
Let us begin by assuming $\mu_{D,n}\neq 0$. By further differentiating the two equations, in the intervals $y>0$ and $y<0$, the problem reduces to the decoupled equations
\begin{align}
    &\partial_5^2 \chi_n =-(\mu_{D,n}^2-M_D^2)\chi_n\equiv -\mu_n^2\chi_n\,,\\
    &\partial_5^2 \xi_n \equiv-(\mu_{D,n}^2-M_D^2)\xi_n=-\mu_n^2\xi_n\,,
\end{align}
where we defined $\mu_n$ via
\begin{equation}
    \label{eq:mass-bulk-dirac}\mu_{D,n}^2=\mu_n^2+M_D^2\,.
\end{equation}
The most general solution for both of them is a linear combination of sines and cosines. Given the discontinuity of $\sign{y}$ at the origin, we rely on different boundary conditions. We can use parity to fix the odd solution and the spectrum, and then derive the even solution by means of the EOMs.
General solutions to the above equations read 
\begin{align}
    &\xi_{n}(y)=c_{\xi,n}\sin(\mu_n y)\,, &\chi_n(y)=c_{\chi,n} \cos(\mu_n y)+d_{\chi,n} \sin(\mu_n |y|)\,.
\end{align}
Boundary conditions at $y=\pi R$ force $\mu_n=n/R$.
By matching to the EOMs one can fix $c_{\chi,n}$ and $d_{\chi,n}$ and find
\begin{equation}
    \chi_n(y)=- \frac{c_{\xi,n}}{\sqrt{\mu_n^2+M_D^2}}\left[\mu_n\cos(\mu_n y)-M_D \sin(\mu_n |y|)\right]\,.
\end{equation}
The normalisation condition of Eq.~\eqref{eq:normalisation} fixes the coefficient to 
\begin{equation}
    c_{\xi,n} =(-1)^n\sqrt{2}\,.
\end{equation}
Therefore, the wavefunctions on the brane read
\begin{equation}
    \chi_n(\pi R)=\sqrt{2}\times \frac{\mu_n}{\mu_{D,n}}=\sqrt{2}\times \frac{\mu_n}{\sqrt{\mu_n^2+M_D^2}}\,.
\end{equation}
Finally, let us turn to the zero mode, which we identify with $\mu_0=0$. In such a case, the solutions are at most first-order polynomials in $y$, possibly depending on its absolute value. However, the Dirichlet conditions for $\xi$ force it to $\xi_0(y)=0$. The even mode can then be determined by
\begin{align}
    &\partial_5 \chi_0 = -M_D \sign{y}\chi_0\,, &\chi_0(y)=c_{\chi,0}e^{-|y|M_D }\,.
\end{align}
The normalisation fixes $c_{\chi,0}$
\begin{equation}
    \chi_0(y)=\left(\frac{2\pi M_D R}{e^{2\pi M_D R}-1}\right)^{1/2}\,e^{M_D(\pi R- |y|)}\,.
\end{equation}
The presence of a $y$-dependent Dirac mass term acts as a potential, changing the wavefunctions and causing an exponential localisation of the zero mode. 
For instance, in our conventions, if $M_D>0$, then the potential is minimised when $\sign{y}<0$. Therefore, the zero mode WF gets localised near $y=0$ and exponentially suppressed at $y=\pi R$. If $M_D<0$, the opposite situation applies. Finally, the WFs of the massive KK-modes get affected by $M_D\neq 0$, but read the same at $y=0,\pi R$ as in the previous cases.
\section{LED Phenomenology in Neutrino Oscillations}
\label{sec:phenomenology}
In the following, we consider how the presence of extra dimensions can modify how neutrinos propagate. We study the scenario in which the SM is localised on the orbifold fixed point brane at $y=\pi R$
\begin{equation}
    S_{\text{brane}}=\int d^4x\int \limits_{-\pi R}^{\pi R}dy\,\sqrt{-G}\,\mathcal{L}_\text{brane}\,\delta(y-\pi R)\,,
\end{equation}
where $g$ is now the determinant of the 4D induced metric on the brane, and the delta function localises interactions on the SM brane. Notice that the fixed point $y=0$, would have been an equally good possibility, and our choice does not affect the results.
Because $\psi_R$ is even under $y\!\to\!-y$ while $\psi_L$ is odd, one has $\psi_L(0)=\psi_L(\pi R)=0$. Hence, only $\psi_R$ can couple to operators localised on the SM brane at $y=\pi R$.~\footnote{Notice that fluctuations of the brane position would also impact neutrino oscillations~\cite{Dienes:1998sb,Pas:2005rb}.}
From now onwards, we will work only with the 4D EFT on the brane; we hence will omit the localisation of the functions at $y=\pi R$, unless specified otherwise. 

The most generic interaction term reads
\begin{equation}
    -\mathcal{L}_{\text{brane}}\supset\ov{L_L}\tilde Y_D\widetilde{H}\psi_{R}+\frac{1}{2}\ov{\psi_{R}^c} \tilde B\psi_{R}+\text{h.c.}\,,
\end{equation}
where $H$ and $L_L$ denote the Higgs and leptonic $SU(2)$ doublets with $H^T=\left(H^{+}, H^0\right)$ and $\tilde{H}=i \sigma_2 H^*$. The couplings $\tilde Y_D$ and $\tilde B$ are generically matrices and encode non-trivial flavour structure. 
Upon performing the KK-decomposition of Eq.~\eqref{eq:KK-decomposition}, in terms of the KK tower, it reads
\begin{equation}
   \label{eq:lag-brane} -\mathcal{L}_{\text{brane}}\supset \ov{L_L}\left(\frac{\tilde Y_D}{\sqrt{V}}\right)\widetilde{H}\sum\limits_{n=0} \psi_{R,n}\chi_{n}+\frac{1}{2}\sum\limits_{n,m=0}\ov{\psi_{R,n}^c} \left(\frac{\tilde B}{V}\right)\psi_{R,m} (\chi_n\chi_m)+\text{h.c.}\,.
\end{equation}
We will denote
\begin{align}
    &m_D\equiv  \frac{v}{\sqrt{2}}\times\left(\frac{\tilde Y_D}{\sqrt{V}}\right)\,, &B\equiv \frac{\tilde B}{V}\,,
\end{align}
defined to absorb the Higgs vacuum expectation value for later convenience.
In the following analysis, we work under the simplifying assumption that all the mass matrices become \textit{simultaneously} diagonal when the matrix $Y_D$ is rotated, generating the PMNS matrix. 
While for our purposes, such an assumption is necessary to obtain analytical results, such a construction could be realised by imposing some form of minimal flavour violation~\cite{DAmbrosio:2002vsn} in the UV.  We comment on the effects of departing from such an assumption in~\ref{app:matrices_not_diagonalized}. In this work, we keep the $m_D$ matrix as a free parameter, and we do not impose any bias on the size of its entries, in line with other works in the literature, e.g. Refs.~\cite{Carena:2017qhd,Antoniadis:2025rck}.
\begin{table}[]
    \centering
    \begin{tabular}{l|lccc}
    \toprule
         Model&Extra params. &$\chi_0(\pi R)$&$\chi_n(\pi R)$&KK-masses \\
         \midrule
         Brane Dirac& $-$ &  $1$&$\sqrt{2}$&$n\mu_1$ \\
         Bulk Dirac& $M_D$ &  $\left[\frac{2\pi RM_D}{\left(e^{2\pi RM_D}-1\right)}\right]^{1/2}$&$\frac{\sqrt{2}\mu_n}{\sqrt{\mu_n^2+M_D^2}}$&$\sqrt{M_D^2+(n\mu_1)^2}$ \\
         Bulk Majorana& $M_J$ &$1$&$\sqrt{2}$&$M_J\pm n\mu_1$\\
         Brane Majorana& $B$ &$1$&$\sqrt{2}$&$\pm n\mu_1$ \\
         \bottomrule
    \end{tabular}
    \caption{Summary of the localised WFs and KK-spectrum ($n\geq 1$) for the case studies of this work. The parameters $m_D$ and $\mu_1=R^{-1}$ are common to all scenarios. The $\pm$ for the Majorana cases stems from the choice of basis used to describe the mass matrix. For all models, the lightest KK mode is massless, $\mu_0=0$, and thus it is not reported in the table.}
    \label{tab:benchmarks}
\end{table}

A summary of the models studied so far, along with the relevant parameters for neutrino phenomenology, can be found in Table~\ref{tab:benchmarks}. All models share different features in the shape of the KK-spectrum, in the relative size of the zero WFs to the massive ones and, most importantly, in the shape of the mass matrix. All of these features will affect how neutrinos oscillate, and as we will explore in the next Section.

\subsection{KK Neutrino Oscillations}
The inclusion of a large number of sterile states that mix with the SM neutrinos can modify the SM prediction for neutrino oscillation.\footnote{As a complementary probe, it was recently proposed to study neutron oscillations induced by mixing with bulk fields~\cite{Dvali:2023zww}.}
We begin by reviewing the theoretical formalism of neutrino oscillations, both in vacuum and in matter. We then apply it to the different case studies, with brane and bulk Dirac and Majorana masses.
\subsubsection{Oscillations in Vacuum}
In this section, we develop the formalism for vacuum neutrino oscillations with KK modes. We derive analytical formulas for oscillation probabilities in the long-baseline limit, which will guide us in the interpretation of the full numerical results later in this work.

Let us denote the flavour eigenstates by $\nu_{\alpha,i}$ where  $\alpha \in \{e, \mu, \tau\}$ represent flavour and the Roman index $i,j,\dots\in \{1,2,3\}$ and $n,m,\dots\in \{0,1,2...,N\}$ labels the mass eigenstates and the KK modes, respectively. These flavour states are related to the  mass eigenstates $\hat{\nu}_{i,n}$ via the unitary transformation:
\begin{equation}
\label{eq:basis-definition}
    \ket{\nu_{\alpha,n}}= \sum\limits_{i}\sum\limits_{m} U_{\alpha i} V^{i}_{nm}\ket{\hat{\nu}_{i,m}}\,,
\end{equation}
where $V^{i}_{nm}$ is the matrix related to the second rotation involving the KK-modes and $U_{\alpha i}$ is the PMNS matrix that rotates the three flavour states and diagonalises $Y_D$. The flavour index of $V$ reflects the fact that in our limit of simultaneous diagonalisation of the different mass matrices, a different $V_{nm}$ exists for each flavour.
The exact structure of $V^{i}_{nm}$ has to be determined case by case. The procedure to obtain $V^{i}_{nm}$ is straightforward but tedious and amounts to the diagonalisation of the full mass matrix. We will provide explicit solutions for all the cases listed in Table~\ref{tab:benchmarks} in the following Sections.

Let us consider a SM flavour eigenstate $\ket{\nu_{\alpha,0}}$.
The oscillation probability of such a state in time is given by
\begin{align}
    &P_{\alpha\to\beta}(t) \equiv|\braket{\nu_{\beta,0}(t)|\nu_{\alpha,0}(0)}|^2=\left|\sum\limits_{in}\left(U_{\alpha i}V^{i}_{0n}\right)\left(U_{\beta i}V^{i}_{0n}\right)^\star e^{-i E_{in}t}\right|^2 \,,
\end{align}
where $E_{in}\simeq E+\frac{m_{in}^2}{2E}$ is the energy of the mass
eigenstate $\ket{\hat\nu_{i,n}}$ in the ultra-relativistic limit. 
This equation will be used to derive all constraints in Sec.~\ref{sec:phenomenology}.

For the sake of intuition, it is convenient to study a simplified limit. In long-baseline experiments where the distance travelled by the neutrino is large relative to its energy, oscillation between different eigenstates is rapid and averages out. As a result, the oscillation probability becomes an incoherent sum,
\begin{align}
\label{eq:oscillation_probability}
    P_{\alpha\to \beta}(t\to\infty)&\approx \sum\limits_{i}|U_{\alpha i}|^2|U_{\beta i}|^2\times\left(\sum\limits_{n}|V^{i}_{0n}|^4\right) \,.
\end{align}
Such a formula will be used to estimate long baseline expectations and better understand numerical results. In this limit, the deviation from the SM prediction is entirely encoded in the $0$-eigenvector of the neutrinos' mass matrix. This will turn out to be particularly enlightening when trying to understand many of the numerical results presented in Section~\ref{sec:dirac-brane-osc} and onwards.
\subsubsection{Oscillations in Matter}
In this Section, we focus on neutrino oscillations in matter. We define the notation and the formalism which will be employed for the different models.
As a reference, we follow the notation of Ref.~\cite{Giunti:2007ry} and the notation of Ref.~\cite{Machado:2011jt} for the matter effects.

Let us begin by considering the original Lagrangian of Eq.~\eqref{eq:lag-brane}. Before applying any rotation to the fields, the Yukawa term involving $Y_D$ includes the interaction term
\begin{equation}
    \mathcal{L}\supset \ov{L_L^\alpha} \tilde{Y}_D^{\alpha\beta} \widetilde{H}\psi_R^\beta \supset - \ov{\nu_{L,0}^\alpha}m_D^{\alpha\beta}\left(\sum\limits_{n=0}^\infty \chi_n\psi_{R,n}^\beta\right)\,,
\end{equation}
where the fields indicate the original fields in the interaction basis, and we denote $\tilde\nu_{L,0}$ the neutrinos in the basis where $Y_D$ is diagonal.
The above term can be diagonalised by a bi-unitary rotation
\begin{align}
    &\nu_{L,0}^\alpha=U^{\alpha i} \tilde\nu_{L,0}^i\,, &\psi_{L,R}^\alpha=R^{\alpha i}\tilde\psi_{L,R}^i\,,
\end{align}
such that
\begin{equation}
    U^\dagger m_D R = m_{D,\text{diag}}\,.
\end{equation}
The rotation $U$ is the PMNS already presented in Eq.~\eqref{eq:basis-definition}. The rotation $R$ is applied equally to all KK modes that stem from $\psi_R$ upon dimensional reduction. Notice that the tilded basis $\ket{\tilde\nu}$ corresponds to the basis where $\tilde Y_D$ is diagonal, while $\ket{\nu}$ and $\ket{\hat\nu}$ correspond to the interaction and mass basis used in the previous section, respectively.

Since the matter affects the propagation of only $\nu_{L,0}$, it is convenient to make explicit the flavour index
\begin{align}
    &\nu_L^\alpha \equiv \begin{pmatrix}
        \nu_{L,0}^\alpha  & X_{1}^\alpha &  X_{2}^\alpha &\dots & X_{N}^\alpha 
    \end{pmatrix}^T\,, &\nu_L=\begin{pmatrix}
        \nu_L^e\\
        \nu_L^\mu\\
        \nu_L^\tau
    \end{pmatrix}\,,
\end{align}
where $X_{i}^\alpha$ could be any left-handed field that appears in the Lagrangian. 
In this notation, the change of basis can be conveniently written as
\begin{align}
    &\begin{pmatrix}
        \nu_L^e\\
        \nu_L^\mu\\
        \nu_L^\tau
    \end{pmatrix}^\alpha=\mathcal{U}^{\alpha i}\begin{pmatrix}
        \tilde\nu_L^1\\
       \tilde\nu_L^2\\
        \tilde\nu_L^3
    \end{pmatrix}^i\,, & \left(\mathcal{U}\right)_{\alpha i}=\begin{pmatrix}
        U_{\alpha i} & 0\\
        0 & R_{\alpha i}
    \end{pmatrix}\,.
\end{align}
This way, the Hamiltonian of the system can be conveniently written as
\begin{align}
    &H=H_0+\mathcal{V}_m\,, &&\mathcal{V}_m=\begin{pmatrix}
      \mathcal{V}_e & 0 & 0\\
      0 & \mathcal{V}_\mu & 0 \\
      0 & 0 & \mathcal{V}_\tau 
    \end{pmatrix}\,, &&\mathcal{V}_\alpha=\begin{pmatrix}
        \delta_{e\alpha }V_{CC}+V_{NC} & \mathbf{0}\\
        \mathbf{0} & \mathbf{0}\\
    \end{pmatrix}\,,
\end{align}
where $H_0$ is the vacuum part and $\mathcal{V}_m$ is the matter potential
\begin{align}
    &V_{CC}=\sqrt{2}G_F n_e\,, &V_{NC}=-\frac{1}{\sqrt{2}}G_F n_n\,.
\end{align}
with $G_F$ is Fermi's constant and $n_{e,n}$ is the number density of electrons and neutrons, respectively.
Schr{\"o}dinger equation dictates the time evolution of $\ket{\tilde\nu_i}$
\begin{align}
    &i\frac{d}{dt}\ket{\tilde\nu}=\mathcal{U}^\dagger\left( H_0 + \mathcal{V}_m\right)\mathcal{U}\ket{\tilde\nu}\approx\left(\frac{\mathbf{M}\mathbf{M}^\dagger}{2E}+\mathcal{U}^\dagger \mathcal{V}_m\mathcal{U}\right)\ket{\tilde\nu}\equiv \frac{1}{2E}\left(\mathbf{M}\mathbf{M}^\dagger+2E\mathcal{V}\right)\ket{\tilde\nu}\,,
\end{align}
where
\begin{equation}
    \mathcal{V}\equiv \mathcal{U}^\dagger \mathcal{V}_m\mathcal{U}= \mathcal{U}^\dagger\begin{pmatrix}
      \mathcal{V}_e & 0 & 0\\
      0 & \mathcal{V}_\mu & 0 \\
      0 & 0 & \mathcal{V}_\tau 
    \end{pmatrix}\mathcal{U}\,.
\end{equation}
If all mass matrices can be made simultaneously diagonal,
\begin{align}
    &\mathbf{M}=\begin{pmatrix}
        M_e & 0 & 0\\
        0 & M_\mu & 0\\
        0 & 0 & M_\tau
    \end{pmatrix}\,, &\mathbf{M}\mathbf{M}^\dagger=\begin{pmatrix}
        M_eM_e^\dagger & 0 & 0\\
        0 & M_\mu M_\mu^\dagger & 0\\
        0 & 0 & M_\tau M_\tau^\dagger
    \end{pmatrix}\,.
\end{align}
The above formulas will be implemented numerically.
In the following Sections, we analyse the spectrum of the case studies of this work listed in Table~\ref{tab:benchmarks} employing the above discussed formalism. We will express the masses and mixing factors in terms of the parameters of the benchmark model and of the inverse size of the extra dimension, $\mu_1=1/R$. Finally, we will denote the eigenvalues of the mass matrices by $\{m_{\lambda_n}\}_n$.
\subsection{Dirac Brane Term}
\label{sec:dirac-brane-osc}
In this section, the analysis is restricted to the minimal setup with $M_J=M_D=0$ and the field couples to the SM only via the Yukawa portal, hereafter referred to as the ``\textit{Dirac brane}'' scenario. This benchmark has been the focus of a substantial body of work, which has examined the phenomenology of bulk neutrinos across a wide range of observables and energy scales~\cite{Mohapatra:1999zd,McLaughlin:1999br,Barbieri:2000mg,McLaughlin:2000iq,Mohapatra:2000wn,DeGouvea:2001mz,Davoudiasl:2002fq,Cao:2003yx,Machado:2011jt,Esmaili:2014esa,Rodejohann:2014eka,MINOS:2016vvv,Berryman:2016szd,Stenico:2018jpl,Basto-Gonzalez:2021aus,Forero:2022skg,Siyeon:2024pte,Panda:2024ioo,Elacmaz:2025ihm,Franklin:2025muw}. In the following, we will review and comment about the spectrum of the theory, which will then be applied to neutrino oscillations. We will employ previously derived results in \secref{sec:theory-dirac-brane}. 

Under our current assumptions, the Dirac brane Yukawa $Y_D$ is already diagonal, therefore we can treat each flavour independently and omit the flavour indices. The $4D$ effective Lagrangian in the flavour basis reads
\begin{equation}
    -\mathcal{L}\supset \ov{L_L}\tilde{Y}_D\widetilde{H}\left(\sum\limits_{n=0}^N \psi_{R,n}\chi_{n}\right)+
    \sum\limits_{n=1}^N\,\mu_n\ov{\psi_{L,n}}\psi_{R,n} +\text{h.c.}\,.
\end{equation}
For convenience of notation and without loss of generality, we relabel the fields as
\begin{align}
    &&\nu_\text{SM}\equiv \nu_{L,0}\,, &&\psi_{L,n}\equiv\nu_{L,n}\,, &&\psi_{R,n}\equiv \nu_{R,n}\,.
\end{align}
We can group into a more compact notation $\nu_L$ and $\nu_R$ all fields, and write the mass matrix as
\begin{equation}
\label{eq:lag:mass-dirac}
    -\mathcal{L}\supset \sum\limits_{n=1}^N\,\mu_n\ov{\nu_{L,n}}\nu_{R,n}+\ov{\nu_{L,0}}(m_D\chi_0)\left[\nu_{R,0}+\sqrt{2}\sum\limits_{n=1}^N \nu_{R,n}\right]+\text{h.c.}\,.
\end{equation}
Since $\chi_{n\geq 1}=\sqrt{2}\chi_0$, only one combination $(m_D\chi_0)$ appears, one could replace
\begin{equation}
    (m_D\chi_0)\to m_D\,.
\end{equation}
Such an observation is not relevant in the flat case as $\chi_0=1$, but has an interesting impact for the warped scenario, as we will discuss more in detail later on.
The associated mass matrix reads
\begin{equation}
    \mathcal{L}\supset -\ov{\nu_L}\begin{pmatrix}
        m_D & \sqrt{2}m_D &\sqrt{2}m_D   &\dots& \sqrt{2}m_D\\
        0 & \mu_1 &0&\dots & 0\\
        0 & 0 & \mu_2 &\dots & 0\\
        \dots&\dots&\dots&\dots&\dots\\
        0 & 0 & 0 &\dots & \mu_N
    \end{pmatrix}\nu_R\equiv -\ov{\nu_L}\mathbf{M}\nu_R\,.
\end{equation}
The upper-triangular structure of the mass matrix comes from
the brane Yukawa coupling of $\nu_{L,0}$ to every $\nu_{R,n}$ with
weights $\chi_n(\pi R)$ and that the KK Dirac masses pair $\nu_{L,n}$ with $\nu_{R,n}$.
This yields a democratic mixing of the SM state with the tower.
The mass spectrum is obtained by diagonalising $\mathbf{M}$, or equivalently 
$\mathbf{M}\mathbf{M}^\dagger$ for the left-handed fields. Since only $\nu_L$ 
participates in weak interactions, we focus on the left-handed rotation $\mathbf{L}$.
The diagonalisation of the mass matrix can be achieved by performing independent rotations of $\nu_{L,R}$
\begin{align}
    &\nu_L\to \mathbf{L}~\nu_L\,, &\nu_R\to \mathbf{R}~\nu_R\,.
\end{align}
Since only $\nu_L$ enters the charged interactions, we focus only on the left-handed rotation, which can be computed by noticing that
\begin{align}
    \mathbf{L}^\dagger (\mathbf{M}\mathbf{M}^\dagger) \mathbf{L}= \text{diag}(\mathbf{M}^2)\,,
\end{align}
where
\begin{equation}
\label{eq:Dirac-massmatrixsquared}
    \mathbf{M}\mathbf{M}^\dagger=\begin{pmatrix}
       m_D^2(1+2N) & \sqrt{2}\mu_1 m_D & \sqrt{2}\mu_2m_D &\dots & \sqrt{2}\mu_N m_D\\
        \sqrt{2}\mu_1 m_D & \mu_1^2 & 0 &\dots & 0\\
        \sqrt{2}\mu_2 m_D & 0&\mu_2^2  &\dots & 0\\
        \dots & \dots& \dots & \dots & \dots\\
        \sqrt{2}\mu_N m_D & 0 & 0&\dots &\mu_N^2
    \end{pmatrix}\,.
\end{equation}
The eigenvalues can be obtained by computing the solutions to the following equation
\begin{align}
    \label{eq:DBrane-eigenvalues}\sum\limits_{n=0}^N\frac{(\chi_n m_D)^2}{m_\lambda^2-\mu_n^2}=1 
   \qquad \overset{N\to\infty}{\Rightarrow} \qquad \pi\cot \left(\frac{\pi  m_\lambda}{\mu_1}\right)= \frac{\mu_1 m_\lambda}{m_D^2}\,,
\end{align}
where the closed form of the summation is obtained by taking the $N\to\infty$ limit.
The result is in agreement with what had been found in previous works, e.g. in Ref.~\cite{Dienes:1998sb}. The normalised eigenvectors $u_\lambda$ relative to the eigenvalue $m_\lambda$ are found to be
\begin{align}
    \label{eq:brane-dirac-N}&u_\lambda
=\mathcal{N}_\lambda\begin{pmatrix}
    1\\
    \vdots\\
    \frac{\sqrt{2}\mu_n m_D}{m_\lambda^2-\mu_n^2}\\
    \vdots
\end{pmatrix}\,, &&\mathcal{N}_\lambda=\left[1+2\sum\limits_{n=1}^N\left(\frac{\mu_n m_D}{m_\lambda^2-\mu_n^2}\right)^2\right]^{-1/2}\overset{N\to\infty}{=}\left[\frac{1}{2} \left(\frac{\pi ^2  m_D^2}{\mu_1^2}+\frac{m_\lambda^2}{m_D^2}+1\right)\right]^{-1/2}\,.
\end{align}
The above results are in agreement with previous works~\cite{Arkani-Hamed:1998wuz,Mohapatra:1999af}.
The normalisation factor $\mathcal{N}_\lambda$ is larger for lighter modes, thus they couple more to the SM as expected. Interestingly, its functional shape is analogous to the one derived for the extra-dimensional axion case~\cite{deGiorgi:2024elx}.

The left-handed rotation $\mathbf{L}$ can be reconstructed starting from the above eigenvectors.
Notice that in terms of the notation employed in Eq.~\eqref{eq:basis-definition} to describe neutrino oscillations $\mathbf{L}_{nm}^i=V_{nm}^i$. Finally, the above results imply $(u_\lambda)_0=\mathcal{N}_\lambda$, and thus all relevant information for neutrino oscillations, besides the mass spectrum, is encoded in the normalisation coefficients $\mathcal{N}_\lambda$\,. 
Finally, unitarity enforces sum rules involving powers of $\mathcal{N}_\lambda$, the most relevant being
\begin{align}
    &\label{eq:unitarity}\sum\limits_\lambda \mathcal{N}_\lambda^2=1\,.
\end{align}
Such a relation can be used to assess the number of neutrinos relevant for each observable quantitatively and as a valuable countercheck in numerical calculations when truncating the spectrum.

We now investigate the structure of the spectrum of the theory. As it will turn out, states can be conveniently labelled by integers $n$, so we will use the notation $m_{\lambda_n}\equiv m_n$. 
In the limit $\mu_1\gg m_D$, the extra dimension decouples. On the other hand, the zero mode of the tower remains massless and can form a Dirac pair with the SM neutrino, yielding
\begin{align}
    &m_\lambda^\text{lightest} =m_D \left(1-\frac{\pi ^2 m_D^2}{6 \mu_1^2}+\dots\right)\,, &\label{eq:db-asym}m_{\lambda_{n\geq 1}}\approx \mu_n\,.
\end{align}

In the opposite limit, $\mu_1\ll m_D$, the mixing of the SM neutrino with the KK-tower gets large, and the spectrum is found to be
\begin{equation}
    \label{eq:dirac-brane-lightest2}m_{\lambda_n} = \left(n+\frac{1}{2}\right) \mu_1\left(1-\frac{\mu_1^2}{\pi ^2 m_D^2}+\dots\right)\,.
\end{equation}
Notice how in this limit the masses of the lightest modes do scale with $\mu_1$, not $m_D$. 
The Dirac Yukawa regulates the number of modes for which the approximation is valid before the decoupling part of the matrix takes over. In fact, the existence of a mode $n_\star$ such that $\mu_{n_\star}\gg m_D$ is unavoidable. When such a limit is reached, the result of Eq.~\eqref{eq:db-asym} applies and $m_{\lambda_{n\geq n_\star}}\approx \mu_{n_\star}$. 
Interestingly, the mixing $\mathcal{N}_\lambda$ of the modes found in Eq.~\eqref{eq:brane-dirac-N} becomes constant if $m_\lambda \sim \mathcal{O}(\mu_1) \ll m_D$ yielding
\begin{equation}
    \mathcal{N}_{\lambda_{n\leq n_\star}}(\mu_1\ll m_D)\approx \frac{\sqrt{2}}{\pi}\frac{\mu_1}{m_D}\,.
\end{equation}
The value of $n_\star$ at which this happens can be qualitatively estimated by considering when the ratio of the diagonal entries of the mass matrix becomes comparable with the off-diagonal ones. A more precise statement can be obtained by studying the normalisation factor of Eq.~\eqref{eq:brane-dirac-N}, realising that the KK tower contribution becomes dominant over $m_D$ when $\mathcal{N}_\lambda$ stops being constant, yielding 
\begin{equation}
    \label{eq:nstar-dirac-brane}n_\star \approx \left(\frac{ \pi m_D}{\sqrt{2}\mu_1}\right)^2\,.
\end{equation}

\begin{figure}
        \centering
        \includegraphics[width=1\linewidth]{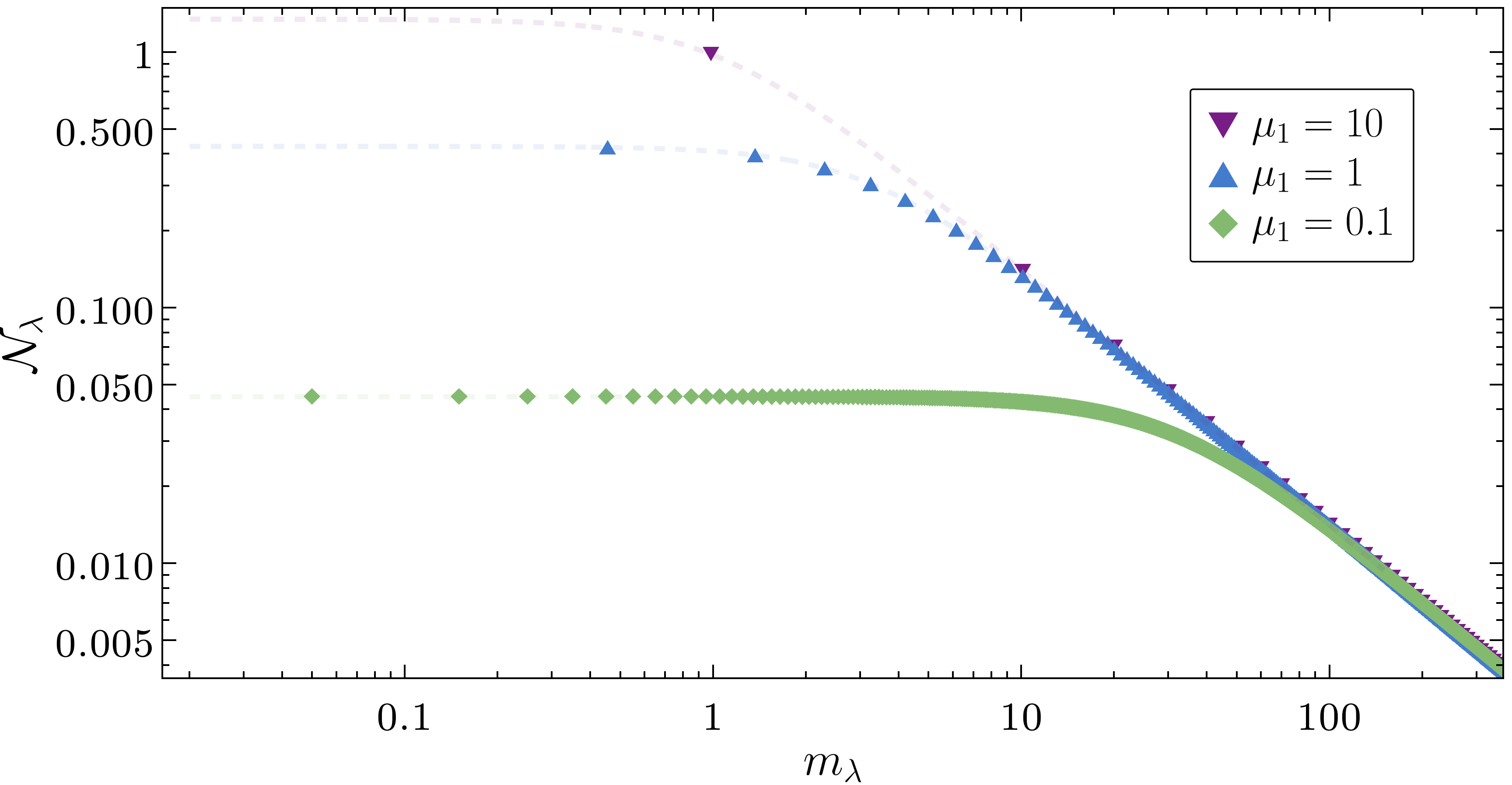}
        \caption{Brane-Dirac spectrum for representative values of $\mu_1$ in units of $m_D=1$. The dashed lines and the dots represent $\mathcal{N}_\lambda$ of Eq.~\eqref{eq:brane-dirac-N} as a continuous function of $m_\lambda$ and the physical masses stemming from Eq.~\eqref{eq:DBrane-eigenvalues}, respectively.}
        \label{fig:brane-dirac-spectrum}
\end{figure}

The spectrum and the normalisation factors can be visualised for a representative choice of the parameters in Fig.~\ref{fig:brane-dirac-spectrum}. The plot shows the mixing $\mathcal N_\lambda$ versus mass eigenvalue $m_\lambda$ (in units of $m_D$) for $\mu_1=\{10,1,0.1\}$ shown in purple, blue and green, respectively. For large KK spacing ($\mu_1 \gg m_D$, purple), the lightest state carries $\mathcal N_\lambda \simeq 1$ while higher modes have negligible overlap, indicating KK decoupling and SM-like oscillations. As $\mu_1$ decreases (blue~$\to$~green), the SM zero mode spreads quasi-democratically over many KK states. Finally, the estimation of the number of modes with equal mixing estimated in Eq.~\eqref{eq:nstar-dirac-brane} is confirmed by the exact numerical results.

We delve now into the impact of such a model on neutrino observables. 
A back of the envelope estimation can be made from Eq.~\eqref{eq:DBrane-eigenvalues} realising that the lightest mass eigenvalue $m_\lambda^\text{lightest}$ will lie within the first quadrant of $\cot\left(x\right),\,~x\in[0,\pi/2]$:
\begin{equation}
    0<m_\lambda^\text{lightest}<\frac{\mu_1}{2}\,,
\end{equation}
and using the best fit value of $\Delta m_{3l}^2 \approx 2.5 \times10^{-3}\mathrm{eV}^2$~($l$=1(2) for NO(IO))~\cite{Esteban:2024eli}, and assuming the lightest neutrino among all flavours to be massless, one obtains $\mu_1\gtrsim 0.1 \mathrm{eV}$.

Similarly, a bound can be extracted by studying the correction factor to the oscillation probability of Eq.~\eqref{eq:oscillation_probability}. Since $V_{0n}^i=\mathcal{N}^i_{\lambda_n}$ (where here we restored the flavour index $i$ for clarity), and since no significant deviation is found in neutrino data so far, the deviation in the neutrino oscillation can be qualitatively captured by requiring
\begin{equation}
    \sum\limits_n |V_{0n}^i|^4= \sum\limits_\lambda (\mathcal{N}_\lambda^i)^4 \overset{!}{\approx} 1\,.
\end{equation}
By employing the asymptotic results previously derived, one finds
\begin{align}
  \sum\limits_\lambda \mathcal{N}_\lambda^4&\approx \begin{cases}
        1& \mu_1 \gg m_D\,,\\
        \left(\sqrt{2}\mu_1/(\pi m_D)\right)^4 & \mu_1 \ll m_D\,,
    \end{cases}
\end{align}
 from which one can immediately extract a rough bound

\begin{equation}
    \label{eq:naive-dirac-brane}\mu_1 \gtrsim (\pi/\sqrt{2}) m_D\,.
\end{equation}
We note that $\mu_1 \ll m_D$, the mixing between active and the sterile states can be larger, and hence $\sum\limits_\lambda \mathcal{N}_\lambda^4 \ll 1$, reducing the overall survival probability of the active neutrinos as they oscillate into steriles.
In the following Section, we explore the more complicated case of a bulk Dirac mass term.
\subsection{Dirac Bulk Term}
\label{sec:dirac-bulk}
In this Section, we consider the case in which the bulk fields have a Dirac mass, $M_J=0$ and $M_D\neq 0$. We will refer hereafter to this case study as the ``\textit{Dirac bulk}'' scenario.
The case study has been the object of several works which explore its phenomenology across different observables and energies~\cite{Ioannisian:1999cw,Lukas:2000wn,Carena:2017qhd,Anchordoqui:2023wkm,Antoniadis:2025rck,Eller:2025lsh}.
We will study the mass matrix, the spectrum and the mixing structure employing the results previously derived in Section~\ref{sec:theory-dirac-bulk}.

Before starting the discussion, it is relevant to point out that the sign of $M_D$ has physical consequences. First, its sign flips the potential of $\ov{\Psi}\Psi$, and thus the exponential localisation of the zero mode. In turn, as we will see, this heavily affects the mixing with the SM neutrinos. Secondly, even though less dramatic, it can appear jointly with $m_D$ in observables. Let us explore it more in detail by considering the relevant terms
\begin{equation}
    -S\supset \int d^4x\int\limits_{-\pi R}^{\pi R} dy \sqrt{G}\left\{\sign{y}M_D\ov{\psi_L}\psi_R+\delta(y-\pi R)\left[\ov{L_L}\tilde{Y}_D\widetilde{H} \psi_{R}+\ov{L_L}\tilde{Y}_e{H}e_R \right]+\text{h.c.}\right\}\,,
\end{equation}
assuming without loss of generality that $\tilde{Y}_D>0$ and that $Y_e$ is already diagonal and positive. If $M_D$ were negative, its minus sign could not be rotated away by redefining $\psi_R$, as it would appear in front of $\tilde{Y}_D$. The converse is also true. It follows that the sign of the product $\tilde{Y}_D M_D$ is physical.

The mass terms of the Lagrangian in the 4D EFT follow the same structure previously discussed for the Dirac brane case in Eq.~\eqref{eq:lag:mass-dirac}.
The Yukawa part of the Lagrangian reads
\begin{equation}
    -\mathcal{L}\supset \sum\limits_{n=1}^N\,\mu_{D,n}\ov{\psi_{L,n}}\psi_{R,n}+\ov{\nu_{L}}m_D\left[\chi_0\nu_{R,0}+\sum\limits_{n=1}^N \chi_n\nu_{R,n}\right]+\text{h.c.}\,,
\end{equation}
where $\mu_{D,n}$ is the KK mass stemming from dimensional reduction~(cf. Eq.~\eqref{eq:mass-bulk-dirac})
\begin{equation}
    \mu_{D,n}=\sqrt{M_D^2+\mu_n^2}\,.
\end{equation}
On the brane $y=\pi R$, the wavefunctions reduce to
\begin{align}
    &\chi_0(\pi R)=\left(\frac{2\pi RM_D}{e^{2\pi R M_D}-1}\right)^{1/2}\,, &\chi_{n\geq 1}(\pi R)=\sqrt{2}\times \frac{\mu_n}{\sqrt{\mu_n^2+M_D^2}}\,.
\end{align}
The main difference compared to the brane Dirac case is 
(i) the different value of $\chi_0\neq 1$ and $\chi_{n\geq 1}\neq \sqrt{2}$, which reproduces the known case in the limit $M_D\to 0$, and (ii) the different KK-mass spectrum, which is now affected by $M_D$.
About the former point, depending on the sign of $M_D$, the magnitude of $\chi_0$ can significantly change
\begin{equation}
    \label{eq:chi0-limits} \chi_0(\pi R)\approx\begin{cases}
       \sqrt{2\pi RM_D}~e^{-\pi RM_D}\,,  & M_D R\gg 1\,,\\
       1\,,  & M_D R \sim 0\,,\\
       \sqrt{2\pi R|M_D|}\,,  & M_D R \ll 1\,.
    \end{cases}
\end{equation}
For large positive values, it gets exponentially small, thus suppressing the mixing between the zero mode and the rest of the states. Conversely, if $M_DR\ll 1$, the mixing increases as $|M_D|^{1/2}$.

By adopting the notation
\begin{align}
    &&\nu_\text{SM}\equiv \nu_{L,0}\,, &&\psi_{L,n}\equiv\nu_{L,n}\,, &&\psi_{R,n}\equiv \nu_{R,n}\,,
\end{align}
we can write the mass matrix as
\begin{equation}
    \mathcal{L}\supset -\ov{\nu_L}\begin{pmatrix}
        \chi_0 m_D  & \chi_1m_D &\chi_2m_D   &\dots& \chi_Nm_D\\
        0 & \mu_{D,1} &0&\dots & 0\\
        0 & 0 & \mu_{D,2} &\dots & 0\\
        \dots&\dots&\dots&\dots&\dots\\
        0 & 0 & 0 &\dots & \mu_{D,N}
    \end{pmatrix}\nu_R\equiv -\ov{\nu_L}\mathbf{M}\nu_R\,.
\end{equation}
The mass matrix is upper-triangular and formally looks as in the Dirac-brane case. However, the zero mode coupling can be substantially smaller or larger than unity, and all KK diagonals are lifted by $M_D$.
The eigenvalues and the left-handed fields rotation can be found by studying 
\begin{equation}
\label{eq:bulk-Dirac-massmatrixsquared}
    \mathbf{M}\mathbf{M}^\dagger=\begin{pmatrix}
       m_D^2\sum\limits_{n=0}^N\chi_n^2 & \chi_1\mu_{D,1}m_D &  \chi_2\mu_{D,2}m_D &\dots &  \chi_N\mu_{D,N}m_D\\
         \chi_1\mu_{D,1}m_D & \mu_{D,1}^2 & 0 &\dots & 0\\
         \chi_2\mu_{D,2}m_D& 0&\mu_{D,2}^2  &\dots & 0\\
        \dots & \dots& \dots & \dots & \dots\\
         \chi_N\mu_{D,N}m_D & 0 & 0&\dots &\mu_{D,N}^2
    \end{pmatrix}\,,
\end{equation}
as previously done for the brane Dirac case in \secref{sec:dirac-brane-osc}.
Its eigenvalues $m_\lambda^2$ can be obtained by solving
\begin{align}
    &\sum\limits_{n=0}^N\frac{(\chi_n m_D)^2}{m_\lambda^2-\mu_{D,n}^2}=1\,, &\mu_{D,n}=\begin{cases}
        0\,, &n=0\,,\\
        \sqrt{\mu_n^2+M_D^2}\,, &n\geq 1\,.
    \end{cases}
\end{align}
In the limit $N\to\infty$, the above expression translates to the transcendental equation
\begin{equation}
\label{eq:masses-bulk-dirac}\frac{\pi }{\mu_1 }\left[\sqrt{m_\lambda^2-M_D^2} \cot \left(\frac{\pi  \sqrt{m_\lambda^2-M_D^2}}{\mu_1}\right)-M_D \coth \left(\frac{\pi  M_D}{\mu_1}\right)\right]+ \chi_0^2=\frac{m_\lambda^2}{m_D^2}\,.
\end{equation}
In the limit $M_D\to 0$, one has $\chi_0\to 1$ and $\tilde{m}_\lambda\to m_\lambda$, thus encountering the result previously found.
The normalised eigenvectors and the mixings are found to be
\begin{align}
    &u_\lambda
=\mathcal{N}_\lambda\begin{pmatrix}
    1\\
    \vdots\\
    \frac{\chi_n\mu_{D,n} m_D}{m_\lambda^2-\mu_{D,n}^2}\\
    \vdots
\end{pmatrix} \,, &\mathcal{N}_\lambda=\left[1+\sum\limits_{n=1}^N\left(\frac{\chi_n\mu_{D,n}m_D}{m_\lambda^2-\mu_{D,n}^2}\right)^2\right]^{-1/2}\,.
\end{align}
Upon taking the limit $N\to\infty$, the normalisation coefficients read
\begin{align}
\label{eq:normalisation-bulk-dirac}\mathcal{N}_\lambda=\left[1-\frac{\pi  m_D^2 }{4 \mu_1^2 } \left(\frac{\mu_1} {\sqrt{m_\lambda^2-M_D^2}} \sin \left(\frac{2 \pi  \sqrt{m_\lambda^2-M_D^2}}{\mu_1}\right)-2 \pi \right)\csc ^2\left(\frac{\pi  \sqrt{m_\lambda^2-M_D^2}}{\mu_1}\right)\right]^{-1/2}\,.
\end{align}
The structure of the mixing is significantly more complicated than in the Dirac brane case. 

Let us now study the spectrum of the theory in some interesting limit cases. We can start by examining how the lightest mode behaves as a function of $M_D$. If $M_D$ is positive and very large, then $\chi_0$ gets exponentially suppressed (cf. Eq.~\eqref{eq:chi0-limits}), and thus the lightest mode gets exponentially lighter. If, on the other hand, $M_D$ is very large but negative, then the diagonal terms of the mass matrix grow compared to the off-diagonal ones, and thus the lightest mode gets again lighter. All in all, for large values of $M_D$, one expects a hierarchy in masses between the lightest and all the rest of the modes. When one of the parameters is much smaller than the others and $m_D$ is not the largest, its mass can be well approximated by
\begin{equation}
    \label{eq:dirac-bulk-lightest}m_\lambda^\text{lightest}\approx
    \begin{cases}
        \frac{\mu_1 m_D}{\sqrt{\mu_1^2+\pi^2 m_D^2/3}}\,, & M_D \to 0\,;\\
        \chi_0 m_D\,, & m_D \to 0\,;\\
        \sqrt{2}M_D~e^{-\pi M_D/\mu_1}\,; & \mu_1\to 0 , M_D>0\,;\\
       |M_D|& \mu_1\to 0 , M_D<0\,.
    \end{cases}
\end{equation}
The limit $M_D\to 0$ reasonably reproduces the previous results obtained for the Dirac brane case in Eqs.~\eqref{eq:db-asym}-\eqref{eq:dirac-brane-lightest2}. 
If $\mu_1\gg m_D$, the extra dimension decouples, and the spectrum can be conveniently labelled by positive integers
\begin{equation}
    m_{\lambda_{n}}\approx\mu_{D,n}=\sqrt{M_D^2+(n\mu_1)^2}\,.
\end{equation}
This limit is always satisfied in the bottom right corner of the mass matrix, and the above approximation always holds for the heaviest modes of the spectrum.

On the other hand, $m_D\gg M_D,\mu_1$, there exists a regime where the heavier modes are still given by
\begin{equation}
    m_{\lambda_{n\geq 1}}\simeq \sqrt{M_D^2+(n \mu_1)^2}\,,
\end{equation}
but where the mixing does not yet decrease, and it becomes instead constant
\begin{equation}
    \label{eq:bulk-dirac-constant}\mathcal{N}_{\lambda,\text{plateau}}\approx \frac{\sqrt{2}\mu_1}{\pi m_D}\,,
\end{equation}
regardless of $M_D$. The above expression is valid if $|M_D|\sim \mu_1$, and needs further $n$-dependent corrections if otherwise.
If instead $m_D \ll \mu_1,|M_D|$, the lightest mode gets most of the mixing as
\begin{equation}
     \label{eq:bulk-dirac-lightest}\mathcal{N}_{\lambda,\text{lightest}}\approx 1+\frac{\pi  m_D^2 }{4 \mu_1^2}\left[\pi  \text{csch}^2\left(\frac{\pi  M_D}{\mu_1}\right)-\frac{\mu_1 }{M_D}\coth \left(\frac{\pi  M_D}{\mu_1}\right)\right]\,.
\end{equation}
For positive and large $M_D$ the exponential suppression of $\chi_0$ makes implausible any chance of detecting modifications to the SM.

\begin{figure}
        \centering
        \includegraphics[width=1\linewidth]{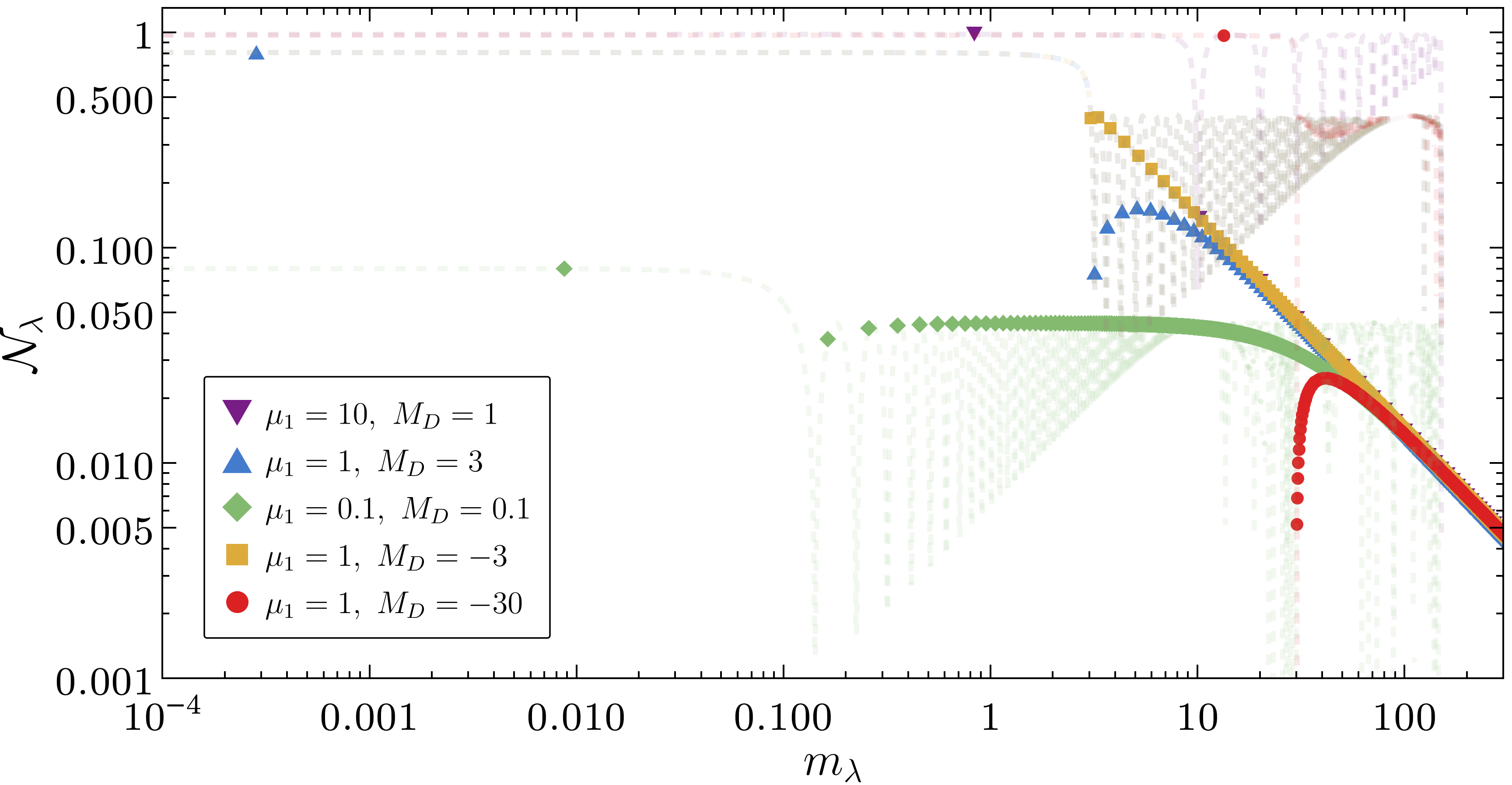}
        \caption{Bulk-Dirac spectrum for representative values of $\mu_1$ in units of $m_D=1$. The dashed lines and the dots represent $\mathcal{N}_\lambda$ of Eq.~\eqref{eq:normalisation-bulk-dirac} as a continuous function of $m_\lambda$ and the physical masses stemming from Eq.~\eqref{eq:masses-bulk-dirac}, respectively.}
        \label{fig:bulk-dirac-spectrum}
\end{figure}

The masses and the normalisation factors can be visualised for a representative choice of the parameters in Fig.~\ref{fig:bulk-dirac-spectrum}.
The spectrum features a lightest mode with the largest mixing. Its mass is exponentially smaller compared to the other modes as $M_D$ grows, in agreement with the previous estimation of Eq.~\eqref{eq:dirac-bulk-lightest}; its mixing is the largest and approaches one as $m_D$ decreases, as previously shown in Eq.~\eqref{eq:bulk-dirac-lightest}. Heavier modes' mixing grows with the mass $m_\lambda$, reaches a peak, and then it starts decreasing. 
It can be shown that the position of the peak in the limit $|M_D|/\mu_1\gg 1$ approximately lies at
\begin{equation}
    m_{\lambda,\text{peak}}\simeq M_D\,.
\end{equation}
The regime where $m_D\gg \mu_1,|M_D|$ shows a plateau structure before the decoupling regimes begin, whose mixing value agrees with the estimation of Eq.~\eqref{eq:bulk-dirac-constant}. 

Let us now turn to the impact on neutrino oscillations. In the scenarios described above, we can derive the corrective factor for the SM probabilities. In the regime of maximal mixing $m_D\gg \mu_1,|M_D|$, we can estimate a rough constraint by assuming the plateau dominates the mixing, and thus the relevant number of modes reads
\begin{equation}
\label{eq:nstar-dirac-bulk}
    n_\star \approx \left(\mathcal{N}_{\lambda,\text{plateau}}^2\right)^{-1}\approx \left(\frac{\pi m_D}{\sqrt{2}\mu_1}\right)^2\,,
\end{equation}
where we used the result of Eq.~\eqref{eq:bulk-dirac-constant}. 
In turn, this implies
\begin{align}
  &\sum\limits_\lambda\mathcal{N}_\lambda^4\approx n_\star \mathcal{N}_{\lambda,\text{plateau}}^4= \left(\frac{\sqrt{2}\mu_1}{\pi m_D}\right)^2\,,  &\mu_1 \gtrsim (\pi/\sqrt{2}) m_D\,.
\end{align}
Interestingly, the number of modes as well as the rough bound are independent of $M_D$ and match the case with $M_D=0$ of Eq.~\eqref{eq:naive-dirac-brane}.
On the other hand, information on $M_D$ can be extracted in the limit $m_D\ll \mu_1,|M_D|$; the mixing factor derived in Eq.~\eqref{eq:bulk-dirac-lightest} gives
\begin{equation}\
\label{eq:nstar_dirac_bulk}
    m_D^2\lesssim \frac{\mu_1^2}{\pi}\left[\pi  \text{csch}^2\left(\frac{\pi  M_D}{\mu_1}\right)-\frac{\mu_1 }{M_D}\coth \left(\frac{\pi  M_D}{\mu_1}\right)\right]^{-1}\approx \begin{cases}
       \frac{3\mu_1^2}{2\pi^2} \,, & \mu_1\gg |M_D|\,,\\
       \frac{\mu_1|M_D|}{\pi} \,, & \mu_1\ll |M_D|\,.
    \end{cases}\,.
\end{equation}
\\

So far, we have considered only Dirac neutrinos. In the next section, we start introducing lepton number violation by means of a Majorana mass. We begin discussing the presence of a bulk Majorana mass, and then move to the case of a brane one.
\subsection{Majorana Bulk Term}
\label{sec:pheno-bulk-majorana}
In this section, we study the impact of a bulk Majorana mass, thus taking $M_J\neq 0$ and $M_D=0$, as introduced in the action of Eq.~\eqref{eq:bulk-action}. We refer hereafter to this case study as the ``\textit{Majorana bulk}'' scenario. Previous works have appeared in the literature focusing on model-building and its phenomenology~\cite{Arkani-Hamed:1998wuz,Dienes:1998sb,Pilaftsis:1999jk,Blennow:2010zu,Garbrecht:2020tng}. We will employ the results previously derived in \secref{sec:theory-majorana-bulk} to analyse the mass matrix and the spectrum of the theory.

In the presence of a Majorana mass term, a convenient basis to write the interactions is defined by Eq.~\eqref{eq:action-smart-n}.
We define the fields vector $X$
\begin{align}
    &X\equiv \begin{pmatrix}
        \nu_L&
        \psi_{R,0}^c&
        \psi_{1,1}^c&
        \psi_{2,1}^c&
        \psi_{1,2}^c&
        \psi_{2,2}^c&
        \dots&
        \psi_{1,N}^c&
        \psi_{2,N}^c
     \end{pmatrix}^T\,,
 \end{align}
such that the mass term can be conveniently written as
\begin{equation}
    -\mathcal{L}\supset \frac{1}{2}\ov{X}\mathbf{M} X^c\,.
\end{equation}
Notice that the zero mode was treated separately. 
As anticipated, in such a basis, the mass matrix $\mathbf{M}$ is almost diagonal
\begin{align}
    \label{eq:Bulk-Majorana-MM}&\mathbf{M}=\begin{pmatrix}
        0 &  \chi_{0}m_D & \tilde\chi_1m_D & \tilde\chi_1m_D  &\dots & \tilde\chi_Nm_D & \tilde\chi_Nm_D\\
        \chi_{0}m_D & M_J & 0 & 0 &\dots & 0 & 0\\
        \tilde\chi_1m_D & 0 & M_J +\mu_1 & 0 & \dots & 0 & 0\\
        \tilde\chi_1m_D & 0 & 0 & M_J -\mu_1 & \dots & 0 & 0\\
        \dots & \dots & \dots & \dots & \dots & \dots & \dots\\
        \tilde\chi_Nm_D & 0 & 0 & 0 & \dots & M_J+\mu_N & 0\\
        \tilde\chi_Nm_D & 0 & 0 & 0 & \dots & 0 & M_J-\mu_N\\
    \end{pmatrix}\,, &\tilde{\chi}\equiv \frac{\chi_n}{\sqrt{2}}\,.
\end{align}

The eigenvalues can be computed as solutions of the following sums, which give a closed result by employing the explicit values of $\chi_n$
\begin{equation}
\label{eq:eigeq-1}
\sum\limits_{n=0}^N\frac{\chi_n^2}{(m_\lambda-M_J)^2-\mu_n^2}=\frac{m_\lambda}{m_\lambda-M_J}\,, \qquad \overset{N\to\infty}{\Rightarrow} \qquad \pi \cot \left(\frac{\pi  (m_\lambda-M_J )}{\mu_1}\right)= \frac{m_\lambda  \mu_1}{m_D^2}\,.
\end{equation}
Similarly, the normalised eigenvectors read
\begin{align}
    &\label{eq:eigeq-2}u_\lambda=\mathcal{N}_\lambda\begin{pmatrix}
       1\\
        \frac{\chi_0}{m_\lambda-M_J} m_D \\
        \frac{\chi_1/\sqrt{2}}{m_\lambda-(M_J-\mu_1)} m_D \\
        \frac{\chi_1/\sqrt{2}}{m_\lambda-(M_J+\mu_1)} m_D \\
        \dots
\end{pmatrix}=\mathcal{N}_\lambda\begin{pmatrix}
       1\\
        \frac{ m_D}{m_\lambda-M_J}  \\
        \frac{ m_D}{m_\lambda-(M_J-\mu_1)} \\
        \frac{ m_D}{m_\lambda-(M_J+\mu_1)} \\
        \dots
    \end{pmatrix}\,, &\mathcal{N}_\lambda=\left[ \frac{m_D^2\pi ^2}{\mu_1^2}+\frac{m_\lambda^2}{m_D^2}+1\right]^{-1/2}\,.
\end{align}
The normalisation $\mathcal{N}_\lambda$ is larger for lighter modes, which then couple more to the SM. Formally, it has the same functional dependence on $m_\lambda$ as in the Dirac brane case, up to a factor of $1/2$ within the square root, which accounts for the different multiplicity of modes in the two cases. 

Let us investigate more in detail how the spectrum looks. 
First of all, notice that if
\begin{equation}
    \label{eq:lightest-special}M_J=\frac{n_\text{odd}}{2}\mu_1\,, \qquad \Rightarrow \qquad m^\text{lightest}=0\,. 
\end{equation}
This is an exact result for the entire range of parameters. Since it is not possible\footnote{It must be noted that one could explain neutrino oscillation even if the masses are 0 by allowing efficient mixing to sterile modes; however, it is impossible to explain every experimental result pertaining to neutrino oscillation in such a framework.} to explain neutrino oscillations with all neutrino masses being zero, one obtains exclusion regions centred at
\begin{equation}
\label{eq:half_integer_resonance}
    \mu_1 \approx\frac{2}{n_\text{odd}} M_J=\left\{2,\frac{2}{3},\frac{2}{5},\dots\right\}M_J\,.
\end{equation}
Let us now study the analytical limits of this case. In the decoupling limit $\mu_1\gg m_D,M_J$, only the zero mode of the tower remains, leaving in the spectrum two lighter modes and a heavy tower: 
\begin{align}
    &m_{\lambda_\pm}^\text{lightest}\approx \frac{1}{2}\left(M_J\pm\sqrt{M_J^2+4m_D^2}\right)\,, &m_{\lambda_{n,\pm}}\approx \pm \mu_n\,.
\end{align}
This case effectively amounts to the well-studied case of SM augmented with a set of right-handed neutrinos with Majorana mass term $M_J$. It is therefore expected in the exclusion plots to find bounds even for large $\mu_1$.

The case, $M_J\gg m_D,\mu_1$ is somewhat more involved. For each flavour, it amounts to a traditional type-I seesaw where the sterile sector is comprised of $2N+1$ sterile neutrinos.
The diagonal entries, which contain $M_J$, are barely affected by the mixing induced by $m_D$, and the corresponding spectrum can be approximated by 
\begin{align}
     &m_{\lambda_n}\approx M_J+\mu_n+\frac{m_D^2}{M_J+\mu_n}\,, &n\in \mathbb{Z}\,.
\end{align}
This is true only if $|M_J\pm\mu_n|\gg m_D$. This is not always necessarily the case, and if it happens, a diagonal entry can become as small as $m_D$, leading to large mixings. 
The mass of the lightest mode of the theory can be approximated for $m_D\ll \mu_1,M_J$ by
\begin{equation}
    \label{eq:lightest-bulk-majorana}m^\text{lightest}\approx -\frac{\pi m_D^2 }{\mu_1}\cot \left(\frac{\pi M_J}{\mu_1}\right)\,.
\end{equation}
The zeros of the cotangent appear at $M_J=(n_\text{odd}/2)\mu_1$ and do match the observation previously made in Eq.~\eqref{eq:lightest-special}.
From such a result, it is also manifest that when $M_J= n_\text{}\mu_1$, something happens. Indeed, in such a special case, the eigenvalue reads
\begin{equation}
    \label{eq:bulk-Majorana-lightest-resonance}m^\text{lightest}(M_J=\mu_n)\approx \mp\frac{m_D\mu_1}{\sqrt{(\pi m_D)^2/3+\mu_1^2}}\overset{m_D\to 0}{=}\mp m_D\,,
\end{equation}
that is, two degenerate eigenmodes appear. Such a feature can be understood in simpler terms.
Intuitively, this can be understood on simpler grounds. The mass matrix $\mathbf{M}$ of Eq.~\eqref{eq:Bulk-Majorana-MM} schematically has the structure:
\begin{equation}
\label{eq:resonance_MM}
\mathbf{M}\sim \begin{pmatrix}
0 & a_n\\
a_n & b_n
\end{pmatrix}\,,
\end{equation}
which can be diagonalised via an orthogonal mixing matrix with
\begin{align}
& \tan(2\theta_n) =\frac{2a_n}{b_n}\,,
\end{align}
where $\theta_n$ is the active sterile mixing. As can be noticed, when $b_n=0$, i.e. when $M_J=\mu_n$, the mixing becomes enhanced.
Let us try to explore this intuition more quantitatively, parametrising
\begin{equation}
    M_J=\mu_{n}+\epsilon m_D\,,
\end{equation}
where $\epsilon$ is a book-keeping index that helps in understanding the width of the effect.
Since all the eigenvalues are known to be heavy, the mass matrix can be effectively reduced to a $2\times 2$ matrix
\begin{align}
    &\mathbf{M}=\begin{pmatrix}
        0 &  m_D\\
        m_D & \epsilon m_D
    \end{pmatrix}\,.
\end{align}
The corresponding eigenvalues read
\begin{equation}
     \label{eq:majorana-bulk-m-degenerate}m_{\pm}=\frac{m_D}{2}\left(\epsilon\pm\sqrt{4+\epsilon^2}\right)\approx \pm m_D+\frac{m_D}{2}\epsilon +\mathcal{O}(\epsilon^2)\,,
\end{equation}
in agreement with the previous, more general result of Eq.~\eqref{eq:bulk-Majorana-lightest-resonance}.
The mixing of the two states gets large and can be approximated by
\begin{equation}
    \label{eq:majorana-bulk-N-degenerate}\mathcal{N}_\lambda \approx \left(1+\frac{m_\lambda^2}{m_D^2}\right)^{-1/2}\approx \frac{1}{\sqrt{2}}\mp \frac{\epsilon}{4\sqrt{2}}+\mathcal{O}(\alpha^2)\,.
\end{equation}
Notice that for $\epsilon \to 0$, the states are degenerate and have maximal mixing.

Finally, in the limit $m_D\gg \mu_1,M_J$ the masses can be labelled by a signed integer $n\in Z$ and read
\begin{align}
    m_{\lambda_n}\approx \left[M_J+ \mu_1 \left(\frac{1}{2}+ n\right)\right]\left[1-\frac{\mu_1^2}{\pi ^2 m_D^2}\right]\,.
\end{align}
The lightest eigenmode does not correspond to $n=0$, but rather to the element that minimises the element in the first squared bracket. As for the Dirac brane case, the mixing becomes constant for modes whose mass is smaller compared to $m_D$
\begin{equation}
    \mathcal{N}_{\lambda}(m_\lambda \ll m_D)\approx \frac{\mu_1}{\pi m_D}\,.
\end{equation}
The approximation breaks whenever $\mathcal{N}_{\lambda}$ stops being constant, and instead holds whenever
\begin{align}
    &-n_\star\lesssim n\lesssim n_\star\,, &n_\star =\pi\left(\frac{m_D}{\mu_1}\right)^2\,.
\end{align}
Interestingly, the number of relevant modes $\sim 2n_\star$, is independent of $M_J$.

\begin{figure}
        \centering
        \includegraphics[width=1\linewidth]{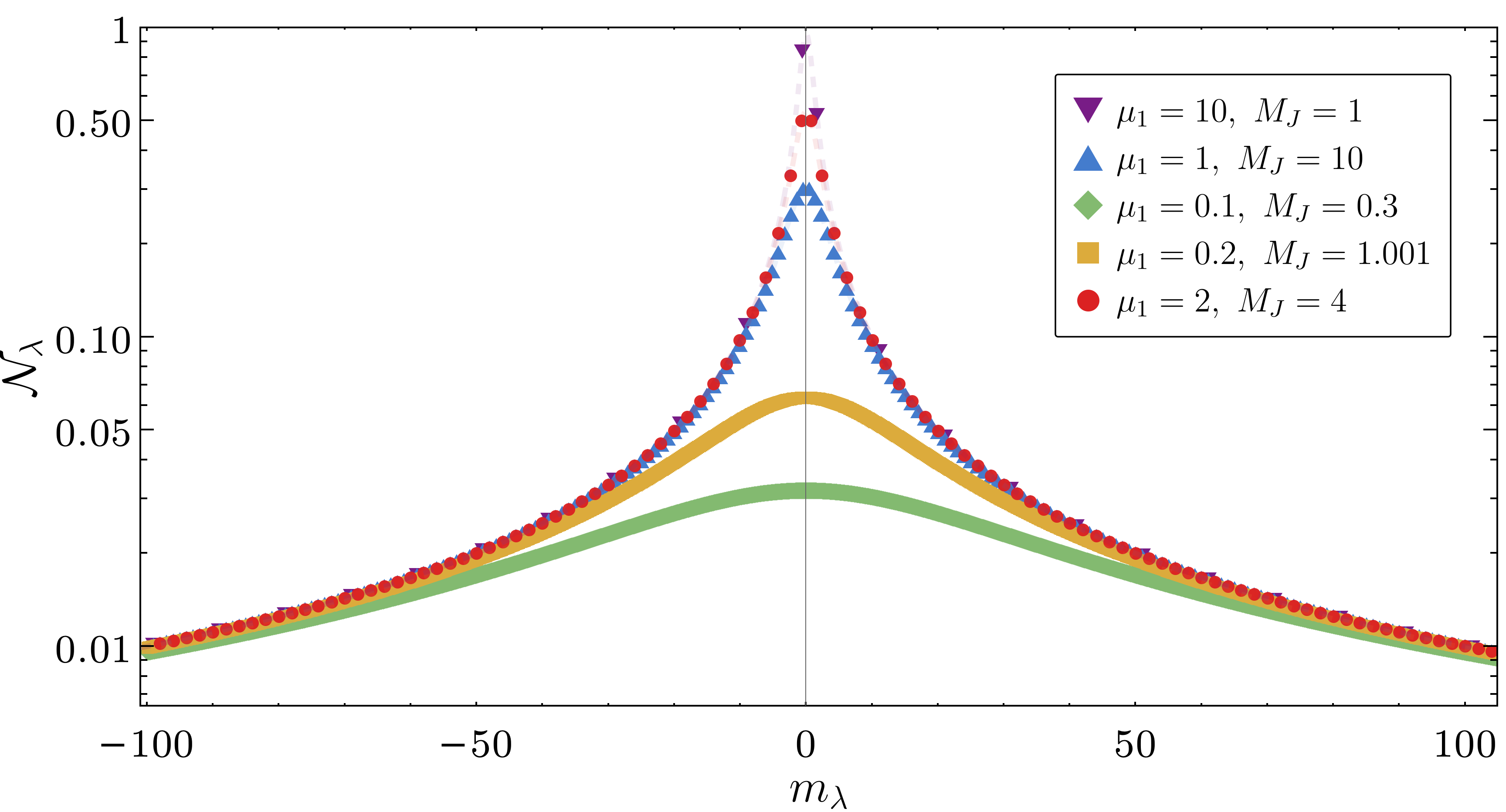}
        \caption{Bulk-Majorana spectrum for representative values of $\mu_1$ in units of $m_D=1$. The dashed lines and the dots represent $\mathcal{N}_\lambda$ of Eq.~\eqref{eq:eigeq-2} as a continuous function of $m_\lambda$ and the physical masses stemming from Eq.~\eqref{eq:eigeq-1}, respectively.}
        \label{fig:bulk-majorana-spectrum}
    \end{figure}

The spectrum and the normalisation factors can be visualised for a representative choice of the parameters in Fig.~\ref{fig:bulk-majorana-spectrum}.
It is composed of almost degenerate positive and negative massive modes. The mixing gets larger for smaller masses, resulting in a symmetric plot peaked around zero since the sign of the masses is not relevant in Eq.~\eqref{eq:eigeq-2}. 

Let us now turn to the qualitative estimation of the correction to neutrino oscillations.
The SM probability stemming from the normalisation factor gives
\begin{equation}
    \sum\limits_\lambda \mathcal{N}_\lambda^4\approx \begin{cases}
       1-\frac{2m_D^2}{4m_D^2+M_J^2} \,, &\mu_1\gg m_D,M_J\,,\\
       1-\frac{2\pi ^2  m_D^2 }{\mu_1^2} \csc^2 \left(\frac{\pi M_J}{\mu_1}\right)\,,&M_J\gg m_D,\mu_1\,,\\
        \frac{2}{\pi^3}\frac{\mu_1^2}{m_D^2} \,, &m_D\gg \mu_1,M_J\,.
    \end{cases}
\end{equation}
The decoupling case has to match current limits on sterile neutrinos, and thus it does not constrain the extra dimension.
All in all, from these estimations, for $M_J\neq \mu_n$, we get the naive constraints
\begin{equation}
    \label{eq:naive-bulk-majorana}\mu_1 \gtrsim m_D\times\begin{cases}
    (2\pi^2)^{1/2}  \left|\csc \left(\frac{\pi M_J}{\mu_1}\right)\right|\,,\\
        (2\pi^3)^{1/2}\,.
    \end{cases}
\end{equation}
The constraints are similar to the ones derived for the brane Dirac case in Eq.~\eqref{eq:naive-dirac-brane}.

The case $M_J\approx\mu_n$ deserves a separate discussion. Let us consider once more the case $M_J=\mu_n+\epsilon m_D$ in which $1\gg|\epsilon|\neq 0$. The correction to the probability is dominated by the lightest two modes, whose contribution can be estimated via the masses and mixings derived in Eqs.~\eqref{eq:majorana-bulk-m-degenerate}-\eqref{eq:majorana-bulk-N-degenerate}. If the fast oscillations can be averaged out, i.e. if $(m_+^2-m_-^2) L\gg 2E$, the survival probability can be approximated as
\begin{equation}
    P\propto \mathcal{N}_+^4+\mathcal{N}_-^4\approx 1-\frac{2}{4+\epsilon^2}\approx \frac{1}{2}+\mathcal{O}(\epsilon^2)\,.
\end{equation}
The probability touches the $95\%$ for $\epsilon\approx \pm 6$. Therefore, a rough bound can be cast: 
\begin{align}
    &\mu_1 \nsim \frac{M_J}{n}\,, &\forall n\geq 1\,.
\end{align}
The width of such a bound is dictated by $\sim 6m_D$.

If instead $\epsilon=0$, that is, when $M_J=\mu_n$, the two degenerate neutrinos can form an exact Dirac pair
leading to $m_+^2-m_-^2=0$. The probability would then read as 
\begin{equation}
    P\left(\nu_{\alpha} \rightarrow \nu_{\alpha}\right) = \left|\sum_{in}\left|U_{\alpha i }\right|^2 |V_{0n}^i|^2e^{\frac{im_{in}^2L}{2E}}\right|^2\approx  P\left(\nu_{\alpha} \rightarrow \nu_{\alpha}\right)_\text{SM}\times  \left(\frac{1}{2}+\frac{1}{2}\right)^2= P\left(\nu_{\alpha} \rightarrow \nu_{\alpha}\right)_\text{SM}\,.
\end{equation}
Therefore, in such a situation, no deviation can be observed. Observation-wise, one would therefore expect an exclusion region for all values of $\mu_1$ such that $M_J\approx \mu_n$ but with an allowed region when the relation is exact $M_J= \mu_n$. In the next Section, we discuss the last case study of this work, where the Majorana mass is localised on the brane.
\subsection{Majorana Brane Term}
Finally, we focus on the last possibility, namely of having all bulk parameters set to zero, $M_J=M_D=0$, while having non-vanishing brane-localised Majorana mass. We will hereafter refer to this case study as the ``\textit{Majorana brane}'' scenario. To the best of our knowledge, this case has not been studied in the literature.  For the characterisation of the mass matrix, we will employ the results previously derived in \secref{sec:theory-majorana-bulk}.
The starting point is the brane Lagrangian
\begin{equation}
    -\mathcal{L}\supset \ov{L_L}Y_D \widetilde{H}\left(\sum\limits_{n=0}^\infty\psi_{R,n}\chi_n\right)+\frac{1}{2}\sum\limits_{ij=0}\ov{\psi^c_{R,i}}B\psi_{R,j}+\text{h.c.}\,,
\end{equation}
where $B$ is the brane Majorana mass already rescaled by the volume of the extra dimension.
Employing the parametrisation used for the bulk Majorana mass of Eq.~\eqref{eq:action-smart-n}, the mass matrix can be written as
\begin{align}
    &\label{eq:mass-matrix-brane}\mathbf{M}=\begin{pmatrix}
        0 &  m_D & m_D & m_D  &\dots & m_D & m_D\\
        m_D &  B&   B&   B&\dots & B& B\\
        m_D & B& B+\mu_1 &  B & \dots &   B &  B\\
         m_D & B &  B& B-\mu_1 & \dots & B & B\\
        \dots & \dots & \dots & \dots & \dots & \dots & \dots\\
         m_D&  B&   B&  B& \dots & B+\mu_N & B \\
        m_D&   B &   B&  B& \dots &   B&  B-\mu_N\\
    \end{pmatrix}\,.
\end{align} 
The eigenvalues $m_\lambda$ can be computed as the solutions of
\begin{equation}
     \label{eq:brane-majorana-eigen1}\sum\limits_{n=0}^N\frac{\chi_n^2}{m_\lambda^2-\mu_n^2}=\frac{1}{m_D^2+m_\lambda B}\,, \qquad \overset{N\to\infty}{\Rightarrow} \qquad \pi\cot\left(\pi\frac{m_\lambda}{\mu_1}\right)=\frac{m_\lambda\mu_1}{m_D^2+m_\lambda B}\,,
\end{equation}
where $\chi_n=\sqrt{2}\chi_0=\sqrt{2}$ and the closed form of the sum was derived taking the limit $N\to \infty$.
The corresponding normalised eigenvectors and normalisation coefficients in the limit $N\to \infty$ are given by
\begin{align}
   &\label{eq:brane-majorana-eigen2}u_\lambda =\mathcal{N}_\lambda\begin{pmatrix}
        1\\
         (m_D^2+m_\lambda B)/(m_D m_\lambda)\\
         (m_D^2+m_\lambda B)/[m_D(m_\lambda-\mu_1)]\\
          (m_D^2+m_\lambda B)/[m_D(m_\lambda+\mu_1)]\\
         \dots
    \end{pmatrix}\,, & \mathcal{N}_\lambda=\left[1+\frac{m_\lambda^2}{m_D^2}+\frac{\pi ^2\left(m_D^2+m_\lambda B\right)^2}{\mu_1^2 m_D^2}\right]^{-1/2}\,.
\end{align}
 Contrary to the case of bulk-Majorana, the mixing factor is sensitive to the sign of $m_\lambda$. This results in a shift of the mixing peak from zero towards
\begin{align}
    \label{eq:peaks}&m_{\lambda,\text{peak}}=-\frac{\pi^2 m_D^2 B}{\mu_1^2+\pi^2 B^2}\,, &\mathcal{N}_{\lambda,\text{peak}}=\sqrt{1-\frac{\pi^2 m_D^2}{\mu_1^2+\pi^2(m_D^2+B^2)}}\,.
\end{align}
Given similar magnitudes among $\mu_1B$, the position of the peak is controlled by $\sim m_D/B$; the maximum mixing is then given to a massive mode, thus strongly impacting neutrino oscillations. One would therefore expect to be able to set a naive constraint on $m_D/B$ for $B\gg \mu_1$.

Let us inspect the spectrum of the theory in some extreme cases. 
In the decoupling limit, $\mu_1\gg m_D,B$, only the zero mode remains, and the model reduces to the SM augmented with a right-handed neutrino with Majorana mass $B$. The lightest mass of the spectrum is then given by
\begin{equation}
\label{eq:lightest_brane_mass}
    m_\lambda^\text{lightest}\approx \frac{1}{2} \left(B-\sqrt{B^2+4 m_D^2}\right)\overset{B\gg m_D}{\approx} -\frac{m_D^2}{B}\,.
\end{equation}
The last approximation of the above expression is nothing but the typical type-I seesaw relation. Heavier modes of the spectrum can be approximated by
\begin{align}
&m_{\lambda_n}\approx B+ \mu_n\,, &n\in\mathbb{Z}\,.
\end{align}
Notice that even if the condition $\mu_1\gg m_D,B$ is not satisfied, it will be for some $n_\star$ such that $\mu_{n_\star}\gg m_D,B$, and the above approximation will be valid.

\begin{figure}
        \centering
        \includegraphics[width=1\linewidth]{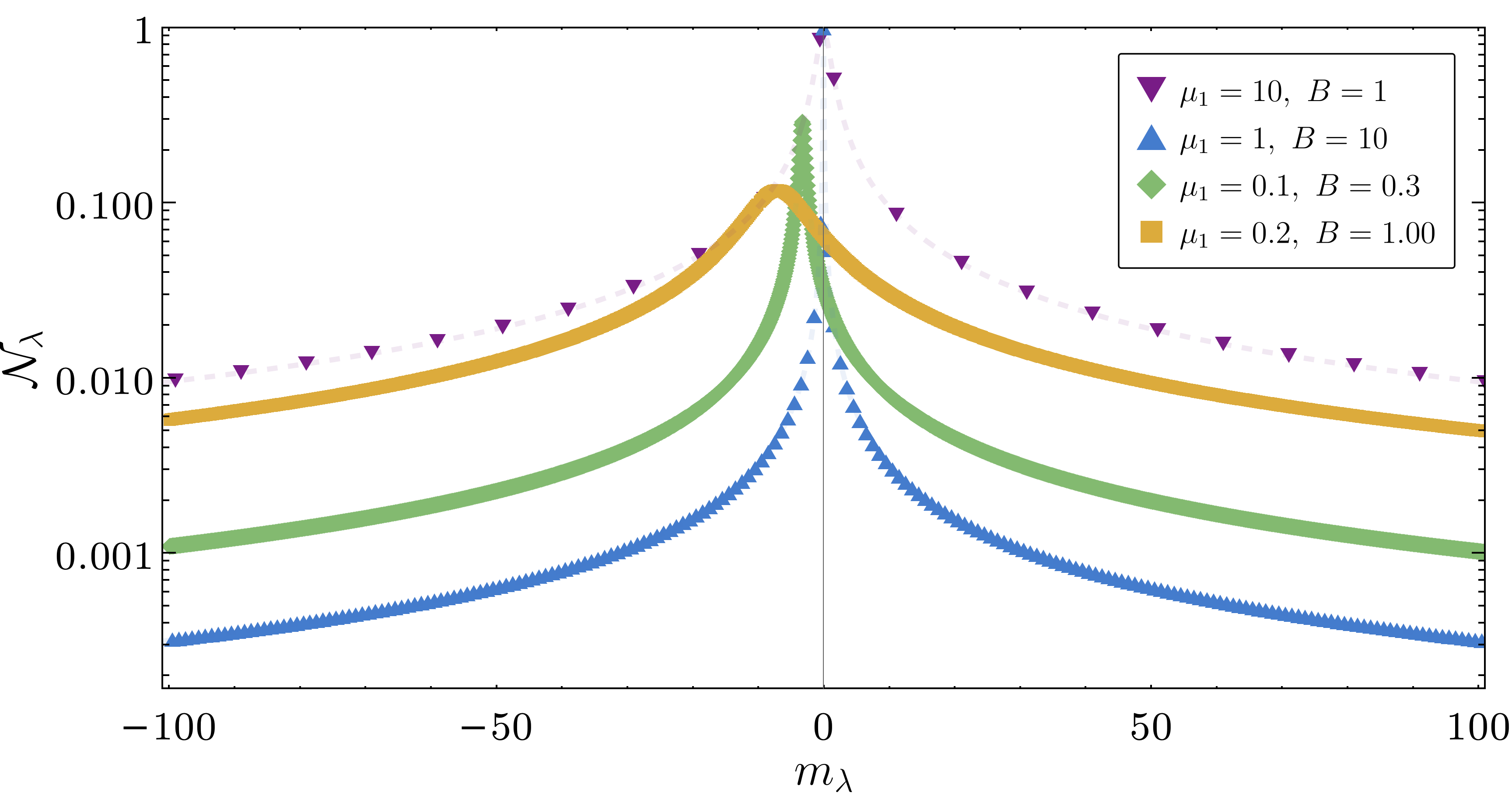}
        \caption{Brane-Majorana spectrum for representative values of $\mu_1$ in units of $m_D=1$. The dashed lines and the dots represent $\mathcal{N}_\lambda$ of Eq.~\eqref{eq:brane-majorana-eigen2} as a continuous function of $m_\lambda$ and the physical masses stemming from Eq.~\eqref{eq:brane-majorana-eigen1}, respectively.}
        \label{fig:brane-majorana-spectrum}
    \end{figure}

If instead $B\gg \mu_1,m_D$, the lightest mode still can be obtained by Eq.~\eqref{eq:lightest_brane_mass}.  This is no surprise. In such a limit, the details of the KK-tower are washed out, and the field $\psi_R$ can be treated as a single field with Majorana mass $B$, leading to the type-I seesaw scenario. The other modes can be conveniently labelled by an integer $n$ and are found to be
\begin{align}
    \label{eq:brane-majorana-light-modes}&m_{\lambda_{n}}\approx \left(\frac{1}{2}+ n\right)\mu_1\,, &n\in\mathbb{Z}\,.
\end{align}
The last possibility is $m_D\gg \mu_1,B$. In this case, the mixing becomes very large, light modes can be approximated by Eq.~\eqref{eq:brane-majorana-light-modes}.
However, these are not the most interesting modes for phenomenology, as the mixing $\mathcal{N}_\lambda$ peaks for heavier modes whose mass grows with $m_D^2$, as derived in Eq.~\eqref{eq:peaks}. If $m_D$ is the largest parameter, the mixing at the peak can be approximated by
\begin{equation}
    \mathcal{N}^\text{peak}_\lambda\approx \frac{\sqrt{\mu_1^2+\pi^2B^2}}{\pi m_D}\,.
\end{equation}
The number of relevant modes for phenomenology is bounded from below
\begin{equation}
    \label{eq:majorana-brane-number} n_\star \gtrsim \frac{1}{\mathcal{N}_{\lambda,\text{peak}}^2}  = \frac{\pi^2 m_D^2}{\mu_1^2+\pi^2 B^2}\,.
\end{equation}
One may wonder if something special happens whenever $B=\mu_n$, as was the case for the Majorana bulk case~(cf.~\secref{sec:pheno-bulk-majorana}). An analogy can be obtained by noticing that the structure of the mass matrix of Eq.~\eqref{eq:mass-matrix-brane} can be simplified by diagonalising the block with $n\geq 1$. The eigenvalues of the initial matrix do not change if unitary rotations are applied. Let us define the new eigenvalues as $X_n$; after taking the limit $N\to\infty$, they are found to satisfy the equation
\begin{equation}
\label{eq:semi_diagonal_brane_transform}
    \tan\left(\frac{\pi X_n}{\mu_1}\right)=\frac{\pi B}{\mu_1}\,.
\end{equation}
Given that the tangent is $\pi$-periodic, it suffices to find the fundamental solution in the range $\pi X/\mu_1\in [0,\pi/2)$ ($B,\mu_1>0$) and then obtain the others via
\begin{align}
    &X_n = X+ n\mu_1\,, &n\in Z\,.
\end{align}
The eigenvalue problem can then be reduced to the study of
\begin{equation}
\label{eq:semi_diagonal_brane_MM}
    \tilde{\mathbf{M}}=\begin{pmatrix}
        0 &  m_D & m_D & m_D  &\dots & m_D & m_D\\
        m_D & X & 0 & 0 &\dots & 0 & 0\\
        m_D & 0 & X+\mu_1 & 0 & \dots & 0 & 0\\
        m_D& 0 & 0 & X-\mu_1 & \dots & 0 & 0\\
        \dots & \dots & \dots & \dots & \dots & \dots & \dots\\
        m_D & 0 & 0 & 0 & \dots & X+\mu_N & 0\\
        m_D & 0 & 0 & 0 & \dots & 0 & X-\mu_N\\
    \end{pmatrix}\,.
\end{equation}
Notice that the structure of such a matrix is the same as that of the bulk case of Eq.~\eqref{eq:Bulk-Majorana-MM} upon identifying $X=M_J$. This allows us to use partially the same intuition developed before, resonance-wise. Since the tangent is a monotonic function, for the study of the fundamental solution, we can focus on two extreme cases:
\begin{itemize}
    \item if $B/\mu_1\to 0$, then $X\to 0$.
     \item if $B/\mu_1\to \infty$, then $X\to \mu_1/2$.
\end{itemize}
Therefore, using the analogy to the bulk case, $X\in[0,1/2)\mu_1$, which amounts to an effective Majorana bulk mass of size at most $\mu_1/2$. Since the exotic features appeared for $M_J=(n/2)\mu_1$ and the effective Majorana mass $X<\mu_1/2$, no value of $B$ can give rise to the effects discussed in \secref{sec:pheno-bulk-majorana}.
The spectrum and normalisation factors can be visualised for a representative choice of the parameters in Fig.~\ref{fig:brane-majorana-spectrum}.
The spectrum comprises positive and negative mass modes. The mixing structure is not symmetric among them, and its peak position and height are well approximated by Eq.~\eqref{eq:peaks}.  Let us now turn to the impact on neutrino oscillations.
By employing the result of Eq.~\eqref{eq:majorana-brane-number}, we can estimate in the limit $m_D\gg \mu_1, B$
\begin{equation}
    \sum\limits_\lambda \mathcal{N}_\lambda^4\approx \frac{\mu_1^2+\pi^2 B^2}{\pi^2 m_D^2}\,,
\end{equation}
which in turn implies
\begin{equation}
\label{eq:contraint_brane}
     m_D \lesssim \left[\frac{\mu_1^2}{\pi^2}+B^2\right]^{1/2}\,.
\end{equation}
For $B\to 0$, it amounts to the same constraint as Eq.~\eqref{eq:naive-dirac-brane}. On the other hand, if $B\gg \mu_1$, the details of the extra dimensions are washed out, and the limit reduces to seesaw-like bounds $m_D\lesssim B$.
\section{Experimental Setup and Analysis}
\label{sec:setup}
We now turn to a more systematic analysis of the phenomenological implications of bulk neutrinos for neutrino oscillations. Following the simplified working assumptions adopted in this work, we restrict ourselves to the four benchmark scenarios introduced in \secref{sec:phenomenology}, each characterised by a single additional mass term. This separation allows us to isolate and better assess the qualitative features of each case. A summary of the relevant parameters and phenomenological characteristics is presented in Table~\ref{tab:summary}. While the simultaneous presence of multiple mass terms would alter the quantitative bounds, it would not qualitatively change the characteristic behaviour of each scenario.

Our statistical analysis employs a $\chi^2$-based test to derive exclusion limits in the $(m^{\text{lightest}}, \mu_1)$ parameter space, where to reiterate, $\mu_1 = 1/R$ and R here is the size of the extra dimension in consideration, and $m^{\text{lightest}}$ is the physical mass of the lightest neutrino. The remaining two physical masses are fixed by the best-fit mass-squared differences reported in Ref.~\cite{Esteban:2024eli}, and all other oscillation parameters are likewise taken from the same global fit. We choose to constrain the physical neutrino mass as a free parameter, as it is the quantity also constrained by terrestrial~\cite{KATRIN:2024cdt}, as well as cosmological probes~\cite{DiValentino:2024xsv,DESI:2025zgx}. However, since from a model-building perspective the parameter $m_D$ is more relevant, we report the same constraints as a function of $\mu_1$ and $m_D$ in App.~\ref{app:lagragian_parameter_scans}.
We do not vary the oscillation parameters but instead fix them to their best-fit values, as our goal here is not to perform a full global fit but rather to highlight the modifications to neutrino phenomenology induced by extra dimensions.  

\begin{table}[t]
    \centering
    \vspace{0.3em}
    \begin{adjustbox}{max width=\textwidth}
    \begin{tabularx}{\textwidth}{@{}l c  X@{}}
        \toprule
        \textbf{Model} & $\left(\mu_1,m_D^{\text{lightest}}\right)$\textbf{+Params.}  & \textbf{Features} \\
        \midrule
        Brane Dirac & $-$ & Standard (vanilla) LED scenario. \\[3pt]
        Bulk Dirac & $M_D$  & Large positive $M_D$ exponentially suppresses the lightest mode mass; negative $M_D$ does not exhibit exponential behaviour. \\[3pt]
        Bulk Majorana & $M_J$  & Oscillation probability strongly dependent on extra dimensional parameters $(M_J,\mu_1)$; $M_J \approx (n/2)\mu_1$ associated with ``resonant" behaviour of flavour oscillation probability. \\[3pt]
        Brane Majorana & $B$  & Weak dependence on $\mu_1$; phenomenology effectively mimics a triple-sterile neutrino setup. \\
        \bottomrule
    \end{tabularx}
    \end{adjustbox}
    \caption{Summary of the benchmark models, their characteristic parameters, and qualitative phenomenological features.}
    \label{tab:summary}
\end{table}

We use data from both MINOS/MINOS+ and Daya Bay, as these experiments currently provide the most stringent constraints on extra-dimensional neutrino oscillation phenomenology in the normal and inverted mass orderings, respectively~\cite{Forero:2022skg}. It is worth noting that several upcoming or proposed detectors, such as DUNE~\cite{Siyeon:2024pte,Panda:2024ioo,Berryman:2016szd}, nuSTORM~\cite{Franklin:2025muw}, JUNO(+TAO)~\cite{Basto-Gonzalez:2021aus}, and T2HK~\cite{Panda:2024ioo}, are expected to offer significantly improved sensitivity to models with large extra dimensions.
The MINOS/MINOS+ datasets, with their long baselines and broad energy coverage, are primarily sensitive to $\nu_\mu$ disappearance, whereas Daya Bay, with its short baselines and high-precision reactor measurements, constrains $\bar{\nu}_e$ disappearance. For MINOS/MINOS+, we use the combined dataset corresponding to $16.36\times10^{20}$~POT, with Near and Far Detectors located at 1.04~km and 735~km, respectively~\cite{MINOS:2017cae}. For Daya Bay, we include 3158~days of operation, during which antineutrinos from six reactor cores were detected at baselines of 360-2000~m~\cite{DayaBay:2022orm}. These data are also compiled in Ref.~\cite{Eller:2025lsh}.

Together, these two experiments provide complementary coverage of the relevant $L/E$ regimes: MINOS/MINOS+ probes the slower oscillation modes driven by lightest KK states at long baselines, while Daya Bay constrains the rapid oscillation patterns induced by heavier modes at short baselines. This complementarity enables a robust and wide-ranging sensitivity to the compactification scale $R^{-1}$ and the lightest physical mass $m^{\text{lightest}}.$
The test statistic is defined as
\begin{equation}
    \Delta \chi^2 = -2\left[\log \mathcal{L}_\text{ED} - \log \mathcal{L}_\text{SM}\right],
    \label{eq:delta-chi2}
\end{equation}
where $\mathcal{L}_\text{ED}$ denotes the likelihood of the extra-dimensional model and $\mathcal{L}_\text{SM}$ that of the standard three-flavour scenario. Independent analyses are performed for each experiment and for both mass orderings to extract the corresponding confidence intervals.

On a technical note, one may wonder how many KK modes are numerically required to obtain stable and meaningful results. This is a subtle issue that strongly depends on the model parameters, as discussed in the previous section for the various scenarios (e.g.~Eq.~\eqref{eq:nstar-dirac-brane} for the Dirac brane case). In practice, a robust criterion for checking the validity of the truncation is to verify unitarity; in terms of the mixing coefficients $\mathcal{N}_\lambda$, this corresponds to the sum rule (cf.~Eq.~\eqref{eq:unitarity})
$\sum_\lambda \mathcal{N}_\lambda^2 = 1.$
In our numerical analysis, we always included enough modes to satisfy
\begin{equation}
    \sum\limits_\lambda \mathcal{N}_\lambda^2 > 99\%\,.
\end{equation}
The exact number of required modes depends on the parameters of the model.

\subsection{Dirac Brane Term}
\label{sec:dirac-brane}
\begin{figure}[t]
    \centering
    \includegraphics[width=0.7\linewidth]{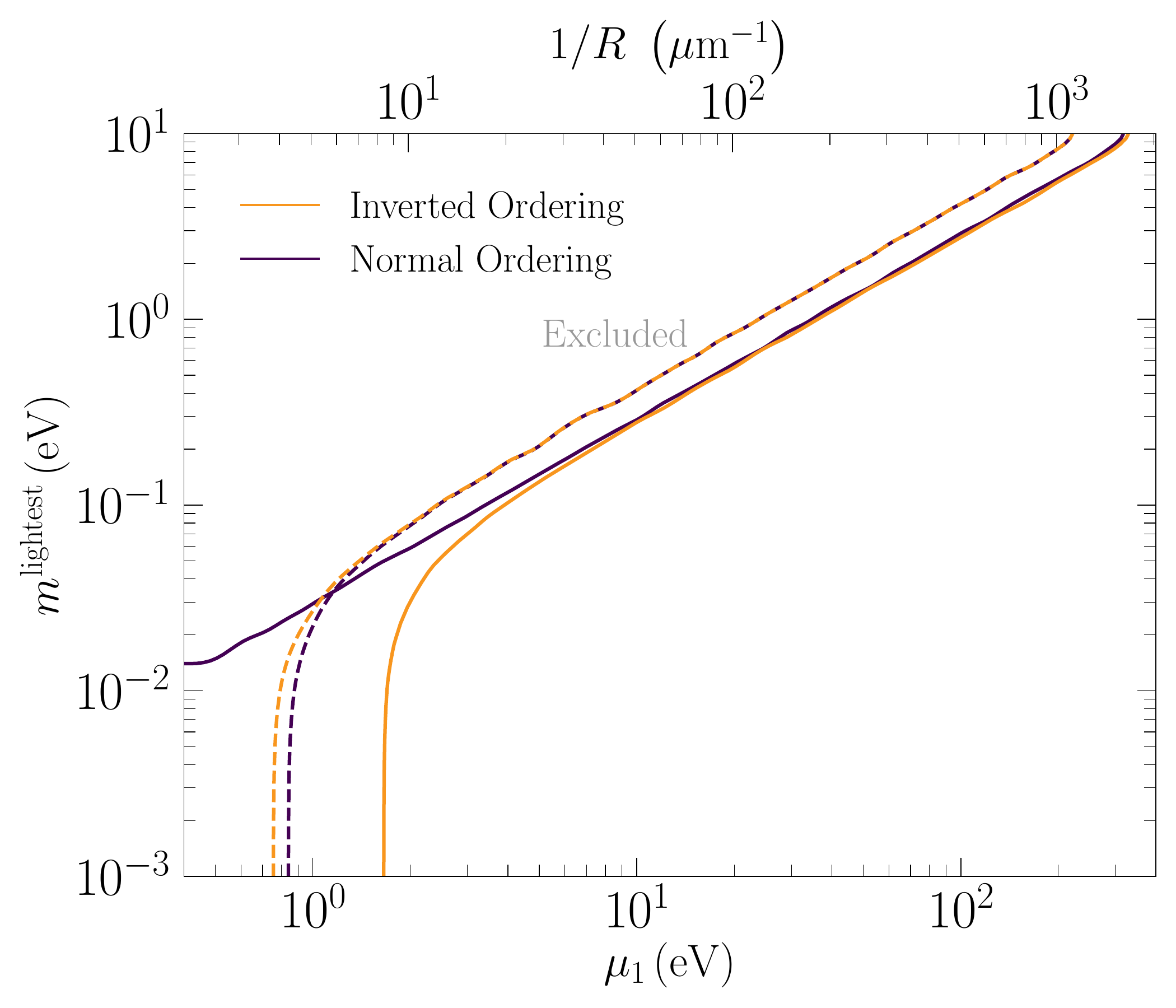}
    \caption{90\%~C.L. exclusion contours for the Dirac brane scenario from MINOS/MINOS+ and Daya Bay. Solid lines correspond to exclusion from Daya Bay, and dotted lines MINOS/MINOS+. An estimate for the number of modes to be included can be found in ~Eq.~\eqref{eq:nstar-dirac-brane}. Small fluctuations are statistical artifacts.}
    \label{fig:Dirac_Brane_MINOS_DayaBay}
\end{figure}

In this section, we focus on the bounds on the size of the extra dimension in the vanilla scenario, obtained from neutrino oscillation data.
Fig.~\ref{fig:Dirac_Brane_MINOS_DayaBay} displays the 90\%~C.L. exclusion contours for the model parameters, showing good agreement with previous studies~\cite{Machado:2011jt,Forero:2022skg,Elacmaz:2025ihm}. 

Upon examining the plots, we notice that Daya Bay yields a stronger constraint in the inverted ordering (IO), while MINOS/MINOS+ provides a tighter limit in the normal ordering (NO). This difference in sensitivity originates from the distinct flavour channels probed by the two experiments. The MINOS/MINOS+ analysis is primarily driven by $\nu_\mu$ disappearance, which depends on terms of the form $|U_{\mu i}|^2 (V_{0i}^n)^2$. Since all $|U_{\mu i}|^2$ are of comparable size, no single mass eigenstate dominates the oscillation behaviour, resulting in similar sensitivities for NO and IO. In contrast, Daya Bay probes $\nu_e$-disappearance, governed by $|U_{e i}|^2 (V_{0i}^n)^2$, where the smallness of $|U_{e3}|$ enhances the relative weight of the $i=1,2$ components, thereby increasing the sensitivity to IO. Numerically, this translates into $\mu_1 \gtrsim 0.7~\mathrm{eV}$ ($R>0.28\,\mu m$) for MINOS/MINOS+, largely independent of the mass ordering, and $\mu_1 \gtrsim 1.8~\mathrm{eV}$ ($R>0.11\,\mu m$) for Daya Bay in the inverted ordering, assuming $m^{\mathrm{lightest}} \sim \mathcal{O}(10^{-3})~\mathrm{eV}$.

\subsection{Dirac Bulk Term}
\label{sec:pheno-dirac-bulk}
\begin{figure}[t]
    \centering
    \includegraphics[width=0.99\linewidth]{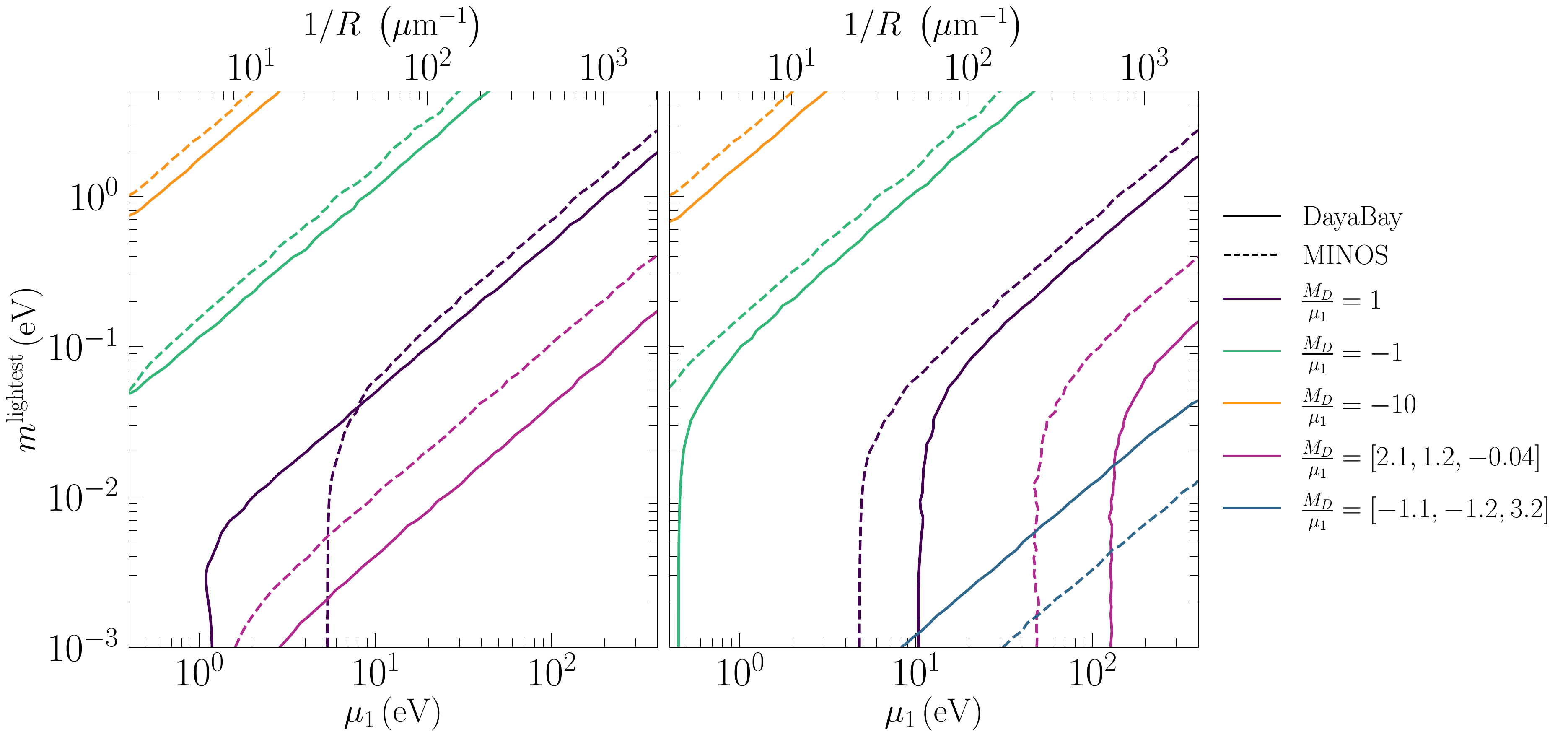}
    \caption{90\%~C.L. exclusion contours for the Dirac bulk scenario in normal ordering (left) and inverted ordering (right), obtained from the two experiments under consideration. The blue line does not appear in the left plot because it is further on the right. An estimate for the number of modes to be included can be found in ~Eq.~\eqref{eq:nstar-dirac-bulk}. Small fluctuations are statistical artifacts.}
    \label{fig:dirac_bulk_90_Cl_contour}
\end{figure}

In this section, we revisit the Dirac bulk mass scenario, both for completeness and to highlight several interesting features that emerge for specific parameter choices. 
Fig.~\ref{fig:dirac_bulk_90_Cl_contour} shows the 90\%~C.L. exclusion contours for both normal (NO) and inverted (IO) mass orderings. The overall pattern of sensitivities mirrors that of the Dirac brane case, with Daya Bay performing better in IO and MINOS/MINOS+ yielding comparable bounds across orderings. 

A particularly noteworthy difference relative to the Dirac brane scenario is the role played by the sign of the bulk mass term $M_D$. For $M_D < 0$, the allowed parameter space becomes broader, whereas for $M_D > 0$, the region is significantly more constrained. The underlying reason is that a negative $M_D$ localises the zero mode wavefunction $\chi_0$ closer to the brane where the Standard Model fields reside (see Eq.~\eqref{eq:chi0-limits}), effectively increasing the active neutrino masses and facilitating the reproduction of the observed mass-squared differences. Although one could, in principle, accommodate oscillation data with slightly shifted splitting, the precision of current multi-baseline measurements makes such deviations tightly constrained.

The sign dependence of $M_D$ also leads to an interesting asymmetry between NO and IO visible in Fig.~\ref{fig:dirac_bulk_90_Cl_contour}. In the IO case (right panel), a particular parameter choice $M_D/\mu_1 = [-1.1,\,-1.2,\,3.2]$ yields an allowed region absent in the NO case (left panel). This arises because, for IO, a negative $M_D^{1,2}$ enhances $m_{1,2}$, improving consistency with the measured spectrum for inverted ordering. Conversely, for NO, the competing tendencies of $M_D^{3}$ (which lower $m_{3}$) and $M_D^{1,2}$ (which increases $m_{1,2}$) create a tension that tightens the bound on $\mu_1$. The complementary choice $M_D/\mu_1 = [2.1,\,1.2,\,-0.04]$ follows the same logic, and hence has a stronger bound for IO in comparison to NO.

\subsection{Majorana Bulk Term}
\label{sec:bulk-majorana}
As discussed in Section~\ref{sec:pheno-bulk-majorana}, the presence of bulk Majorana masses gives rise to a rich resonance structure, with resonances occurring at
\begin{equation}
    M_J = n\,\mu_1 
    \qquad \text{and} \qquad 
    M_J = \left(n + \tfrac{1}{2}\right)\mu_1 \,,
\end{equation}
where $n$ is an integer. Here, by resonance, we mean that the oscillation probability can be strongly enhanced or suppressed at these points. At the integer points $M_J = n\,\mu_1$, the oscillation probabilities in the presence of extra dimensions coincide exactly with those of the Standard Model.  
Small deviations from these points, however, can induce rapid and substantial changes in the oscillation pattern (cf. Eqs.~\eqref{eq:resonance_MM}-\eqref{eq:majorana-bulk-N-degenerate}).
The half-integer points $M_J = \left(n + \tfrac{1}{2}\right)\mu_1$ are particularly interesting, as all three active neutrino masses become strongly suppressed simultaneously (Eq.~\eqref{eq:lightest-special}), making it impossible to reproduce the observed oscillation data.
\begin{figure}[t]
    \centering
    \begin{subfigure}[b]{0.48\textwidth}
        \includegraphics[width=0.95\linewidth]{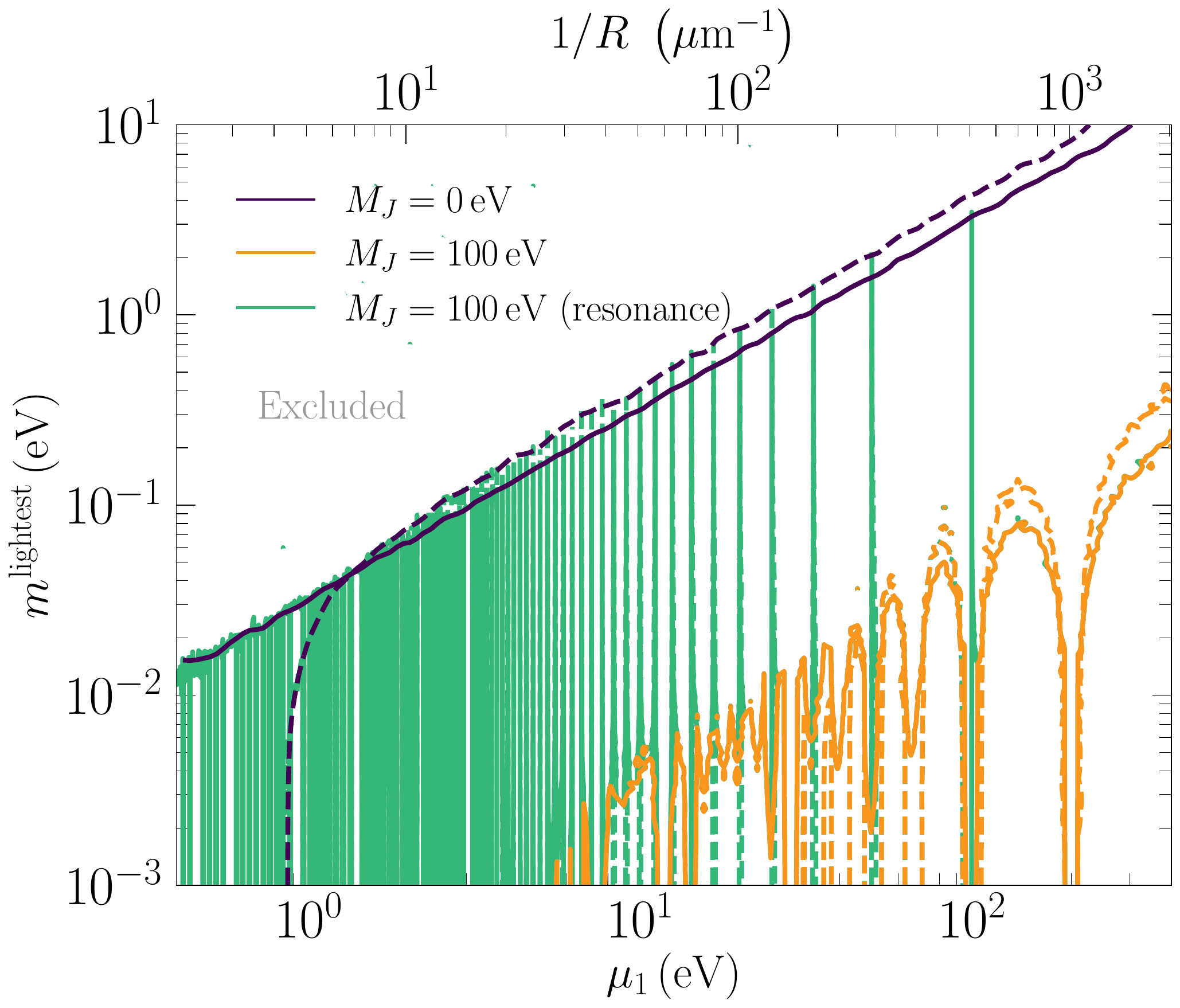}
        \caption{Normal Ordering}
    \end{subfigure}
    \begin{subfigure}[b]{0.48\textwidth}           \includegraphics[width=0.95\linewidth]{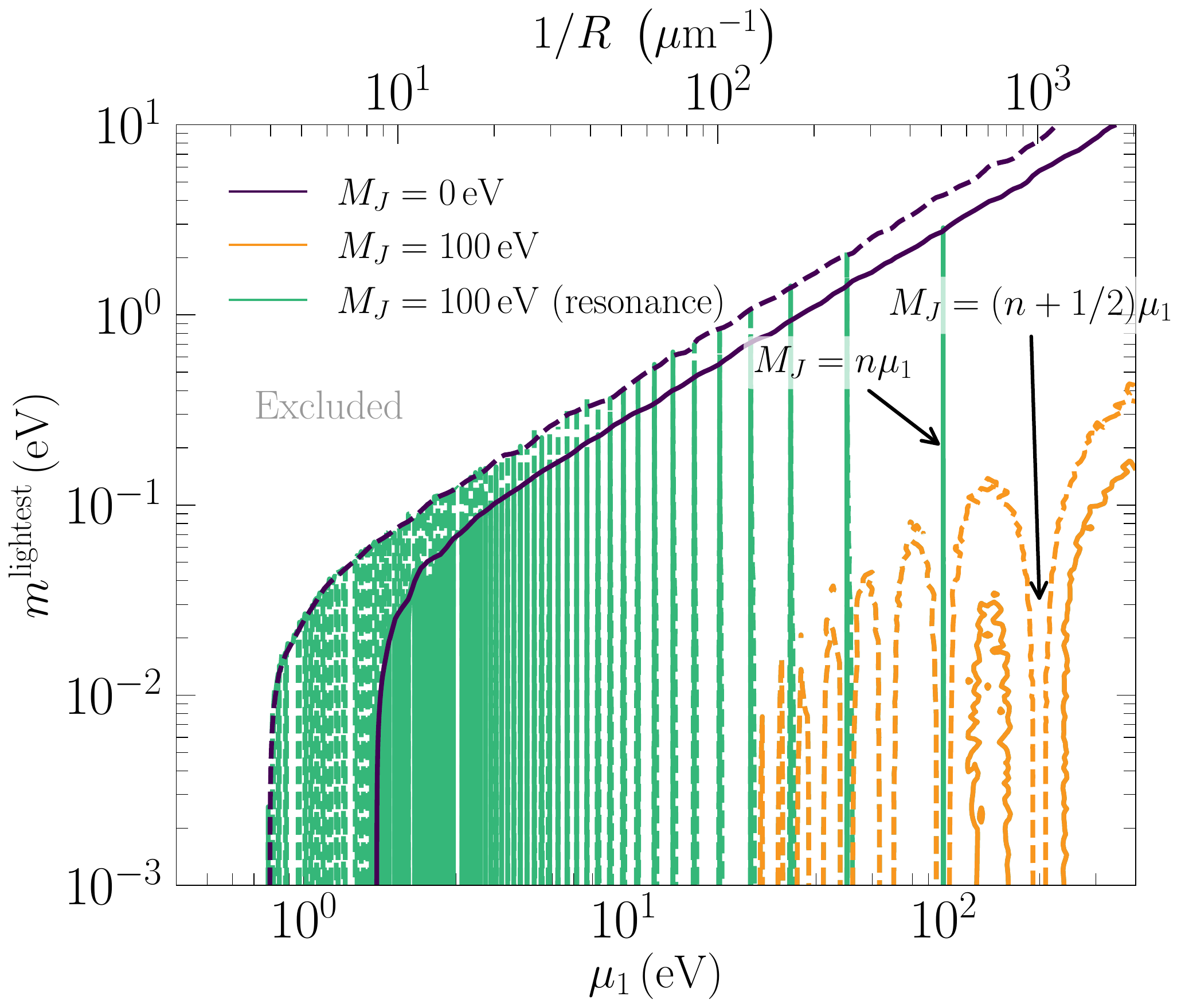}
        \caption{Inverted Ordering}
    \end{subfigure}
    \caption{90\%~C.L. exclusion contours for the Majorana bulk scenario in normal (left) and inverted (right) mass orderings, obtained from the two experiments under consideration. 
    Dashed lines correspond to MINOS/MINOS+, while solid lines denote Daya Bay. To better capture the resonance points, they have been presented in green, while the general contour excluding the resonance points is presented in yellow. It must be noted that the resonances are extremely fine-tuned, i.e. $M_J = n \mu_1$ in the limit, and have been enhanced for the purpose of visibility, but only an extremely fine random sampling can, in theory, encompass all the points. An estimate for the number of modes to be included can be found using the formula $ \lceil M_J/\mu_1\rceil+\left(\frac{ \pi m_D}{\sqrt{2}\mu_1}\right)^2$, where the $\lceil M_J/\mu_1\rceil$ ensures that the resonance structure is resolved, and $\left(\frac{ \pi m_D}{\sqrt{2}\mu_1}\right)^2$ is added to obtain the standard LED effect in limit $\mu_1>M_J$. Small fluctuations are statistical artifacts.}
    \label{fig:bulk_majorana_90_Cl_contour}
\end{figure}
Fig.~\ref{fig:bulk_majorana_90_Cl_contour} illustrates these ``resonance" structures for baselines relevant to the experiments under consideration.  
The behaviour matches the expectations outlined in Section~\ref{sec:pheno-bulk-majorana} remarkably well.  
Even when allowing for non-degenerate Majorana masses, the resonance structure persists across the parameter space.  
This is because there can exist parameter points where $M_J^i \approx n_i \mu_1$ or $M_J^i \approx (n_i + 0.5)\mu_1$, with $n_1,n_2,n_3$ non-degenerate. More on this has been discussed in Section~\ref{app:lagragian_parameter_scans}.

In the regime $\mu_1 \ll M_J$, the resonances become extremely dense. 
However, it must be stressed that, even though the contour in Fig.~\ref{fig:bulk_majorana_90_Cl_contour} would make one believe that in this limit one goes back to the vanilla case this is not the case as the resonances have been particularly enhanced for visibility and being at the special point $M_J = n\mu_1$ it becomes statistically unlikely for random parameter choices to reproduce Standard Model–like oscillations.  
However, one should note that perfect resonances ($M_J = n\,\mu_1$) remain possible and would yield oscillation probabilities identical to those of the Standard Model.  
For this reason, we refrain from quoting definitive bounds on $\mu_1 = 1/R$ in this scenario.
However, one can indeed say that the bound is at least as strong as the vanilla case with $M_J = 0$, and furthermore, if one adopts a statistical interpretation where such fine-tuned resonances are treated as effectively excluded, the resulting bounds would in fact be much stronger than those of the vanilla Dirac brane scenario.  
This follows because, unlike in the Dirac bulk case, there is no analogous mechanism, such as wavefunction localisation, that can extend the viable parameter space.

Finally, we note that parameter points with bulk Majorana masses of order $\mathcal{O}(1\,\mathrm{eV})$ are already excluded by existing sterile neutrino searches.  
As can be seen from the mass matrix in Eq.~\eqref{eq:Bulk-Majorana-MM}, in the limit $\mu_1 \gg m_D, M_J$, the setup effectively reduces to a sterile-neutrino-like scenario, allowing existing bounds to be directly applied.  
A caveat, however, is that specific fine-tuned resonances may still reproduce Standard Model oscillations, and thus cannot be fully ruled out.

\subsection{Majorana Brane Term}
\label{sec:brane-majorana}
\begin{figure}[t]
    \centering
    \begin{subfigure}[b]{0.495\textwidth}
        \includegraphics[width=\linewidth]{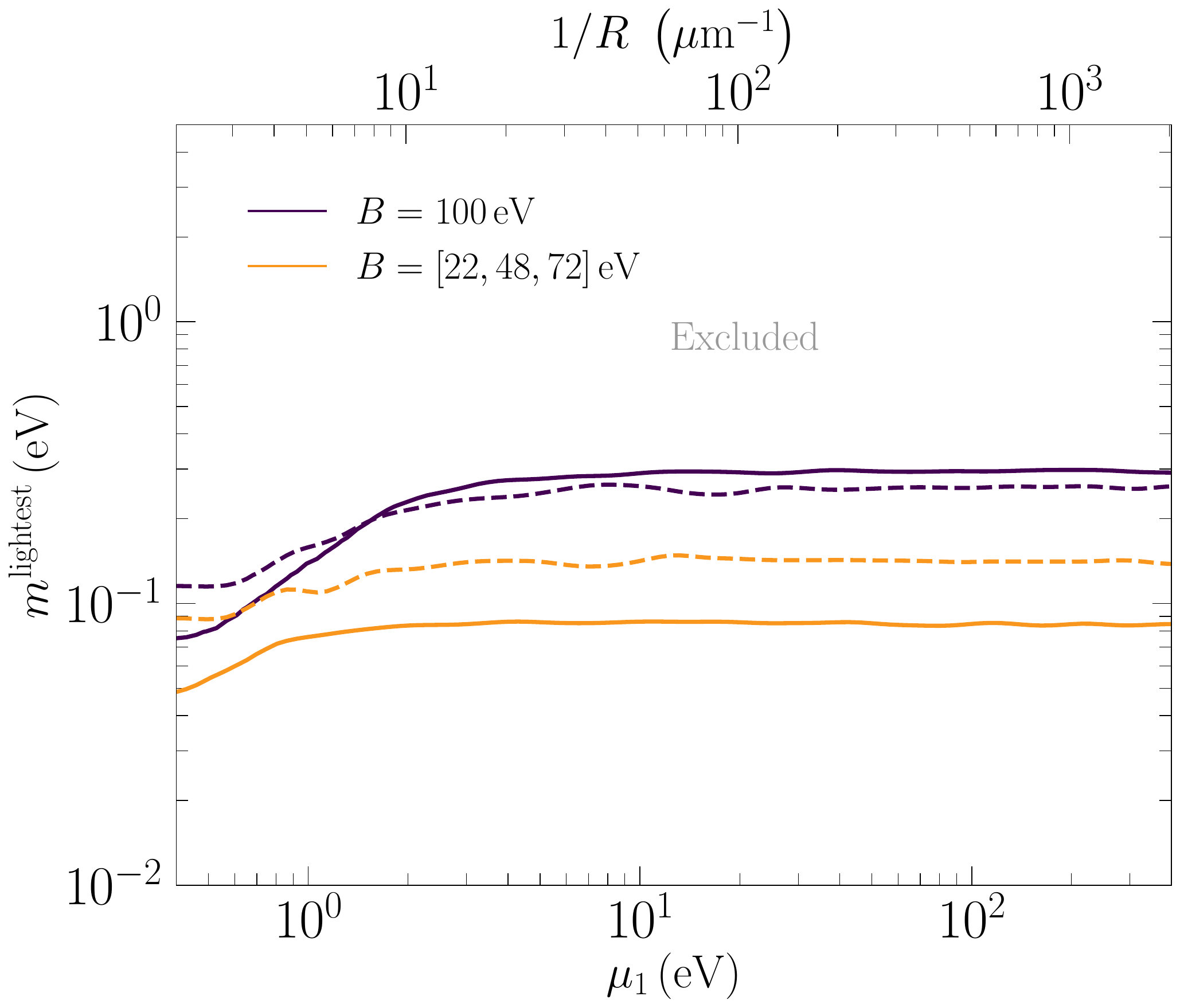}
        \caption{Normal Ordering}
    \end{subfigure}
    \begin{subfigure}[b]{0.495\textwidth}
        \includegraphics[width=\linewidth]{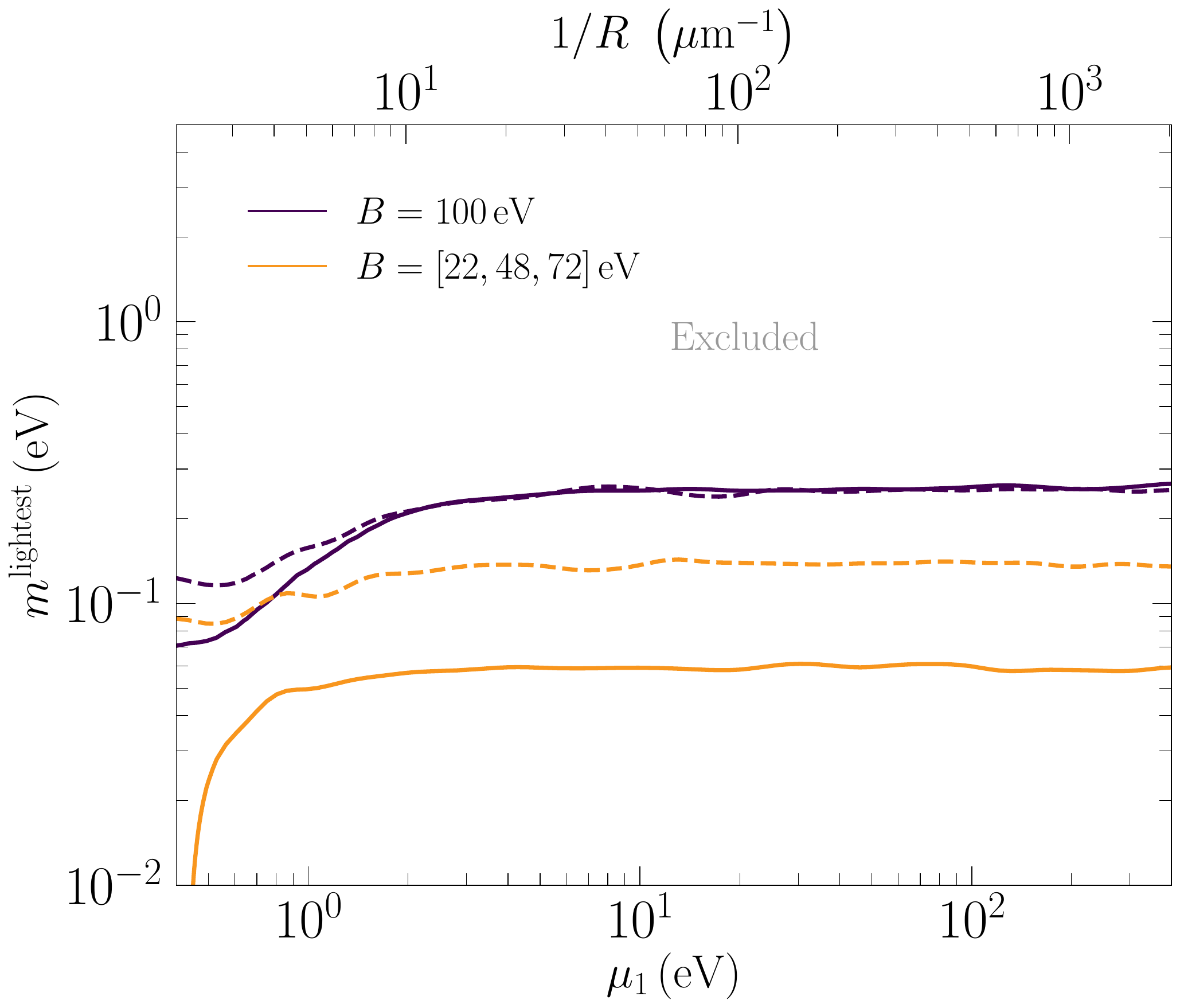}
        \caption{Inverted Ordering}
    \end{subfigure}
    \caption{90\%~C.L. exclusion contours for normal ordering (left) and inverted ordering (right), obtained from the two experiments under consideration. The dotted lines correspond to MINOS/MINOS+, and the solid lines to Daya Bay. An estimate for the minimal number of modes to be included can be found in ~Eq.~\eqref{eq:majorana-brane-number}. Small fluctuations are statistical artifacts.}
    \label{fig:majo_brane_90_Cl_contour}
\end{figure}

This section examines the dynamics in the presence of a brane-localised Majorana term, corresponding to $B \neq 0$ and $M_J = M_D = 0$. 
Fig.~\ref{fig:majo_brane_90_Cl_contour} shows the resulting 90\%~C.L. exclusion contours for illustrative choices of the parameter $B$.
The most striking feature of these contours is their near independence from the extra-dimensional parameter $\mu_1$. As discussed in Eq.~\eqref{eq:lightest_brane_mass}, this behaviour arises because, in the limit $B \gg m_D$, the light neutrino mass spectrum is governed by an effective seesaw-like relation in which the dependence on $\mu_1$ cancels out. Consequently, one recovers an approximately straight contour in the $(m_D,\mu_1)$ plane.
This approximation holds as long as $B \gg m_D$, consistent with the analytic limit derived in Eq.~\eqref{eq:contraint_brane}. For smaller values of $B$, however, sterile neutrino-like oscillations begin to play a role. Thus, as in the bulk Majorana case, phenomenologically viable scenarios typically require relatively large brane Majorana masses. 

Unlike the bulk Majorana scenario, the exclusion plots here do not display a pronounced resonance structure. This behaviour is consistent with the theoretical discussion in Eqs.~\eqref{eq:semi_diagonal_brane_transform}--\eqref{eq:semi_diagonal_brane_MM}, where it was shown that perfect resonances cannot be realised for the dense effective mass matrix associated with the brane Majorana term.
Interestingly, this also highlights how brane-localised Majorana interactions can effectively mask the extra-dimensional nature of neutrino propagation. In this regime, the dominant phenomenology is driven by brane physics, rendering neutrino oscillations almost insensitive to the details of the higher-dimensional geometry.
\section{Summary and Conclusions}
\label{sec:conclusions}
Neutrino oscillations provide some of the most compelling evidence for physics beyond the Standard Model. The observations can be elegantly explained if neutrinos are massive. In the simplest extensions, neutrino masses arise by introducing right-handed neutrinos (RHNs), which are fermion singlets under the SM gauge group. Their singlet nature places RHNs uniquely: they can act as a portal to a dark or hidden sector. This possibility is naturally realised in the context of extra dimensions, where even if the SM gauge fields and matter are localised on a brane, singlet fermions can consistently propagate in the bulk. In such scenarios, the SM left-handed neutrinos mix with an infinite tower of bulk modes, leading to characteristic distortions in neutrino observables. Conversely, an experimental confirmation of the extra-dimensional nature of neutrinos would offer a powerful and complementary probe of the geometry and properties of the extra dimension, alongside gravitational tests.

In this work, we have carried out an in–depth study of such scenarios, assuming the existence of a fifth compact orbifolded dimension through which three bulk neutrinos can propagate. We have considered four minimal realisations, in which neutrinos acquire mass from either a Dirac or a Majorana term localised in the bulk or on the brane. To better isolate the qualitative effect of each structure, in \secref{sec:phenomenology} we analysed these four cases separately under the simplifying assumption that all mass matrices can be simultaneously diagonalised in flavour space. This allows one to derive fully analytical expressions for the KK wavefunctions, masses and mixings, and to understand in detail how the active neutrinos couple to the KK tower in each scenario.
For each benchmark, we obtained closed–form expressions for the mass spectrum and for the active–sterile mixing encoded in the normalisation factors of the lightest modes. We then studied the spectra in several interesting limits: how the KK tower decouples when the compactification scale $\mu_1 = 1/R$ is large;  how the SM–like neutrino mass is recovered or modified depending on whether the dominant mass term is Dirac or Majorana and finally how additional bulk parameters (such as a Dirac mass $M_D$ or Majorana mass $M_J$) reshape the KK spectrum and the localisation of the zero mode. These analytical insights form the backbone of the phenomenological discussion in \secref{sec:phenomenology}.

In \secref{sec:setup}, we carried out a statistical analysis of the MINOS/MINOS+ and Daya Bay datasets in the parameter space $(m^{\rm lightest},\, \mu_1 = 1/R)$ for the scenarios defined in the previous section, with the key features summarised in Table~\ref{tab:summary}. In addition, the corresponding analysis in the $(m_D^{\rm lightest},\, \mu_1)$ plane is presented in App.~\ref{app:lagragian_parameter_scans}.. For clarity, we activated one term at a time in the mass matrix in order to highlight the qualitative behaviour of each contribution independently. In principle, all four terms could be switched on simultaneously, but this would mainly affect the quantitative strength of the bounds rather than the qualitative conclusions, and hence should be considered when conducting a global fit. 
In \secref{sec:dirac-brane}, we reproduced the established constraints for the standard (vanilla) Dirac--brane scenario. For the Dirac bulk case (\secref{sec:pheno-dirac-bulk}), we showed that the sign of the bulk Dirac mass $M_D$ plays a crucial role: positive values of $M_D$ lead to suppression of neutrino masses, resulting in strengthening of the bounds. Whereas, negative values of $M_D$ lead to enhancement of the neutrino masses, which in turn results in weakening of the bounds in the relevant parameter space. 
In scenarios with non-degenerate Dirac bulk masses, the relative signs of the bulk Dirac terms can favour one mass ordering over the other, yielding more stringent bounds on the disfavoured ordering.
The ``Majorana bulk’’ scenario (\secref{sec:bulk-majorana}) exhibits a particularly rich structure. Because the bulk Majorana mass $M_J$ shifts the KK spectrum, the oscillation probabilities develop ``resonances" whenever 
$M_J \simeq (n/2)\mu_1$ with integer $n$. At the half–integer points, the light neutrino masses are strongly suppressed, making it impossible to reproduce the observed oscillation pattern, meaning these regions are robustly excluded. At the integer points, however, the extra–dimensional model becomes exactly degenerate with the three–flavour SM, leading to ``blind spots’’ where oscillation experiments cannot distinguish the two. Away from these finely tuned loci, the resonant enhancement of active–sterile–like mixing typically yields stronger constraints than in the Dirac brane case. From a statistical point of view, one can thus regard the allowed parameter space as being at least as constrained as the vanilla scenario, with only measure–zero resonant lines remaining indistinguishable from the SM. 
Finally, in \secref{sec:brane-majorana}, we examined the brane-localised Majorana term and extracted the resulting bounds for representative choices of parameters. The scenario is a generalisation of a traditional Type-I seesaw.
Interestingly, despite the possibly very large amount of sterile neutrinos participating, the phenomenology is weakly sensitive to the size of the extra dimension. 

One might wonder how the results would be affected if some of the hypotheses of the work were modified; these include the (i) possibility of not being able to simultaneously diagonalise all mass matrices and (ii) the case of a curved extra dimension. We took the first steps in these directions in App.~\ref{app:generalisations}.  About (i), we derived in App.~\ref{app:matrices_not_diagonalized}, we considered the Majorana bulks scenario, assuming $m_D$ is not diagonal.  We derived an analytic formula which generalises the main text findings and shows how this can be obtained in other constructions. Regarding (ii), we discuss in App.~\ref{app:warped} the modifications to the Dirac brane scenario in a warped extra dimension. Interestingly, for such a case, formally, the results follow the same functional structure, so that the constraints derived in the main text can be straightforwardly applied. The same holds for the brane Majorana case. Instead, this seems not to be the case when bulk masses are included. This is left for future work.

Neutrinos offer a powerful and complementary way to test the existence and properties of extra dimensions. While in this work we have focused on oscillation experiments, the analytical results derived here for the KK spectra and mixings can be straightforwardly applied to other observables, ranging from astrophysical neutrino fluxes to laboratory searches and collider signatures.  In combination with precision gravitational tests, future high-precision neutrino experiments such as JUNO, DUNE, IceCube, T2HK and nuSTORM have the potential to place bounds on the size of the extra dimension that are significantly stronger than the current limits and will provide a unique window into new spatial dimensions. 
\section*{Acknowledgments}
We would like to thank Ivan Martinez-Soler for useful comments on the draft.
A.d.G. thanks V. Takhistov and the International Center for Quantum-field Measurement Systems for Studies of the Universe and Particles (QUP/KEK) for their hospitality and the stimulating working environment during which a core part of this work was realised.
This article/publication is based upon work from COST Action COSMIC WISPers CA21106, supported by COST (European Cooperation in Science and Technology). J.T. would like to thank the Quantum Field Theory Centre at the University of Southern Denmark for their hospitality during the completion of this work.
\appendix

\section{Phenomenology with the Lagrangian parameter $m_D$
}
\label{app:lagragian_parameter_scans}

In the main analysis presented in Section~\ref{sec:setup}, we adopted the physical lightest neutrino mass $m^{\rm lightest}$ as the primary parameter, since it is the quantity most directly constrained by terrestrial and astrophysical observations. However, from a model–building perspective, it is often more natural to formulate the parameter space in terms of the underlying Lagrangian Dirac mass $m_D$. To facilitate comparison between these viewpoints, we provide here the corresponding parameter–space scans performed with $m_D^{\rm lightest}$ as the free parameter. The remaining two Dirac masses are fixed using the mass–squared differences at their global–fit best–fit values~\cite{Esteban:2024eli}, with all other aspects of the scan identical to those used in Section~\ref{sec:setup}.

\begin{figure}[t]
    \centering
    \includegraphics[width=0.7\linewidth]{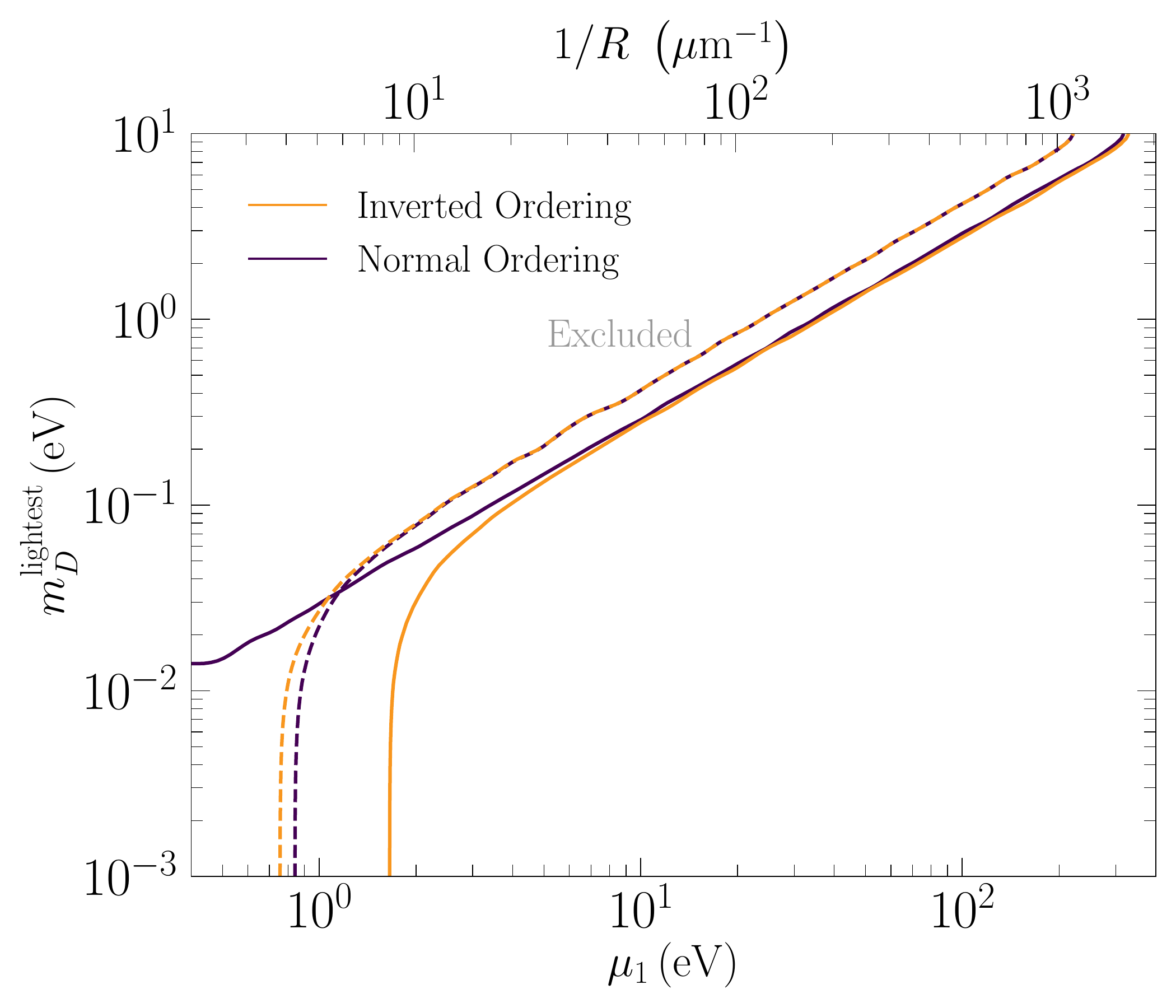}
    \caption{90\%~C.L.\ exclusion contours for the Dirac–brane scenario with the lagrangian parameter $m_D$ in normal (purple) and inverted (orange) mass orderings, obtained from MINOS/MINOS+ and Daya Bay. Solid lines correspond to Daya Bay and dotted lines to MINOS/MINOS+. Small fluctuations are statistical artifacts.}
    \label{fig:Dirac_Brane_MINOS_DayaBay_lagrangian}
\end{figure}

Fig.~\ref{fig:Dirac_Brane_MINOS_DayaBay_lagrangian} shows the results for the vanilla Dirac–brane scenario. The exclusion contours obtained when scanning directly in $m_D^{\rm lightest}$ closely track those in Fig.~\ref{fig:Dirac_Brane_MINOS_DayaBay}, where the physical mass was used. This behaviour is expected: in the region of parameter space consistent with realistic oscillation probabilities, the scans empirically satisfy $m_D \ll \mu_1$, and in this limit the physical lightest mass follows $m_D$ up to small corrections, (cf.\ Eq.~\eqref{eq:db-asym}).

\begin{figure}[t]
    \centering
    \includegraphics[width=0.99\linewidth]{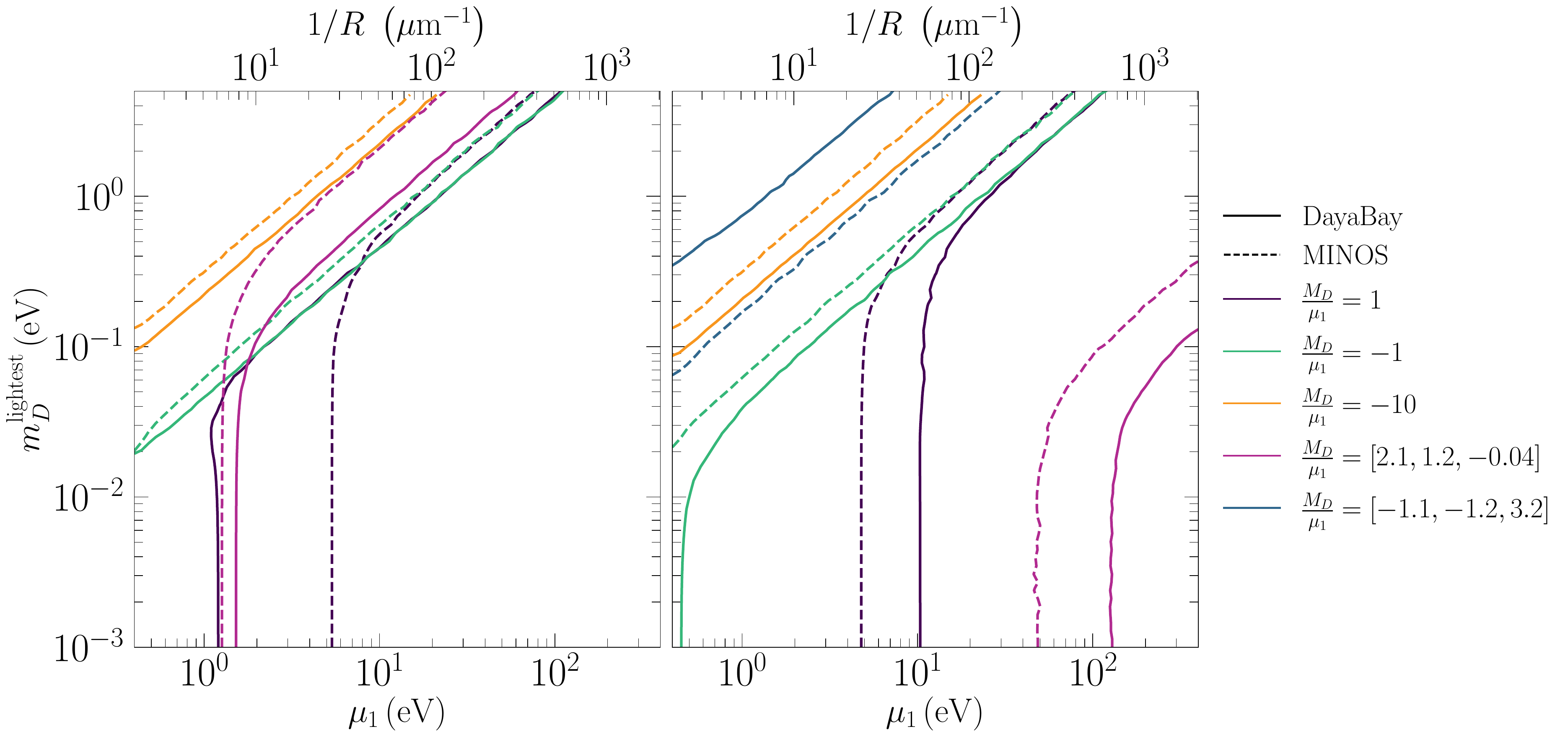}
    \caption{90\%~C.L.\ exclusion contours for the Dirac–bulk scenario in normal ordering (left) and inverted ordering (right). Small fluctuations are statistical artifacts. }
    \label{fig:dirac_bulk_90_Cl_contour_lagrangian}
\end{figure}
\begin{figure}[t]
    \centering
    \begin{subfigure}[b]{0.48\textwidth}
        \includegraphics[width=0.95\linewidth]{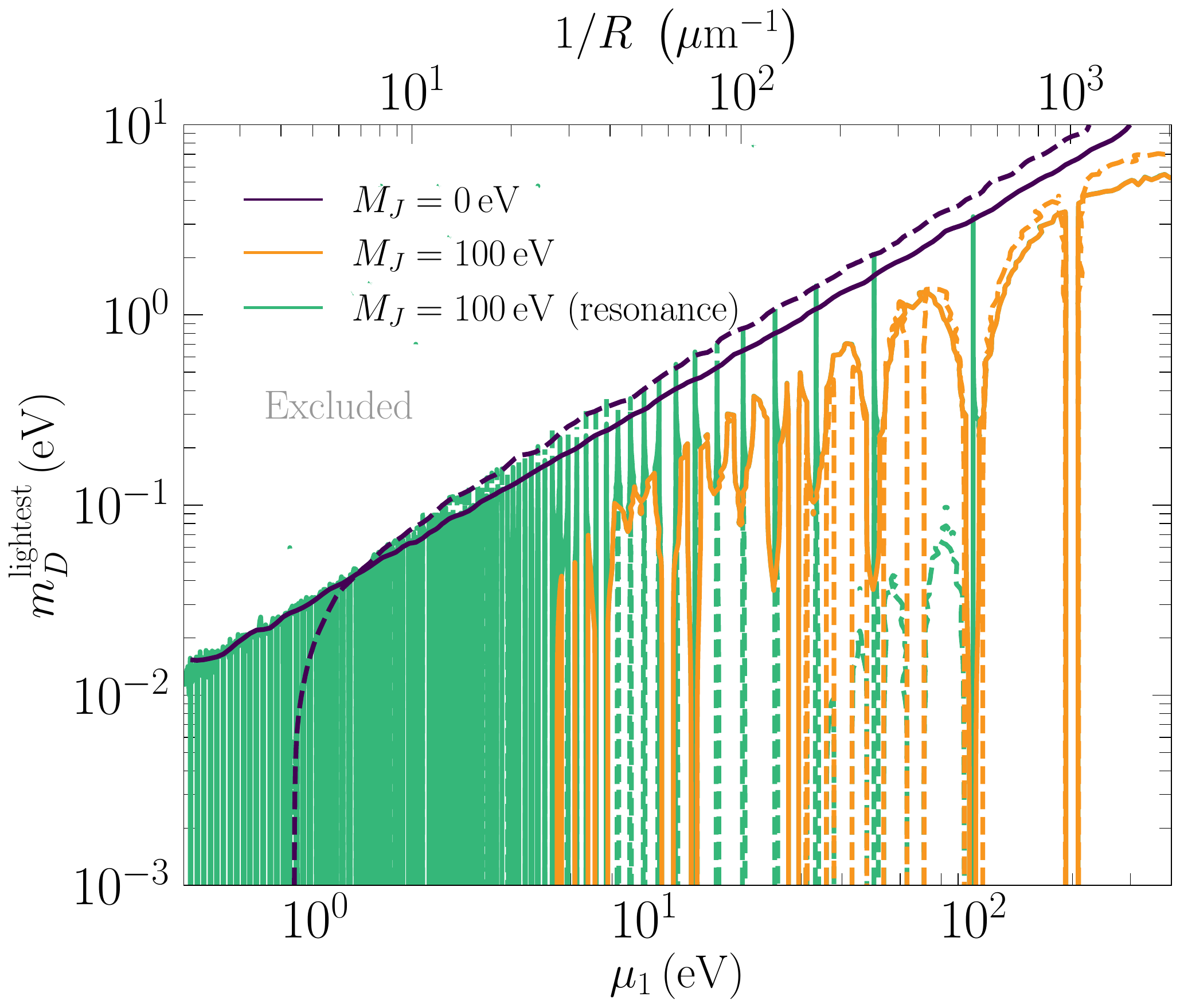}
        \caption{Normal Ordering}
    \end{subfigure}
    \begin{subfigure}[b]{0.48\textwidth}
        \includegraphics[width=0.95\linewidth]{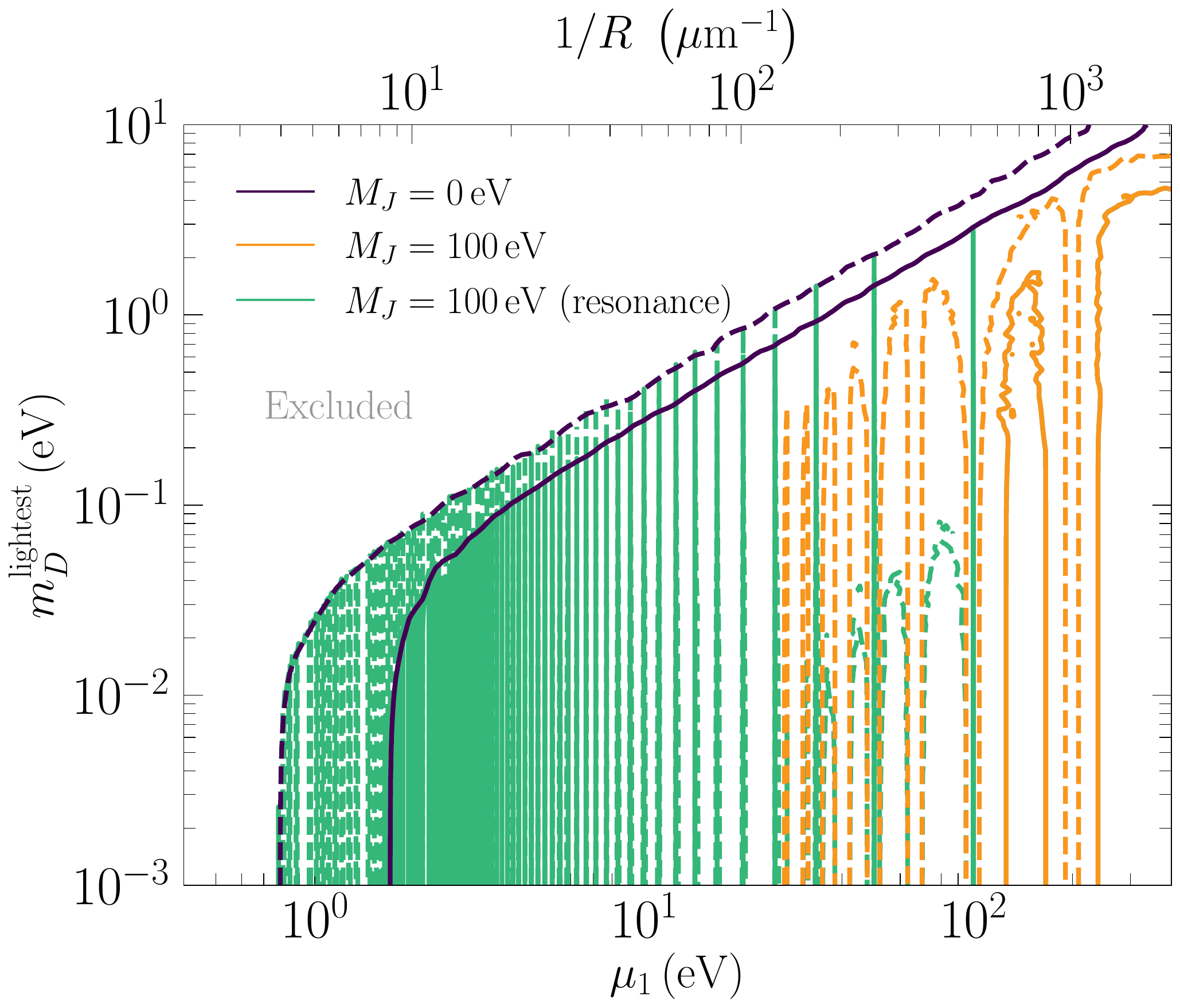}
        \caption{Inverted Ordering}
    \end{subfigure}
    \caption{90\%~C.L.\ exclusion contours for the Majorana–bulk scenario in normal (left) and inverted (right) orderings. Dashed lines correspond to MINOS/MINOS+, and solid lines to Daya Bay. Resonance regions ($M_J = n\mu_1$) are shown in green; the non-resonant regions in yellow. Resonance bands are visually enhanced for clarity. Small fluctuations are statistical artifacts.}
    \label{fig:bulk_majorana_90_Cl_contour_lagrangian}
\end{figure}
The Dirac–bulk case is shown in Fig.~\ref{fig:dirac_bulk_90_Cl_contour_lagrangian}. Here, the mapping between $m_D$ and the physical masses becomes strongly non–linear due to the interplay between the brane–localised Dirac mass and the bulk Dirac mass term. Depending on the sign of the bulk mass, the physical eigenvalues can be either enhanced or suppressed (Eq.~\eqref{eq:dirac-bulk-lightest}, leading to spectra that differ substantially from those in Fig.~\ref{fig:dirac_bulk_90_Cl_contour}. This feature is particularly visible in the non–degenerate contours: when scanning in terms of the physical mass, one obtains much tighter constraints than when scanning directly in $m_D$, reflecting the above non–linearity.

Results for the Majorana–bulk scenario are shown in Fig.~\ref{fig:bulk_majorana_90_Cl_contour_lagrangian}. The green resonance bands align closely with those in Fig.~\ref{fig:bulk_majorana_90_Cl_contour}, since the resonance condition $M_J = n\mu_1$ depends only on the bulk parameters and not on whether the scan is performed in terms of $m_D$ or the physical masses. By contrast, the non–resonant yellow contours shift toward larger values of $m_D$ relative to the corresponding physical–mass scan. This is a consequence of the suppression of the physical neutrino masses by the additional bulk Majorana mass term through a seesaw–like mechanism.

\begin{figure}[t]
    \centering
    \begin{subfigure}[b]{0.48\textwidth}
        \includegraphics[width=0.95\linewidth]{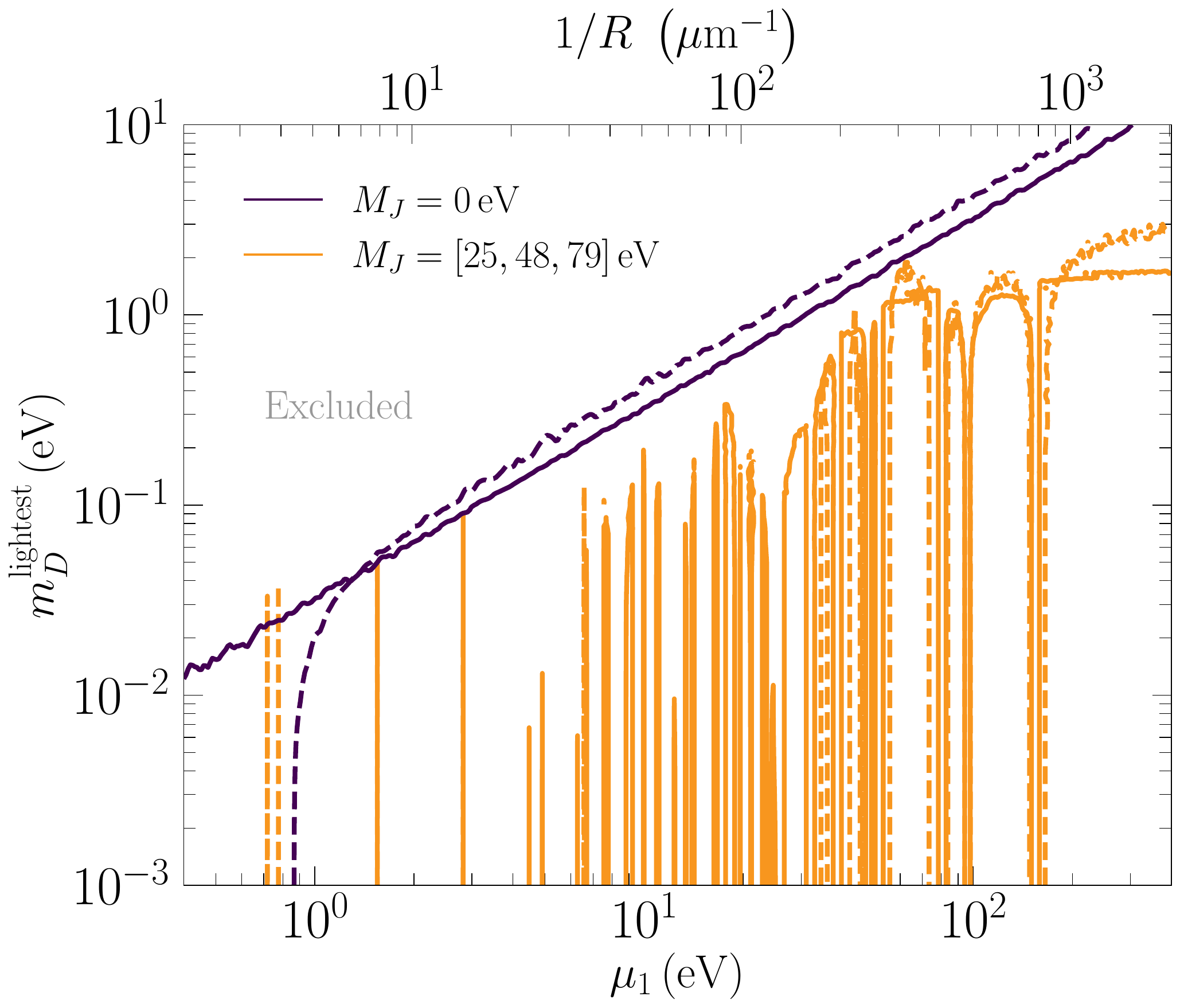}
        \caption{Normal Ordering}
    \end{subfigure}
    \begin{subfigure}[b]{0.48\textwidth}
        \includegraphics[width=0.95\linewidth]{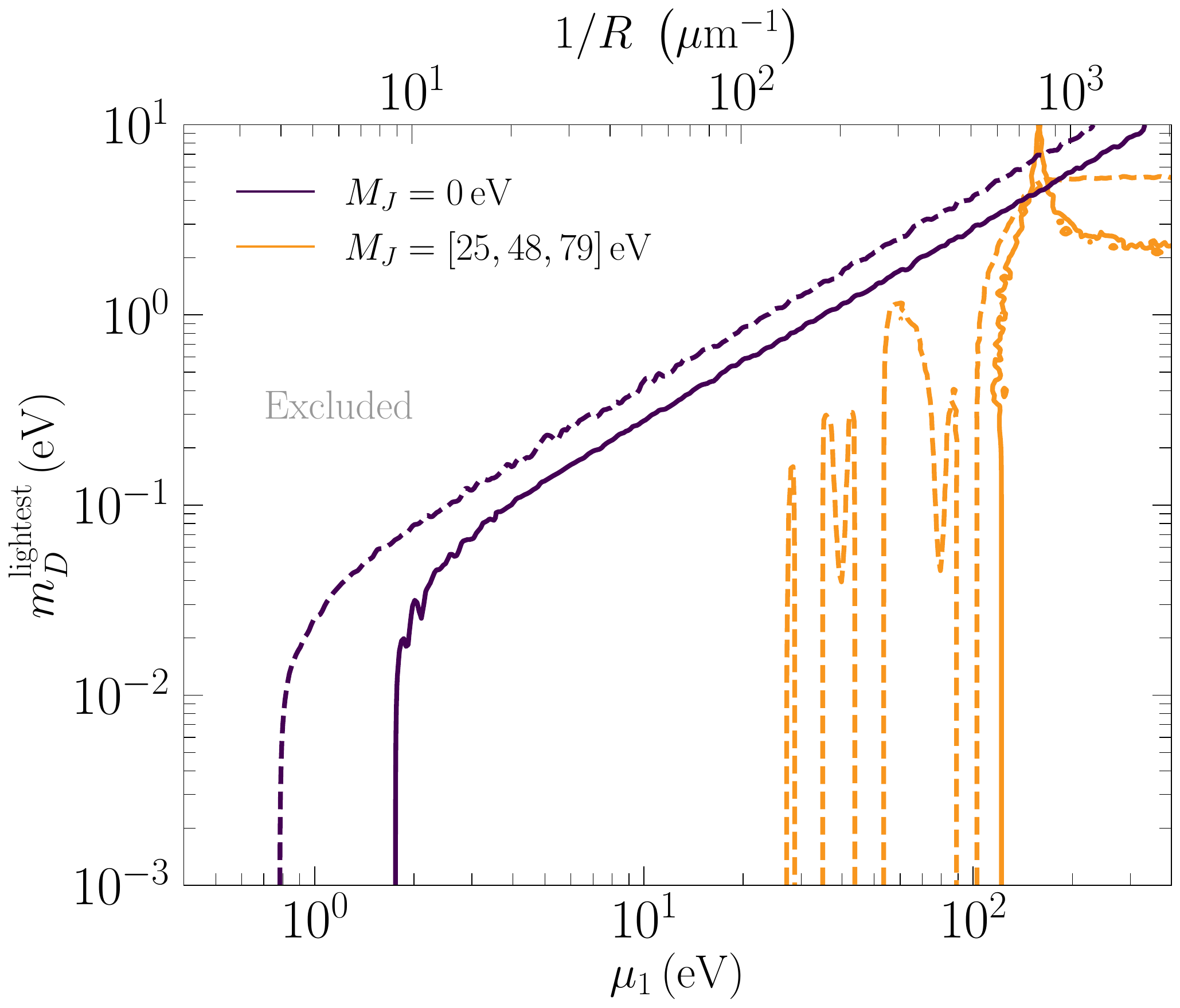}
        \caption{Inverted Ordering}
    \end{subfigure}
    \caption{90\%~C.L.\ exclusion contours for the non-degenerate Majorana–bulk scenario in normal (left) and inverted (right) orderings. Dashed lines correspond to MINOS/MINOS+, and solid lines to Daya Bay. Small fluctuations are statistical artifacts.}
    \label{fig:bulk_majorana_90_Cl_contour_nondeg_lagrangian}
\end{figure}

The non–degenerate Majorana–bulk case is shown in Fig.~\ref{fig:bulk_majorana_90_Cl_contour_nondeg_lagrangian}. When the three bulk Majorana masses $M_J^i$ do not share a common rational ratio, the resonance condition $M_J^i = n_i \mu_1$ cannot be satisfied simultaneously for all three generations. Nevertheless, approximate resonances still arise when the condition is satisfied for at least one $M_J^i$ and approximately for the others, producing the visible structures in the scan. As before, the full set of resonance points would require extremely fine sampling.

\begin{figure}[t]
    \centering
    \begin{subfigure}[b]{0.495\textwidth}
        \includegraphics[width=\linewidth]{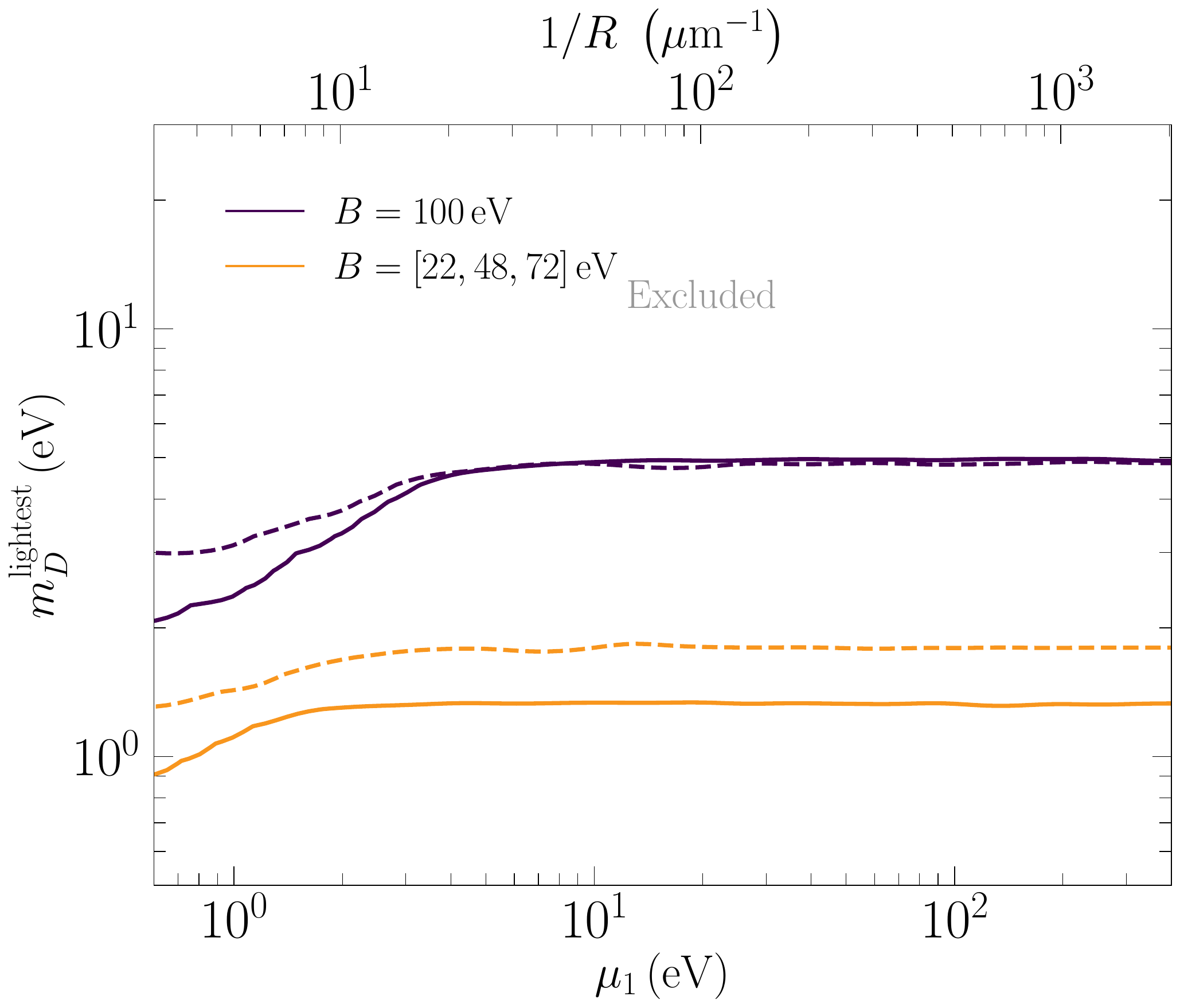}
        \caption{Normal Ordering}
    \end{subfigure}
    \begin{subfigure}[b]{0.495\textwidth}
        \includegraphics[width=\linewidth]{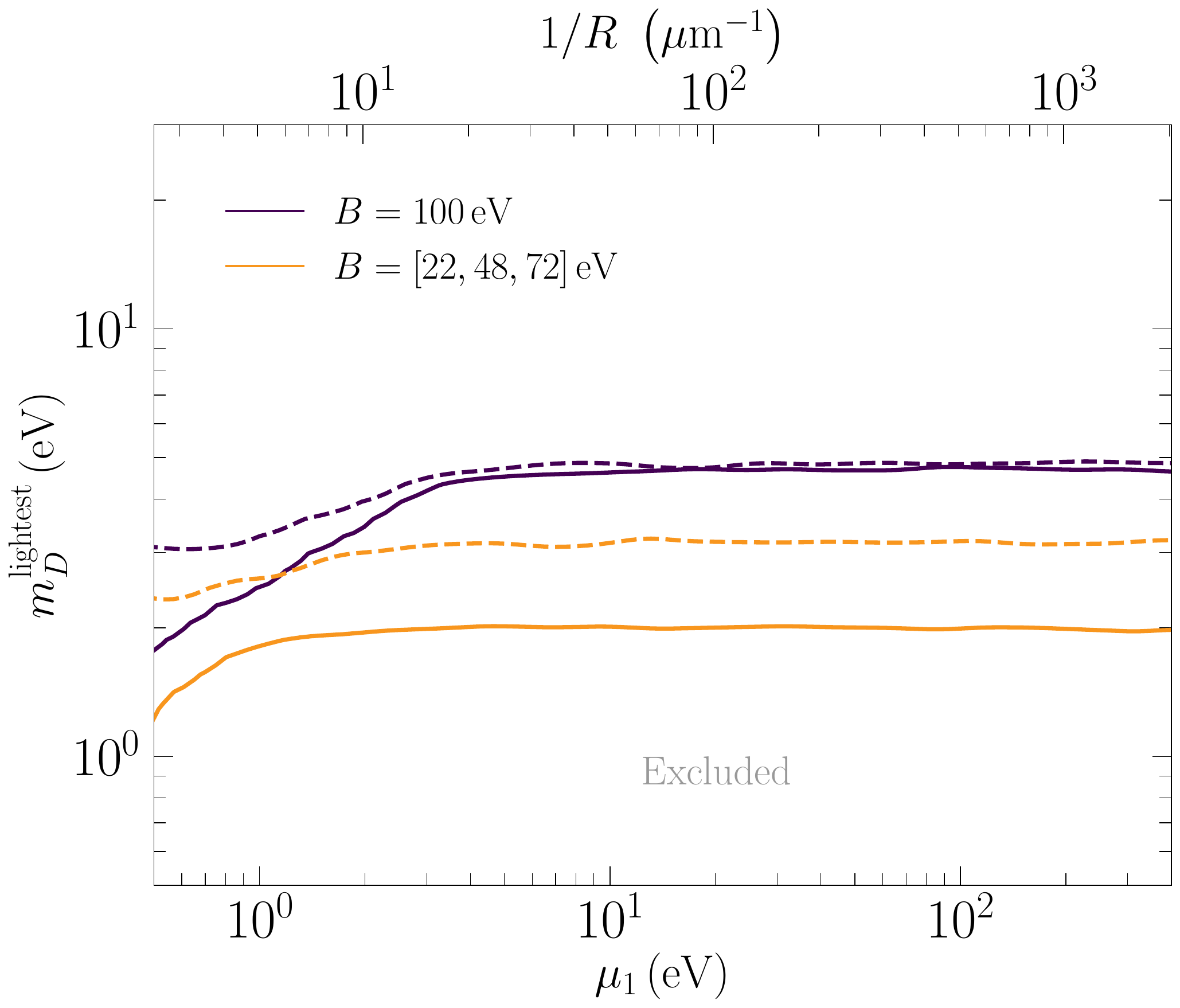}
        \caption{Inverted Ordering}
    \end{subfigure}
    \caption{90\%~C.L.\ exclusion contours for the Brane–Majorana scenario in normal (left) and inverted (right) orderings. Dotted lines correspond to MINOS/MINOS+, and solid lines to Daya Bay. Small fluctuations are statistical artifacts.}
    \label{fig:majo_brane_90_Cl_contour_lagrangian}
\end{figure}

Finally, Fig.~\ref{fig:majo_brane_90_Cl_contour_lagrangian} shows the Brane–Majorana scenario. As in the main analysis, the resulting constraints are largely insensitive to the value of $\mu_1$ and hence largely independent of the extra–dimensional dynamics. The contours shift toward larger $m_D$ values compared to the corresponding physical–mass scan, an effect again attributable to the Majorana mass term suppressing the physical mass through a type-I–like seesaw mechanism.

Across all scenarios, scanning the parameter space using the Lagrangian mass $m_D$ reproduces the qualitative features of the analysis performed using the physical mass $m^{\rm lightest}$, but with notable quantitative differences in cases where bulk or brane Majorana terms induce suppression of the physical mass. 
These observations underscore the importance of performing scans in both parameterisations when assessing the experimental reach of extra–dimensional neutrino models.

\section{Generalisations}
\label{app:generalisations}
In the main text, the analysis was carried out under simplifying assumptions, such as the existence of a single flat extra dimension and the simultaneous diagonalisability of the mass matrices involved in the Lagrangian. In this appendix, we want to comment on some of the possible generalisations.

\subsection{Matrices are Not Simultaneously Diagonalisable}
\label{app:matrices_not_diagonalized}
The hypothesis of having all matrices simultaneously diagonalisable is quite constraining. While it could be realised on the model-building side by imposing flavour symmetries, it can be quite limiting. 
A full treatment of such a system goes beyond the scope of the work, and no closed formulas can be derived for generic matrix structures. 
However, let us briefly discuss here what would change compared with the previous derivations. 

For the sake of conciseness, let us consider the case of a bulk Majorana mass and compare it with the results of section~\ref{sec:pheno-bulk-majorana}. We choose to work in the basis in which the Majorana mass matrix $M_J$ is diagonal, and $m_D$ is not.
One can then show that the eigenvalues of the system can be derived by solving 
\begin{align}
    \label{eq:3x3-eigenvalues}\det\left(\sum\limits_{n=0}^{N}m_D^T\frac{\chi_n^2}{(m_\lambda-M_J)^2- \mu_n^2} m_D-\frac{m_\lambda}{m_\lambda-M_J}\right)=0\,,
\end{align}
where now $m_\lambda$ is a $3\times 3$ matrix proportional to the identity matrix.
Details on the derivation of such a result can be found in App.~\ref{app:3eigen}. 

Notice that, if $M_J$ and $m_D$ are simultaneously diagonal, as in the main text, the determinant of Eq.~\eqref{eq:3x3-eigenvalues} factorises
\begin{align} &\prod\limits_{\alpha=1}^3\left(\sum\limits_{n=0}^{N}\frac{\chi_n^2 m_{D,\alpha}^2}{(m_\lambda-M_{J,\alpha})^2- \mu_n^2}-\frac{m_\lambda}{m_\lambda-M_{J,\alpha}}\right)=0\,.
\end{align}
The above equation allows for three ``copies" of solutions which match those found for the diagonal case in section~\ref{sec:pheno-bulk-majorana}.

An analogous discussion follows for the eigenvectors; they read
\begin{align}
    \label{eq:3x3-eigenvectors}&u^\alpha_\lambda=\mathcal{N}_\lambda\begin{pmatrix}
        u_0^\alpha\\
        \frac{\chi_0}{m_\lambda-M_J} m_D u_0^\alpha\\
        \frac{\chi_1/\sqrt{2}}{m_\lambda-(M_J-\mu_1)} m_D u_0^\alpha\\
        \frac{\chi_1/\sqrt{2}}{m_\lambda-(M_J+\mu_1)} m_D u_0^\alpha\\
        \dots
    \end{pmatrix}\,,    &u_0^1=\begin{pmatrix}
        1\\0\\0
    \end{pmatrix}\,, u_0^2=\begin{pmatrix}
        0\\1\\0
    \end{pmatrix}\,, u_0^3=\begin{pmatrix}
        0\\0\\1
    \end{pmatrix}\,.
\end{align}
More details on the derivation can be found in App.~\ref{app:3eigen}. The choice of $\alpha$ determines the flavour. Once again, if all matrices are diagonal, then for each $\alpha$ there will be several zeros, and by rearranging the rows, one can always get back to the eigenvectors derived in the all-diagonalisable case.

The results of Eqs.~\eqref{eq:3x3-eigenvalues} and \eqref{eq:3x3-eigenvectors} can be derived in a similar fashion for all the other cases. The sum can be formally performed. All in all, they ought to speed up numerical calculations, allowing the evaluation of more intricate cases.

\subsection{Warped Extra Dimension}
\label{app:warped}
We extend the discussion to the Randall-Sundrum~(RS) warped extra dimension~\cite{Randall:1999ee,Randall:1999vf}. The RS metric is given by
\begin{equation}
    ds^2=A(y)^2dx^2-dy^2\,,
\end{equation}
where $dx^2\equiv \eta_{\mu\nu}dx^\mu dx^\nu$, $A(y)\equiv e^{-k|y|}$ is the so-called \textit{warping factor} and $k\sim\sqrt{-\Lambda_B}$ parametrizes the curvature of the extra dimension. We define now, for convenience, the dimensionless parameter
\begin{equation}
    \mu\equiv kR\,,
\end{equation}
which controls the warping effect on observables.
They are given by
\begin{align}
    &\Omega_\mu=\frac{1}{2}A'(y) \gamma_5 \gamma_\mu\,, &\Omega_5=0\,.
\end{align}
The kinetic part of the Lagrangian then reads
\begin{align}
\label{eq:kinetic-RS}
    S\supset&\int dy\,\sqrt{g}\, i\ov{\Psi}e_a^M \Gamma^a\nabla_M\Psi= \int dyA^4\left[A^{-1}i\ov{\Psi}\Gamma^\mu\partial_\mu\Psi+i\ov{\Psi}\Gamma^5\left(\partial_5+2A'A^{-1}\right)\Psi\right]\,,\nonumber\\
    &=\int dy\left[A^{3}i\ov{\Psi}\gamma^\mu\partial_\mu\Psi-A^4\ov{\Psi}\gamma^5\left(\partial_5+2A^{-1}A'\right)\Psi\right]\,.
\end{align}
With respect to the flat case, the derivative of the warp-factor $A'$ enters as a contribution to the KK WFs and masses.
The expression can be simplified by performing a rotation of the field $\Psi = A^{-3/2}\hat\Psi$, leading to the simplified Lagrangian
\begin{equation}
    S\supset\int dy\left[i\ov{\hat\Psi}\gamma^\mu\partial_\mu\hat\Psi-\ov{\hat\Psi}\gamma^5\left(A\,\partial_5+\frac{1}{2}A'\right)\hat\Psi-\sign{y} A M_D \ov{\hat\Psi}\hat\Psi-A\frac{M_J}{2}\ov{\hat\Psi}\hat\Psi^c\right]\,.
\end{equation}
Since all mass terms in the bulk receive a $A^4$ contribution from the determinant of the metric and a $A^{-3/2}$ from each fermion field due to the field redefinition. All in all, all dimensionful quantities are also redefined
\begin{equation}
    M\to A(y) M\,.
\end{equation}
In such a basis, the EOMs in a warped background read
\begin{align}
\label{eq:EOMs-warped}
    &i\slashed{\partial}\Psi-\gamma^5\left(A\,\partial_5+\frac{1}{2}A'\right) \hat\Psi=A\left(\sign{y} M_D \Psi+\frac{M_J}{2}\hat\Psi^c\right)\,,
\end{align}
matching preciously found results~\cite{Huber:2003sf}.

In the massless bulk case, Eq.~\eqref{eq:EOMs-warped} reduces to the system of equations
\begin{align}
    &A(y)\hat\xi_n'+\frac{1}{2}A'(y)\hat\xi_n=-\mu_n \hat\chi_n\,, &A(y)\hat\chi_n'+\frac{1}{2}A'(y)\hat\chi_n=+\mu_n \hat\xi_n\,.
\end{align}
which, in the interval $y\in [0,\pi]$, yield for $n>0$
\begin{align}
    &\hat\chi_n(y)=(-1)^n \sqrt{\frac{2\pi \mu }{e^{\pi \mu}-1}}\cos \left( n\pi\,\frac{e^{k y}-1}{e^{\pi\mu }-1}\right)e^{ky/2}\,,\\
    &\hat\xi(y)=(-1)^n \sqrt{\frac{2\pi \mu }{e^{\pi \mu}-1}}\sin \left( n\pi\,\frac{e^{k y}-1}{e^{\pi\mu }-1}\right)e^{ky/2} \,,
\end{align}
where the sign $(-1)^n$ has been chosen as such for convenience.
The KK masses are given by
\begin{equation}
    \mu_n=k\,\frac{n\pi}{e^{\mu\pi}-1}\approx (n\pi) k e^{-\mu\pi}\,,
\end{equation}
where the $\approx$ sign corresponds to the leading $\mu\gg1$ limit.
On the brane at $y=\pi R$, the WFs read
\begin{align}
    &\hat\chi_n(\pi R)=\sqrt{\frac{2\pi \mu }{e^{\pi \mu}-1}}e^{\pi\mu/2}\,,\\
    &\hat\xi_n(\pi R)=0 \,.
\end{align}
The zero modes corresponding to $\mu_0=0$ read
\begin{align}
    &\hat\xi_0(y)=0\,, &\hat\chi_0(y)=\sqrt{\frac{\pi\mu}{e^{\pi\mu}-1}}e^{ky/2}=\frac{\hat\chi_n(\pi R)}{\sqrt{2}}\,.
\end{align}
The results match the previous findings of Ref.~\cite{Chang:1999nh}.

The WFs on the brane appear in the same ratio as in the flat case $\chi_n=\sqrt{2}\chi_0$, and the same goes for the masses $\mu_n=n\mu_1$. From the phenomenological perspective, this implies that bounds from neutrino physics can only be cast on the combination $(m_D \chi_0)$, and thus predictions for the observables remain formally the same. This implies that neutrino oscillations alone cannot distinguish warped and flat scenarios.
\section{Three Flavours Eigensystem}
\label{app:3eigen}
In this Appendix, we derive the formula for the three-flavour Majorana bulk case (see Section~\ref{sec:pheno-bulk-majorana} system assuming $M_J$ and $m_D$ are diagonal and non-diagonal matrices, respectively. Analogous results for the other case studies can be derived employing the same techniques. We discuss the eigenvalues and the eigenvectors separately.

\subsection{Eigenvalues}
We are interested in the eigenvalues of the block matrix of Eq.~\eqref{eq:Bulk-Majorana-MM}
\begin{align}
    \begin{vmatrix}
        -m_\lambda &  m_D^T\chi_{0} & m_D^T\tilde\chi_1 & m_D^T\tilde\chi_1  &\dots & m_D^T\tilde\chi_N & m_D^T\tilde\chi_N\\
        m_D\chi_{0} & M_J-m_\lambda & 0 & 0 &\dots & 0 & 0\\
        m_D\tilde\chi_1 & 0 & M_J+\mu_1-m_\lambda & 0 & \dots & 0 & 0\\
        m_D\tilde\chi_1 & 0 & 0 & M_J-\mu_1-m_\lambda & \dots & 0 & 0\\
        \dots & \dots & \dots & \dots & \dots & \dots & \dots\\
        m_D\tilde\chi_N & 0 & 0 & 0 & \dots & M_J+\mu_N-m_\lambda & 0\\
        m_D\tilde\chi_N & 0 & 0 & 0 & \dots & 0 & M_J-\mu_N-m_\lambda\\
    \end{vmatrix}=0\,.
\end{align}
We use the property of the determinant, for which rescaling rows or columns by some number just rescales the whole determinant to find
\begin{equation}
\resizebox{.9\hsize}{!}{$
    \begin{vmatrix}
        -m_\lambda &  m_D^T & m_D^T & m_D^T  &\dots & m_D^T & m_D^T\\
        m_D & (M_J-m_\lambda)/(\chi_{0})^2 & 0 & 0 &\dots & 0 & 0\\
        m_D & 0 & (M_J+\mu_1-m_\lambda)/(\tilde\chi_1)^2 & 0 & \dots & 0 & 0\\
        m_D & 0 & 0 & (M_J-\mu_1-m_\lambda)/(\tilde\chi_1)^2 & \dots & 0 & 0\\
        \dots & \dots & \dots & \dots & \dots & \dots & \dots\\
        m_D & 0 & 0 & 0 & \dots & (M_J+\mu_N-m_\lambda)/(\tilde\chi_N)^2 & 0\\
        m_D & 0 & 0 & 0 & \dots & 0 & (M_J\mu_N-m_\lambda)/(\tilde\chi_N)^2\\
    \end{vmatrix}=0\,.$}
\end{equation}
We will denote these diagonal elements later on as $-C_n$.
We can factorise out $m_D$ so that
\begin{equation}
\resizebox{.9\hsize}{!}{$
    \begin{vmatrix}
        -\tilde\lambda &  \mathbf{1} & \mathbf{1} &\mathbf{1}  &\dots & \mathbf{1} & \mathbf{1}\\
        \mathbf{1} & (\Rho-\lambda)/(\chi_{0})^2 & 0 & 0 &\dots & 0 & 0\\
        \mathbf{1} & 0 & (\Rho+\rho_1-\lambda)/(\tilde\chi_1)^2 & 0 & \dots & 0 & 0\\
        \mathbf{1} & 0 & 0 & (\Rho-\rho_1-\lambda)/(\tilde\chi_1)^2 & \dots & 0 & 0\\
        \dots & \dots & \dots & \dots & \dots & \dots & \dots\\
        \mathbf{1} & 0 & 0 & 0 & \dots & (\Rho+\rho_N-\lambda)/(\tilde\chi_N)^2 & 0\\
        \mathbf{1} & 0 & 0 & 0 & \dots & 0 & (\Rho-\rho_N-\lambda)/(\tilde\chi_N)^2\\
    \end{vmatrix}=0\,,$}
\end{equation}
where we defined
\begin{align}
    &\lambda\equiv m_D^{-1}m_\lambda\,, &&\tilde\lambda\equiv m_D^{-T}m_\lambda\,, &&\Rho\equiv m_D^{-1}M_J\,, &&\rho_n\equiv m_D^{-1}\mu_n\,,
\end{align}
with $m^{-T}\equiv(m^T)^{-1}$.
Notice that now $\lambda$ is not diagonal anymore. 
The problem reduces to a matrix with a block shape
\begin{equation}
    \begin{vmatrix}
    \label{eq:middlestep}
        -S_{0} & \mathbf{1} & \mathbf{1} &\dots &\mathbf{1}&\mathbf{1}&\mathbf{1}\\
        \mathbf{1} & -S_1 & 0 & \dots&0 & 0&0\\
        \mathbf{1} & 0 &-S_2 &\dots &0& 0&0\\
        \dots & \dots & \dots & \dots & \dots& \dots& \dots \\
        \mathbf{1} & 0  &0&\dots&-S_{N-2} & 0&0\\
        \mathbf{1} & 0 & 0 & \dots&0&-S_{N-1} & 0\\
        \mathbf{1} & 0 & 0 & \dots&0&0 & -S_{N}
    \end{vmatrix}=0\,.
\end{equation}
The goal is to make the matrix upper triangular. We start by subtracting from each block-row (starting from the top) the block-row below; this leaves us with
\begin{equation}
    \begin{vmatrix}
        -S_{0}\equiv T_0 & \mathbf{1} & \mathbf{1} &\dots &\mathbf{1}&\mathbf{1}&\mathbf{1}\\
        0 & -S_1 & S_2 & \dots&0 & 0&0\\
        0 & 0 &-S_2 &\dots &0& 0&0\\
        \dots & \dots & \dots & \dots & \dots& \dots& \dots \\
        0 & 0  &0&\dots&-S_{N-2} & S_{N-1}&0\\
        0 & 0 & 0 & \dots&0&-S_{N-1} & S_{N}\\
        \mathbf{1} & 0 & 0 & \dots&0&0 & -S_{N}
    \end{vmatrix}=0\,.
\end{equation}
Assuming that each $S_n$ is invertible (and thus $\det(S_n)\neq 0$, we can multiply to the right the first block-column by $S_N$ and add the last block-column to get
\begin{equation}
    \begin{vmatrix}
        T_0S_N+\mathbf{1}\equiv T_1 & \mathbf{1} & \mathbf{1} &\dots &\mathbf{1}&\mathbf{1}&\mathbf{1}\\
        0 & -S_1 & S_2 & \dots&0 & 0&0\\
        0 & 0 &-S_2 &\dots &0& 0&0\\
        \dots & \dots & \dots & \dots & \dots& \dots& \dots \\
        0 & 0  &0&\dots&-S_{N-2} & S_{N-1}&0\\
        S_N & 0 & 0 & \dots&0&-S_{N-1} & S_{N}\\
        0 & 0 & 0 & \dots&0&0 & -S_{N}
    \end{vmatrix}=0\,.
\end{equation}
We do the same, but this time multiplying by $S_N^{-1}S_{N_1}$
\begin{equation}
    \begin{vmatrix}
        T_1S_N^{-1}S_{N-1}+\mathbf{1}\equiv T_2 & \mathbf{1} & \mathbf{1} &\dots &\mathbf{1}&\mathbf{1}&\mathbf{1}\\
        0 & -S_1 & S_2 & \dots&0 & 0&0\\
        0 & 0 &-S_2 &\dots &0& 0&0\\
        \dots & \dots & \dots & \dots & \dots& \dots& \dots \\
        S_{N-1} & 0  &0&\dots&-S_{N-2} & S_{N-1}&0\\
        0 & 0 & 0 & \dots&0&-S_{N-1} & S_{N}\\
        0 & 0 & 0 & \dots&0&0 & -S_{N}
    \end{vmatrix}=0\,.
\end{equation}
We therefore find a recursive relation
\begin{align}
    &T_n=T_{n-1}S_{N-n+2}^{-1}S_{N-n+1}+\mathbf{1}\,, n\in[2,N]\,.
\end{align}
All in all
\begin{equation}
    T_N=\mathbf{1}+T_{N-1}S_{2}^{-1}S_{1}=\mathbf{1}+S_{2}^{-1}S_{1}+T_{N-2}S_{3}^{-1}S_{1}\,,
\end{equation}
and therefore
\begin{equation}
    T_N=\left(\sum\limits_{n=1}^{N}S_{n}^{-1}-S_0\right)S_1\,.
\end{equation}
We are left with the matrix
\begin{equation}
    \begin{vmatrix}
        T_N & \mathbf{1} & \mathbf{1} &\dots &\mathbf{1}&\mathbf{1}&\mathbf{1}\\
        0 & -S_1 & S_2 & \dots&0 & 0&0\\
        0 & 0 &-S_2 &\dots &0& 0&0\\
        \dots & \dots & \dots & \dots & \dots& \dots& \dots \\
        0 & 0  &0&\dots&-S_{N-2} & S_{N-1}&0\\
        0 & 0 & 0 & \dots&0&-S_{N-1} & S_{N}\\
        0 & 0 & 0 & \dots&0&0 & -S_{N}
    \end{vmatrix}=0\,.
\end{equation}
Since the matrix is now block diagonal, one can compute the determinant as the product of the determinants of the blocks. However, by assumption $\det(S_n)\neq 0$, so this implies
\begin{equation}
    |T_N|=\left|\sum\limits_{n=1}^{N}S_{n}^{-1}-S_0\right|=0\,.
\end{equation}
The result matches formulas previously found in the case of scalar entries. Recall that
\begin{align}
    &S_n=m_D^{-1} C_n\,, &&S_n^{-1}=C_n^{-1}m_D\,, &&S_0=m_D^{-T}C_0=m_D^{-T}m_\lambda\,,
\end{align}
and hence
\begin{align}
    &\left|\sum\limits_{n=1}^{N}C_{n}^{-1}m_D-m_D^{-T}C_0\right|=0\,, &&\left|\sum\limits_{n=0}^{N}\frac{\tilde\chi^\ast_n{}^2}{m_\lambda-(M_J\pm \mu_n)}-m_D^{-T}m_\lambda m_D^{-1}\right|=0\,,
\end{align}
where the $\ast$ is to remind to distinguish $\chi_0$ and $\tilde\chi_n$. In compact notation, this reads
\begin{align}
    &\left|\sum\limits_{n=0}^{N}m_D^T\frac{\chi_n^2}{(m_\lambda-M_J)^2- \mu_n^2} m_D-\frac{m_\lambda}{m_\lambda-M_J}\right|=0\,.
\end{align}

\subsection{Eigenvectors}
\label{app:3eigenvectors}
One can obtain the eigenvectors by reducing the problem to the matrix shown in Eq.~\eqref{eq:middlestep}.  One can then show that the eigenvectors read
\begin{equation}
    \Vec{u}_\lambda=\mathcal{N}_\lambda\begin{pmatrix}
        u_0\\
        \frac{\chi^\ast_i}{m_\lambda-M_i^\ast} m_D u_0
    \end{pmatrix}=\mathcal{N}_\lambda\begin{pmatrix}
        u_0\\
        \frac{\chi_0}{m_\lambda-M} m_D u_0\\
        \frac{\chi_1/\sqrt{2}}{m_\lambda-(M_J-\mu_1)} m_D u_0\\
        \frac{\chi_1/\sqrt{2}}{m_\lambda-(M_J+\mu_1)} m_D u_0\\
        \dots
    \end{pmatrix}\,,
\end{equation}
where when shortened the notation and added the $\ast$ to highlight the need to sum over all modes with appropriate $\chi_i$ and masses. The vector $u_0$ is constrained by
\begin{equation}
    \left(\sum\limits_{i=1}^Nm_D^T\frac{\chi_i^\ast{}^2}{m_\lambda-M_i^\ast}m_D-m_\lambda\right)u_0=0\,.
\end{equation}
The equation can be written by opening the modes as
\begin{equation}
    \left[m_D^T\left(\frac{\chi_0^2}{m_\lambda-M_J}+(m_\lambda-M_J)\sum\limits_{n=1}^N\frac{\chi_n^2}{(m_\lambda-M_J)^2-\mu_n^2}\right)m_D-m_\lambda\right]u_0=0\,.
\end{equation}

Finally, $\mathcal{N}_\lambda$ is a normalisation vector such that
\begin{equation}
    \mathcal{N}_\lambda\equiv\left[1+u_0^T m_D^T\sum\limits_{i=1}^N\left(\frac{\chi_i^\ast}{m_\lambda-M_i^\ast}\right)^2m_Du_0\right]^{-1/2}\,,
\end{equation}
where, for convenience, we chose $u_0^Tu_0=1$.
Per flavour, one finds
\begin{equation}
    \mathcal{N}_\lambda=\left[ \frac{m_D^2\pi ^2}{\mu_1^2}+\frac{m_\lambda^2}{m_D^2}+1\right]^{-1/2}\,,
\end{equation}
and we recover the results derived in the main text.

In the case $M_J=0$, we recover the results previously found for the single-flavour case
\begin{equation}
      \left[m_D^T\left(\sum\limits_{n=0}^N\frac{\chi_n^2}{m_\lambda^2-\mu_n^2}\right)m_D-1\right]u_0=0\,.
\end{equation}

\footnotesize

\bibliographystyle{BiblioStyle}
\bibliography{Bibliography}
\end{document}